\documentclass{JHEP3}
\usepackage{amsfonts}
\usepackage{amsmath}
\usepackage{amssymb}
\usepackage{latexsym}
\usepackage{graphicx}
\usepackage{cite}
 
\newcommand\no{\nonumber\\{}}

\newcommand\eqnb{\begin{eqnarray}}
\newcommand\eqne{\end{eqnarray}}

\newcommand\void[1]{}   
\newcommand{\Tr}{{\mathrm{Tr}}}

\usepackage{latexsym}
\numberwithin{equation}{section}

\def\cN{\mathcal{N}}
\def\cD{\mathcal{D}}

\def\cF{\mathcal{F}}
\def\cO{\mathcal{O}}

\def\cR{\mathcal{R}}

\newcommand{\be}{\begin{equation}}
\newcommand{\ee}{\end{equation}}
\newcommand{\bea}{\begin{eqnarray}}
\newcommand{\eea}{\end{eqnarray}}

\preprint{ DAMTP-2010-65}

\title{Multi-loop amplitudes in maximally supersymmetric pure spinor field theory}

\author{Jonas Bj\"ornsson\\
Department of Applied Mathematics and
Theoretical Physics\\
Wilberforce Road, Cambridge CB3 0WA, UK\\
\email{\tt j.bjornsson@damtp.cam.ac.uk}}

\abstract{ 

This paper provides a more detailed background to the results of  {\tt arXiv:1004.2692} concerning properties of multi-loop amplitudes in a pure spinor formulation of field  theories with maximal supersymmetry. This involves the development of a first quantised field theory version of the non-minimal pure spinor formalism originally designed for describing superstring amplitudes. In addition to superspace world-line fields, the formalism involves a set of world-line ghost fields that are required for implementing BRST invariance with a composite $b$ ghost. In particular, we show that BRST invariance requires the presence of certain contact terms. For four-point amplitudes, these are important beyond two loops. We also present an alternative proof of the ``no-triangle hypothesis" and the vanishing of amplitudes with fewer than four external particles.

}


\begin{document}

%\maketitle

%%%%%%%%%%%%%%%%%%%%%%%%%
\section{Introduction and overview}
\label{sec:intro}
%%%%%%%%%%%%%%%%%%%%%%%%%

Maximally supersymmetric quantum field theories have been the subject of great interest in recent times. Although both maximal Yang--Mills (which has a 16-component supercharge) and maximal supergravity (which has a 32-component supercharge)  are not of direct experimental relevance their special properties make them of considerable theoretical interest.  Much information has been obtained, in particular, concerning the perturbative expansions of scattering amplitudes in these theories.  Thus, four-dimensional $\mathcal{N}=4$ Yang--Mills theory has long been known to be free of ultraviolet divergences to all orders in perturbation theory  and is integrable in the planar limit (the large-$N_c$ limit with $SU(N_c)$ ¤gauge group). It is of interest to determine the behaviour of contributions to the amplitude that are sub-leading in the large-$N_c$ limit. Maximal supergravity, which is the low energy limit of M-theory, is non-renormalisable and superficially possesses ultraviolet divergences. However, explicit calculations of the four-graviton amplitude in four-dimensional $\mathcal{N}=8$ supergravity demonstrate that there are no ultraviolet divergences  up to four loops \cite{Green:1982sw,Bern:1998ug,Bern:2007hh,Bern:2009kd}. It is a challenge to determine the order in the perturbation expansion at which UV divergences first appear.

Although sophisticated techniques have been used to address these perturbative field theory issues, such methods generally fail to use the full power of space-time supersymmetry, which severely constrains the structure of the amplitudes.   However, supersymmetry is manifest in the world-line pure spinor formalism recently introduced in \cite{Bjornsson:2010wm}. This is a field theory formalism that is modelled on the pure spinor formalism introduced by  Berkovits in order to quantise the superstring in a manifestly supersymmetric manner \cite{Berkovits:2000fe,Berkovits:2005bt}. This was used in \cite{Bjornsson:2010wm} to demonstrate how supersymmetry restricts the topology of the perturbative diagrams for four-particle amplitudes in maximally supersymmetric Yang--Mills and supergravity in general space-time dimensions. It provides a very efficient procedure for determining the degree of ultraviolet divergence order by order in perturbation theory, reproducing the results up to four loops that were obtained by more standard means for the Yang--Mills case in \cite{Green:1982sw,Bern:1997nh,Bern:2007hh,Bern:2009kd,Dixon:2009tk,Bern:2010tq} and the supergravity case in \cite{Green:1982sw,Bern:1998ug,Bern:2007hh,Bern:2009kd}. Since supersymmetry is manifest at every stage, the results follow without encountering any subtle cancellations between different diagrams.

Furthermore this pure spinor world-line formalism is able to pinpoint the diagrams at higher loops that contribute to the leading ultraviolet divergence.  Particularly intriguing are the indications of a five-loop contribution to the $\partial^8 R^4$ interaction, which is the first explicit indication that this interaction is not protected by supersymmetry. This is in line with many other arguments \cite{Howe:1980th,Vanhove:2010aa,Green:2010sp,BerkovitsICTP,Bossard:2009mn,Elvang:2010jv,Elvang:2010kc,Drummond:2010fp,Beisert:2010jx,Bossard:2010bd}, which indicate that there should be a logarithmic divergence of this form in $\mathcal{N}=8$ supergravity in $D=4$ dimensions, which would there arise at seven loops. However, since the coefficient of this divergence was not evaluated in \cite{Bjornsson:2010wm} it might possibly vanish, but this would require a cancellation that cannot be explained by conventional effects of supersymmetry. 

The world-line formalism was presented in  \cite{Bjornsson:2010wm}  as a series of seemingly ad hoc rules that were based on corresponding rules in the non-minimal pure spinor  string world-sheet formalism.  The main purpose of this paper is to present the formalism in a more coherent manner and demonstrate the consistency of the approach.  Most notably, we will study the BRST invariance of multi-loop amplitudes and in this manner we will determine the rules in a much more precise manner.  Among other things, this will determine the way in which certain contact terms arise in higher-loop diagrams.  The results of  \cite{Bjornsson:2010wm} will also be reviewed.

\subsection{Outline of paper}

In order to motivate the structure of the pure spinor world-line formalism we will first, in section~\ref{sec:firstquant}, describe the world-line formalism for perturbative scalar field theory with cubic vertices. This will generalise the discussion of \cite{Dai:2006vj} to also include the $(b,c)$ ghost system. The rules for constructing vertices and gluing them together with propagators, and the r{\^{o}}le of the $b$ ghost will provide guidance for the later construction of the pure spinor amplitudes.

In section \ref{sec:purespinor} we consider maximally supersymmetric Yang--Mills and gravity, modelled on the non-minimal formalism of the pure spinor string \cite{Berkovits:2005bt}.  This is based on the dynamics of the classical superspace coordinates ($X^m, \theta^\alpha$) and their conjugate momenta ($P_m,p_\alpha$) in a fixed gauge.  However, the standard  $(b,c)$ ghosts are absent and instead there are a number of other bosonic spinor coordinates, $\lambda^{\alpha}$ and $\bar{\lambda}_{\alpha}$ and their conjugate momenta, $w_{\alpha}$ and $\bar{w}^{\alpha}$, as well as fermionic spinor coordinates $r_{\alpha}$ and their conjugate momenta $s_{\alpha}$. These additional world-line fields satisfy ten-dimensional pure spinor constraints.  Since the formulation of the theory is based on the string, the theory uses the terminology of ten-dimensional supersymmetry although the expressions may be evaluated in  an arbitrary space-time dimension reached by supersymmetric dimensional continuation. Although there are no $(b,c)$ ghosts,  we will be able to construct a composite $b$ ghost by mimicking the string construction by Berkovits in \cite{Berkovits:2005bt}. We also construct the three-point vertices that describe how off-shell states absorbs a physical state, which are basically the point particle versions of the string vertices in \cite{Berkovits:2000fe}.

The amplitude prescriptions of Yang--Mills and supergravity are introduced in section~\ref{sec:defamp}. The Yang--Mills propagator involves an insertion of the composite $b$ ghost and regulator. The off-shell Yang--Mills vertex is constructed by imposing locality, which is compatible with BRST invariance. The vertices are glued together with propagators to construct amplitudes. The construction of the supergravity amplitude mimics that of the closed string, doubling the world-line fields (apart from $X$ and $P$). For example, the propagator involves two insertions of the $b$ ghost (one for each sector of the theory).

In section~\ref{sec:BRST} we consider the BRST properties of these amplitudes. Central to the discussion are four-point amplitudes (constructed from three-point vertices) where states can be off-shell. We obtain amplitudes that are compatible with BRST invariance if the three channels of the four-point amplitude are included in all sub-diagrams. 

Important features of the multi-loop amplitudes are encoded in the zero modes of the world-line fields. In section \ref{sec:propamp} we will enlarge the discussion of \cite{Bjornsson:2010wm} concerning the saturation of fermionic zero modes and the need for a regulator to deal with large-$\lambda$ divergencies. This regulator is closely related to the one used in pure spinor string theory \cite{Berkovits:2005bt}.

In section \ref{sec:four-point} we review the ultraviolet properties of the four-point amplitudes that were presented in \cite{Bjornsson:2010wm}. Furthermore, we enlarge the discussion and obtain the one- and two-loop amplitude up to an overall constant using \cite{Berkovits:2005ng,Mafra:2008ar} (see also \cite{Berkovits:2005df,Mafra:2009wq,Berkovits:2006bk}). For the one-loop case, we also study $N$-point amplitudes and give an alternative proof of the ``no-triangle hypothesis'' of supergravity \cite{BjerrumBohr:2006yw,BjerrumBohr:2008vc,BjerrumBohr:2008ji,ArkaniHamed:2008gz}. A consequence of the theorem is that there are no sub-diagrams with bubbles or triangles in supergravity (and Yang--Mills) at any loop. This property is a manifest consequence of maximal supersymmetry. We also show that loop amplitudes with fewer than four points vanish.

In section \ref{sec:disc}, we summarise the main points and briefly consider the connection between the first- and second-quantised pure spinor approaches to theories with maximal supersymmetry. In the appendix the equations of motion of the different component fields are shown to follow from the on-shell constraints and Bianchi identities for Yang--Mills in $D=10$.

%%%%%%%%%%%%%%%%%%%%%%%%%
\section{First-quantised scalar field theory}
\label{sec:firstquant}
%%%%%%%%%%%%%%%%%%%%%%%%%

Before treating the pure spinor case, we will discuss scalar field theory in the first-quantised, or world-line, approach with cubic vertices.  We will obtain the amplitude prescription of the theory by considering free propagation of particles and constructing the three-point vertex. We will extend the discussion of \cite{Dai:2006vj} to include the $(b,c)$  world-line ghosts. The approach to computing loop amplitudes for the scalar theory  may be modelled on bosonic string theory. Although the ghost system in this case has rather trivial effects, it motivates the later use of a $b$ ghost when considering the pure spinor particle.  The relationship between the structure of loop amplitudes in bosonic string theory and the pure spinor string, through the $\cN=2$ topological string \cite{Berkovits:2005bt},  will be used in later sections to motivate a set of rules for constructing loop amplitudes for the pure spinor particle.
 
The starting point is the reparameterisation invariant action for a massless scalar particle\footnote{We will in this paper work in Euclidian signature, by analytically continue from Lorentzian signature.}
\eqnb
S_B &=& \int d\tau \frac{\dot{X}^2}{2e}\,,
\label{eq:SBaction}
\eqne
where $X^m$ are the classical coordinate fields in $D$-dimensions, $m=1,\ldots,D$. The action is invariant under reparameterisations of the world-line, $\tau \rightarrow \tau + \sigma\left(\tau\right)$ 
\eqnb
\dot{X}^m &\rightarrow& \dot{X}^m\left(1+\partial_\tau\sigma\left(\tau\right)\right) \no
e &\rightarrow& e\left(1+\partial_\tau\sigma\left(\tau\right)\right)\,.
\eqne
where $\sigma(\tau)$ and $\partial_\tau \sigma(\tau)$ are infinitesimal. The canonical momentum is defined in the usual way
\eqnb
P_m = \frac{\delta S}{\delta \dot{X}^m} 	= \frac{\dot{X}_m}{e}\,.
\eqne
From the definition of the momenta one can determine the action in phase-space
\eqnb
S_B &=& \int d\tau\left( P\dot{X}-\frac{e}{2}P^2\right)\,.
\label{Actionscalar}
\eqne
Using standard methods, one can obtain the contraction between two $X$'s in, for example\footnote{In this and forthcoming formulas, an overall normalisation will be ignored.}, the gauge $e=1$,
\eqnb
\langle X^m(\tau)X^n(\tau')\rangle &=& \delta^{mn}G(\tau,\tau')
\label{Greensfunction},
\eqne
and the equal time commutator
\eqnb
\left[P^m, X_n\right] &=& -{\delta^m}_n\,.
\eqne
The Green function, $G(\tau,\tau')$ in (\ref{Greensfunction}), satisfies
\eqnb
\partial^{2}_\tau G(\tau,\tau')
	&=&
		-\delta(\tau-\tau') + \rho,
\label{DefeqnGreen}
\eqne
where the constant $\rho$ arises on a compact ``skeleton'' diagram in order to cancel a zero mode of the delta function.

We need to take care of the infinite dimensional gauge group generated by the reparameterisation invariance. This can be achieved in a covariant way by introducing a set of fermionic $(b,c)$ ghosts. Consider therefore the partition function 
\eqnb
Z[J_{m}]
	&=&
		\int \frac{\mathcal{D}X \mathcal{D}P \mathcal{D}e}{V_{rep}} e^{-S_B + \int d\tau\,  X^{m} J_{m} },
\eqne
where $V_{rep}$ is the volume of the gauge group.  The moduli dependence is incorporated in a world-line einbein $e$. The integration over this field can be written as $\mathcal{D}e=dT \mathcal{D}\sigma$, where $\mathcal{D}\sigma$ is the integration over the reparameterisation group and $dT$ the integration over the moduli. Fixing the gauge $e=T$ and parameterising the line as $0\leq \tau\leq 1$ one gets
\eqnb
Z[J_{m}]
	&=&
		\int dT \int \mathcal{D}X\, \mathcal{D}P\, \mathcal{D}b\, \mathcal{D}c
		\left(\int_0^{1} d\tau\, b \partial_{T}\,\left(e\right)\right)\; e^{-S_B-\int d\tau\, T\, b\dot {c} + \int d\tau\,  X^{m} J_{m}}
	\no
	&=&
	\int dT \int\mathcal{D}X \mathcal{D}P\,\mathcal{D}b\, \mathcal{D}c\, \left(\int_0^1 d\tau\, b\right) \, e^{-S_B-\int d\tau\, T\, b\dot {c} + \int d\tau\,  X^{m} J_{m}}.
\eqne
Later expressions are simplified by making a reparameterisation of the line so that the dependence on the modulus is incorporated in the integration limits
\eqnb
Z[J_{m}]
	&=&
		\int dT \int\mathcal{D}X \mathcal{D}P\,\mathcal{D}b\, \mathcal{D}c\, \left(\int_{0}^{T} \frac{d\tau} {T}\, b\right) \, e^{-S_B-S_{gh} + \int d\tau\,  X^{m} J_{m}},
\label{Partitionscalar}
\eqne
where $S_{gh} \equiv \int d\tau\, b\dot {c}$. From the action one can obtain the equal time commutator\footnote{$[,]$ denotes the graded commutator.}
\eqnb
[\hat{b},\hat{c}] &=& 1,
\eqne
where $\hat{b},\hat{c}$ are fermionic operators\footnote{In the following the hats will be suppressed when this cause no confusion}. The $(b,c)$ ghosts can be thought of as the zero mode components  of the ghost system for the bosonic string. The vacuum is two-fold degenerate and satisfies
\eqnb
c\left|\downarrow\right> = \left|\uparrow\right> &\,& c\left|\uparrow\right>=0 \no
b\left|\uparrow\right> = \left|\downarrow\right> &\,& b\left|\downarrow\right> =0 \no
\left<\uparrow\!\left|\downarrow\right>\right. = 1.
\eqne
The theory is invariant under BRST transformations generated by a BRST charge proportional to the Hamiltonian 
\eqnb
Q 
	&=& 
		cH \no
	&=&
		\frac{1}{2}cP^2.
\label{BRSTscalar}
\eqne
The states in the cohomology of (\ref{BRSTscalar}) are
\eqnb
\left|k^m,\downarrow\right> &=& e^{ikX}\left|0,\downarrow\right> \;\; k^2 = 0 \no
\left|k^m,\uparrow\right> &=& e^{ikX}\left|0,\uparrow\right>,
\eqne
where the second set of states is projected out using the physical state condition $b\left|phys.\right>=0$. Observe that since we are in an Euclidian signature the eigenvalue of $P^m$ is purely imaginary\footnote{The sign is chosen to match the theory in Lorentzian signature.}
\eqnb
P^m\left|k^m,\downarrow\right> 
	&=&
		-ik^m\left|k^m,\downarrow\right>.
\eqne
The eigenstates above have been described in the Schr{\"o}dinger picture, where operators and eigenstates are time independent. The operators and eigenstates in Heisenberg picture are related by a canonical transformation
\eqnb
\left|A;\tau\right>
	&=&
		e^{-\tau H}\left|A\right>
	\no
A(\tau)
	&=&
		e^{-\tau H} A e^{\tau H}\, .
\eqne

As the Hamiltonian for the ghost system is zero, the ghosts are locally constant and only depend on the order in which the ghosts are inserted on the line. Therefore, integration over ghosts reduces to a straightforward Grassman integral. Furthermore, we will find that although the  ghost insertions  have rather trivial consequences in the case of  the scalar particle they demonstrate the manner in which the number of $b$ ghost insertions is related to  the number of integrated and unintegrated vertex operators. This will be of use later in our discussion of the pure spinor particle. We need to determine the propagator which map a state at proper time $0$ to proper time $T$. The propagator can be obtained from the partition function in (\ref{Partitionscalar}) by evaluating the integrand of the modulus $T$ between two momentum eigenstates and setting $J_m =0$
\eqnb
P\left(p_f,b_f,p_i,b_i;T\right)
	&=&
		\left<p_f,b_f\left|\int_0^T \frac{d\tau}{T} \hat{b}\, e^{T\hat{H}}\right|p_i,b_i\right>
	\no
	&=&
		\left<p_f,b_f;T\left|\, \hat{b}\,\right|p_i,b_i;0\right>
	\no
	&=&
		\delta\left(b_f-b_i\right)\delta\left(p_f-p_i\right) b_i \, e^{-\frac{T}{2}p_i^2}\, ,
\label{propagatorscalar}
\eqne
where we have introduced the state $\left|b\right> = \left|\downarrow\right> + \left|\uparrow\right> b$ and its dual $\left<b\right| = b\left<\uparrow\right| - \left<\downarrow\right|$. Note that these states satisfy
\eqnb
\hat{b}\left|b\right> &=& \left|b\right>b \no
\left<b\right|\hat{b} &=& -b\left<b\right|  \no
\left.\left<b'\right|b\right> &=& b' - b = \delta(b'-b)\,.
\eqne
As an operator, the propagator is\footnote{In the Schr{\"o}dinger picture it is $\mathcal{P}\left(T\right) = \hat{b}e^{T\hat{H}}.$}
\eqnb
\mathcal{P}\left(T\right) &=& \hat{b}
\label{propagatorscalarII}
\eqne
which demonstrates that each propagator has a $b$ insertion. This $b$ ghost is connected with the existence of one modulus,  which is the length of the propagator.

\begin{figure}[t]
\centerline{\rotatebox{0}{\scalebox{0.4}{\includegraphics*{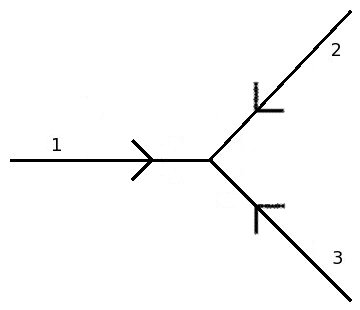}}}}
\caption{The scalar vertex $\left<V_B;\tau\right|$ where the arrows on each line indicate the direction of increasing proper time.}
\label{fig:3vertex}
\end{figure}

The construction of multi-loop amplitudes to be considered later, will involve attaching external vertex operators to propagators in skeleton (or vacuum) diagrams. Each external vertex operator describes absorption of an external physical state carrying momentum $k$ ($k^2 = 0$) by a state propagating in the skeleton. These operators can be obtained from a three particle off-shell vertex $\left<V_B;\tau\right|$ from which the physical three-point amplitude is obtained by contracting with three physical states. Consider now the construction of the three particle off-shell vertex. A natural condition to impose is locality so that the interaction takes place at one point in space-time. This interaction point is integrated leading to conservation of momentum at the vertex. One additional condition is that the vertex should be BRST invariant. Thus, the BRST charge should be conserved at the vertex
\eqnb
\sum_{j=1}^{3} \left<V_B;\tau\right| Q^j  &=& 0\, ,
\eqne
where $Q^{j}$ is the BRST charge acting on the $j$'th leg. In the above equation we have used the sign convention in figure \ref{fig:3vertex}, where arrows denote increasing proper time and the index on the BRST charge labels the different legs of the vertex.

As with the physical states, these conditions are not sufficient  to construct the vertex and one has to impose additional conditions on the ghost coordinates,
\eqnb
\left<V_B;\tau\right| c^j   &=& 0, \phantom{123} j=1,\ldots, 3,
\eqne
which constrain the vertex to be in the up-state on all legs. This implies that $\left<V_B;\tau\right|Q^j  = 0$ for each $j$ separately.

\begin{figure}[t]
\begin{center}
\begin{tabular}{cc}
\rotatebox{0}{\scalebox{0.35}{\includegraphics*{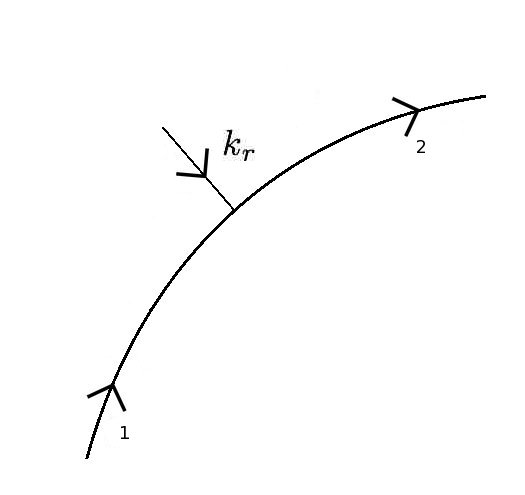}}} &
\rotatebox{0}{\scalebox{0.35}{\includegraphics*{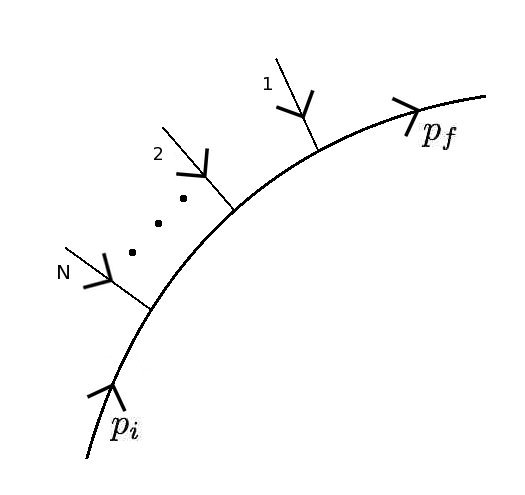}}} \\
(a) & (b)
\end{tabular}
\end{center}
\caption{(a) The function describing absorption of one physical state with momentum $k_r$ ($k^2$) of the three-point vertex. (b) The propagator absorbing $N$ physical particles evaluated between two general states with momenta $p_i$ and $p_f$.}
\label{fig:1absorb}
\end{figure}

The vertex operator describing a physical state of momentum $k_r$ ($k_r^2 = 0$) that is attached to a propagator in a skeleton diagram is obtained by contracting one of the legs of $\left<V_B;\tau\right|$ by $\left|k_r,\downarrow;\tau\right>$  (see figure~\ref{fig:1absorb}), giving
\eqnb
\left.\left<V_B;\tau\right|k_r,\downarrow;\tau\right>
	&=&
		\int \left|x,\uparrow;\tau\right>\;d^Dx\, e^{ik_rx(\tau)}\,\left<x,\uparrow;\tau\right| 
	\no
	&=& 
\int\left|x,c;\tau\right>\;d^Dx\;c\;dc\,e^{ik_rx(\tau)}\left<x,c;\tau\right|,
	\no
	&=&
	\hat{c}e^{ik_rX(\tau)}\int\left|x,c;\tau\right>\;d^Dx\;dc\,\left<x,c;\tau\right|,
	\no
	&=&
	U_0(k_r,\tau)I,
\label{unintscalar}
\eqne
where $I$ is the identity operator. In this expression we have used
\eqnb
\left<x, \uparrow;\tau\left| k_r,\downarrow;\tau\right>\right.
	&=&
		e^{ik_rx},
\eqne
and introduced the state $\left|c\right> = \left|\uparrow\right>+\left|\downarrow\right>c$ and its dual $\left<c\right| = c\left<\downarrow\right|-\left<\uparrow\right|$ satisfying 
\eqnb
\hat{c}\left|c\right> &=& \left|c\right>c \no
\left<c\right|\hat{c} &=& -c\left<c\right| \no
\left<c'\left|\,c\right>\right. &=& \delta(c'-c) = c'- c\, .
\eqne
The operator $U_0(k_r,\tau)$ in (\ref{unintscalar}) is
\eqnb
U_0\left(k_r,\tau\right) = \hat{c}e^{ik_rX(\tau)},
\label{UIscalar}
\eqne
which satisfies $[Q,U_0(k_r,\tau)]=0$.

The integrated vertex operator can be obtained from $U_0(k_r,\tau)$ by attaching propagators to the legs one and three in figure~\ref{fig:1absorb},
\eqnb
\int_{0}^{T} d\tau\,\mathcal{P}(T-\tau)U_0(k_r,\tau)\mathcal{P}(\tau)
	&=&
		\int_{0}^{T}d\tau\, b\, c\,e^{ik_rX(\tau)} b
	\no
	&=&
		\int_{0}^{T}d\tau\, V_0(k_r,\tau)\ b\, .
\label{intVertex}
\eqne
We have now constructed the integrated vertex operator
\eqnb
V_0(k,\tau)
	&=&
		[b,U_0(k,\tau)]
	\no
	&=&
		e^{ikX(\tau)}\, .
\eqne
The integrated vertex also satisfies $[Q,V_0(k,\tau)] = [H,U_0(k,\tau)]$. 

This generalises to the situation in which several external states are absorbed by a propagator in the skeleton. For example, $N$ vertex operators attached to a line of length $T$ in a given order,
\eqnb
\prod_{r=1}^{N}\int_0^{\tau_{r-1}}\!\!\!\!\!d\tau_r\, \mathfrak{B}_j\left(\{k_r\},\{\tau_r\};T\right)
	&=&
		\prod_{r=1}^{N}\int_0^{\tau_{r-1}}\!\!\!\!\!d\tau_r\,\mathcal{P}(T - \tau_{1})\prod_{r=1}^{N}U_0(\tau_r,k_r)\mathcal{P}(\tau_r-\tau_{r+1})
	\no
	&=&
		\prod_{r=1}^{N}\int_0^{\tau_{r-1}}\!\!\!\!\!d\tau_r\, b\, \prod_{r=1}^{N}\left( c\,e^{ikX(\tau_r)} b\right)
	\no
	&=&
		\prod_{r=1}^{N}\int_0^{\tau_{r-1}}\!\!\!\!\!d\tau_r\,\prod_{r=1}^{N}e^{ikX(\tau_r)}\; b\, 
	\no
	&=&
		\prod_{r=1}^{N}\int_0^{\tau_{r-1}}\!\!\!\!\!d\tau_r\,\prod_{r=1}^{N}V_0(\tau_r,k_r)\; b\, ,
\label{Nexternal}
\eqne
where $\tau_{N+1}=0$ and $\tau_0=T$. In the picture where one use the integrated form of the vertices, one can integrate over the whole line, which describe all different distributions of the external vertices.

In constructing loop diagrams it will be useful to define a basic bolding block, $B_j$, describing an internal line, labelled $j$, of length $T_j$ to which $N$ vertex operators are attached at position $\tau_1,\ldots,\tau_N$. This is defined by the matrix element of the integrand of (\ref{Nexternal}) between two general states of momenta $p_i$ and $p_f$ with ghost content $b_i$ and $b_f$,
\eqnb
B_j\left(p_f,b_f,p_i,b_i,\{k_r\},\{\tau_r\};T_j\right)
	&=&
		\left<p_f,b_f;T_j\right|\prod_{r=1}^{N}V_0(\tau_r,k_r)b\left|p_i,b_i;0\right>.
\label{scalarIbuild}	
\eqne
For the corresponding expression in the unintegrated picture, we will choose a different parameterisation. The position of the $r$'th vertex operator is parameterised by $\sum_{s=1}^{r}\tau_s$ and the length of the propagator between the two general momentum states is $\sum_{s=1}^{N+1}\tau_s$. Then the corresponding expression in the unintegrated picture is
\eqnb
B_j\left(p_f,b_f,p_i,b_j,\{k_r\},\{\tau_s\}\right)
	&=&
		\left<p_f,b_f;\sum_{s=1}^{N+1}\tau_s\left|b\, \prod_{r=1}^{N}\left\{U_0\left(\sum_{s=1}^{r}\tau_s,k_{r}\right)\, b\right\}\right|p_i,b_i;0\right>\,. \;\;\;\;\;\;\;\;
\label{scalarUIbuild}
\eqne
Using this parameterisation will make the integration over the moduli more transparent.

We will now give a general discussion of $N$-point amplitudes with $L$ loops.  The computation can be split up into several steps.  First one chooses a particular $L$-loop skeleton diagram, denoted $F_L$, and performs the functional integral with fixed moduli and fixed positions of the external vertices on the lines of the skeleton
\eqnb
I_{F_L}\left(\{k_r\},T_j,\tau_r\right)
&\equiv&
\left<\prod_{r=1}^{N'}V_{0}\left(k_r,\tau_r\right)\prod_{r=N'+1}^{N}U_{0}\left(k_r,\tau_r\right)\right>_{F_L}
\no
&\equiv&
\int \cD X\, \cD P\, \cD c \, \cD b\,\prod_{j=1}^{M} b^j \, \prod_{r=1}^{N'}V_{0}\left(k_r,\tau_r\right)\prod_{r=N'+1}^{N}U_{0} e^{-S_B-S_{gf}},
\label{Ampscalarintegrand}
\eqne
where $N'$ is the number of integrated vertex operators, $N-N'$ is the number of unintegrated vertex operators\footnote{The tree diagrams are special since three-point vertices couple to one or two external states (or three for the three-point amplitude).} and $M$ is the number of moduli. The result of the functional integral is a function, $I_{F_L}$, that is to be integrated over the positions of the integrated vertices and the moduli. This gives the amplitude corresponding to a particular skeleton
\eqnb
\hspace{-.5cm}
A^{(F_L)}_{0}\left(s_{ij}\right)
&\equiv&
\int dT_1\ldots dT_{M} \int_{F_L} \prod_{j=1}^{N'} d\tau_r
\,I_{F_L}\left(\{k_r\},T_j,\tau_r\right),
\label{AmpscalarI}
\eqne
where $\int_{F_L}d\tau_r \equiv \sum_{j=1}^{M} \int_0^{T_j} d\tau_r$ denotes the integral around the whole skeleton. Here $s_{ij}$ are the Mandelstam invariants which can be constructed from the $N$ external momenta. To get the full amplitude with $L$ loops one has to sum the different $L$-loop skeletons
\eqnb
A^{(L)}_0\left(s_{ij}\right)
	&\equiv&
		\sum_{F_L} A^{(F_L)}_{0}\left(s_{ij}\right)\, .
\eqne
The amplitude above is constructed using the picture where as many $b$ ghosts as possible has been contracted, the so-called integrated picture. One can do this in the unintegrated picture as well. In the following we will evaluate the functional integral (\ref{Ampscalarintegrand}). Amplitudes with more than one loop are constructed by gluing together $3L-3$ operators defined in (\ref{scalarIbuild}) with $2L-2$ internal vertices. The one-loop amplitude is special since the ends of the propagator are glued together using an external unintegrated vertex operator.

The functional integral is expressed in terms of the Green function by an expression of the standard form 
\eqnb
\left<\prod_{r=1}^{N'}V_{0}\left(k_r,\tau_r\right)\prod_{r=N'+1}^{N}U_{0}\left(k_r,\tau_r\right)\right>_{F_L}
	&=& 
		\left<1\right>_{F_L}\delta\left(\sum_{r=0}k_r\right)e^{-\sum_{r<s}k_rk_sG\left(\tau_r,\tau_s\right)},
\label{Greenscontr}
\eqne
where $\left<1\right>_{F_L}$ denote the expectation value of the skeleton amplitude.

\subsection{Scalar tree amplitudes}

The simplest amplitude to consider is the three-point tree amplitude. This is the vertex $\left<V_B;\tau \right|$ multiplied with three physical states. As the only non-trivial property of the vertex is momentum conservation, the amplitude equals
\eqnb
A^{(Tree)}_0(\{k_r\}) &=& \delta(\sum_{r=1}^{3} k_r).
\label{3vertexdelta1}
\eqne 
Using the equivalent prescription in (\ref{unintscalar}) one gets
\eqnb
A^{(Tree)}_0(\{k_r\}) &=& \int d^Dx \,dc \,c \,e^{i(\sum_{r=1}^{3} k_{r})x},
\label{3vertexdelta2}
\eqne
which clearly reduces to (\ref{3vertexdelta1}) and also gives some insight into the correspondence with the string expression. A crucial difference from string theory is that here there is only a single mode of $c$, where in string theory $c$ has three zero modes (corresponding to the three conformal killing vectors of the spherical world-sheet)

\begin{figure}[t]
\begin{center}
\begin{tabular}{ccc}
{\rotatebox{0}{\scalebox{0.35}{\includegraphics*{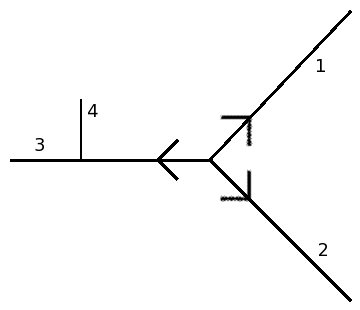}}}}
&
{\rotatebox{0}{\scalebox{0.35}{\includegraphics*{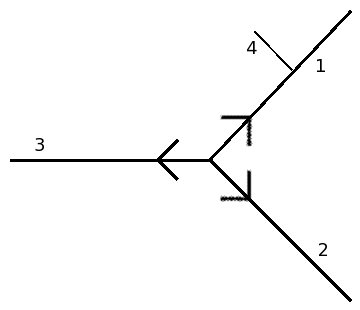}}}}
&
{\rotatebox{0}{\scalebox{0.35}{\includegraphics*{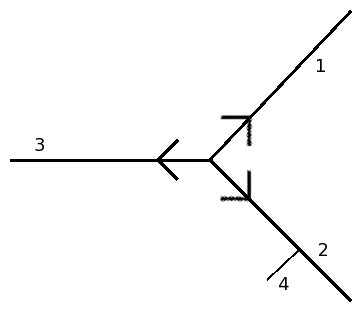}}}}
\\
(a) & (b) & (c)
\end{tabular}
\end{center}
\caption{The figures describing the $s$-, $t$- and $u$-channel of the four-point function. Arrows indicate increasing proper time and the numbers indicate the particle. Figure (a), (b) and (c) illustrate the $s$-, $t$- and $u$-channels, respectively.}
\label{fig:fourpointtree}
\end{figure}

Consider now the four-point tree amplitude. This can be thought of as a three-point amplitude to which a second vertex has been added. The position of the second vertex (involving the fourth particle) should be integrated over all the three legs of the three-point vertex (see figure \ref{fig:fourpointtree})
\eqnb
A^{(Tree)}_0\left(s,t,u\right)
&=&
\lim_{T_j\rightarrow \infty} \sum_{j=1}^{3} \int_0^{T_j} d\tau \left<0\right|U_0\left(k_1,T_1\right)V_0\left(k_2,T_2\right)V_0\left(k_3,T_3\right)V_0\left(k_4,\tau\right) \left|0\right>.
\no
\label{eq:4point}
\eqne
At the vertex, we have $\tau=0$. Observe that the vertices $V_0\left(k_2,T_2\right)$ and $V_0\left(k_3,T_3\right)$ should be thought of as unintegrated vertex operators without a $c$ ghost. In the above equation we have defined the Mandelstam invariants $s = -k_1\cdot k_2$, $t = -k_1\cdot k_4$ and $u = -k_1\cdot k_3$. These satisfy $s+t+u=0$ as the particles involved are massless. The world-line Green function for tree diagrams has the form
\eqnb
G_{Tree}\left(\tau,\tau'\right)
	&=&
		-\frac{1}{2}\left|\tau-\tau'\right|.
\eqne
Using this expression, the matrix element (\ref{eq:4point}) can be reduced straightforwardly to\footnote{As usual, we are ignoring overall constants.}
\eqnb
\no
A^{(Tree)}_0\left(s,t,u\right)
&=&
\int_0^{\infty} d\tau \int d^Dx\, dc\, c\, e^{i\left(k_1+k_2+k_3+k_4\right)x}\left[e^{\tau t}+e^{\tau u}+e^{\tau s}\right]
\no
&=&
-\delta\left(\sum_{r=1}^{4}k_r\right)\left(1/s+1/t+1/u\right)
\no
&=&
\frac{s^2+t^2+u^2}{2stu}\delta\left(\sum_{r=1}^{4}k_r\right).
\eqne
Note that in order for the $\tau$ integral to converge the three terms in the integrand have been defined by separate analytic continuations to  the regions $s<0$, $t<0$ and $u<0$. 

For tree amplitudes with more external particles, one can proceed iteratively. The $N$-point amplitude is constructed by adding a vertex to a $(N-1)$-point tree and summing over all non-equivalent possible positions of the vertex. As we are not interested in tree amplitudes with more than four external particles, we will not construct explicit expressions for these amplitudes.

\begin{figure}[t]
\begin{center}
\begin{tabular}{ccc}
{\rotatebox{0}{\scalebox{0.24}{\includegraphics*{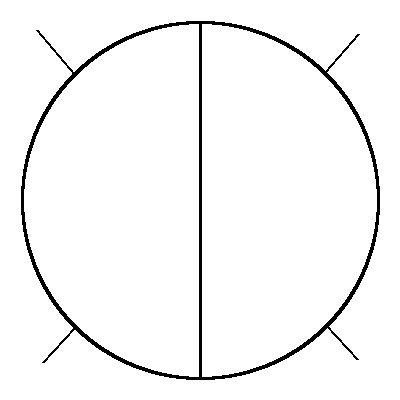}}}}
&
{\rotatebox{0}{\scalebox{0.30}{\includegraphics*{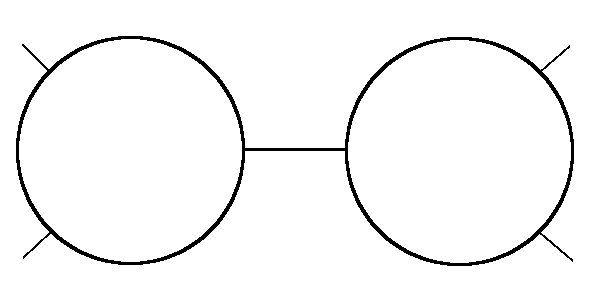}}}}
&
{\rotatebox{0}{\scalebox{0.24}{\includegraphics*{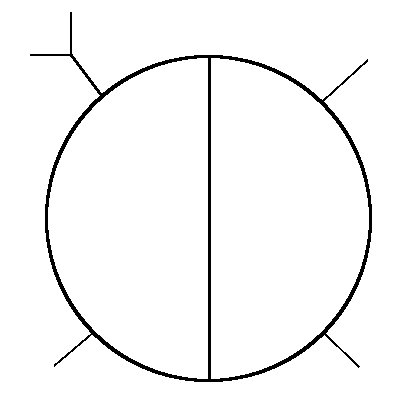}}}}
\\
(a) & (b) & (c)
\end{tabular}
\end{center}
\caption{(a) A one-particle irreducible skeleton.  (b) A one-particle reducible skeleton that consists of two sub-diagrams with loops.  (c) A one-particle irreducible skeleton with a three-point tree sub-diagram attached.  Only  (a) and (b) will be considered in this paper.}
\label{fig:twoloopexamples}
\end{figure}

\subsection{Scalar loop amplitudes}

Our discussion of scalar loop amplitudes will be restricted to diagrams which are either one-particle irreducible (as in figure~\ref{fig:twoloopexamples}(a)) or, diagrams that consists of loop diagrams joined by a particle propagator (as in figure~\ref{fig:twoloopexamples}(b)). We will not consider diagrams in which a tree diagram is attached to a loop (as in figure~\ref{fig:twoloopexamples}(c)).

The one-loop amplitude is special and will be discussed first. This amplitude can be constructed straightforwardly from (\ref{scalarIbuild}) involving $N-1$ external physical particles. The loop is obtained by periodically identifying $0$ and $T$ and gluing together the ends with an unintegrated vertex operator coupled to the $N$'th external particle,
\eqnb
A^{(1)}_0\left(s_{ij}\right)
	&=&
		\int_0^{\infty} dT \int_0^{T} \prod_{r=1}^{N-1}d\tau_r \int d^Dp\,db\, 
	\no
	&\times&
		\left<p,b;T\right|\prod_{r=1}^{N-1}V_{0}\left(k_r,\tau_r\right)b\,U_0\left(k_r,0\right)\left|p,b;0\right>
	\no
	&=&
		\int_0^{\infty} dT \int_0^{T} \prod_{r=1}^{N-1}d\tau_r\int d^Dp\,d^Dx\,dc\,db\, b\,c\,e^{i\left(\sum_{r=1}^{N}k_r\right)x}e^{-\frac{p^2}{2}T}e^{-\sum_{r<s}k_rk_sG\left(\tau_r,\tau_s\right)}
	\no
	&=&
		\delta\left(\sum_{r=1}^{N}k_r\right) \int_0^{\infty} \frac{dT}{T^{D/2}} \int_0^{T} \prod_{r=1}^{N-1}d\tau_r\,e^{-\sum_{r<s}k_rk_sG\left(\tau_r,\tau_s\right)}\, .
\label{scalaroneloop}
\eqne 
Here we have used the integrated form of the vertices. The Green function for the one-loop amplitude satisfies
\eqnb
\partial_\tau^2 G(\tau,\tau') &=& -\delta(\tau,\tau') + 1/T,
\eqne
which has the solution
\eqnb
G(\tau,\tau') &=& -\frac{1}{2}\left|\tau-\tau'\right| + \frac{\left(\tau-\tau'\right)^2}{2T}.
\eqne 

\begin{figure}[t]
\centerline{\rotatebox{0}{\scalebox{0.2}{\includegraphics*{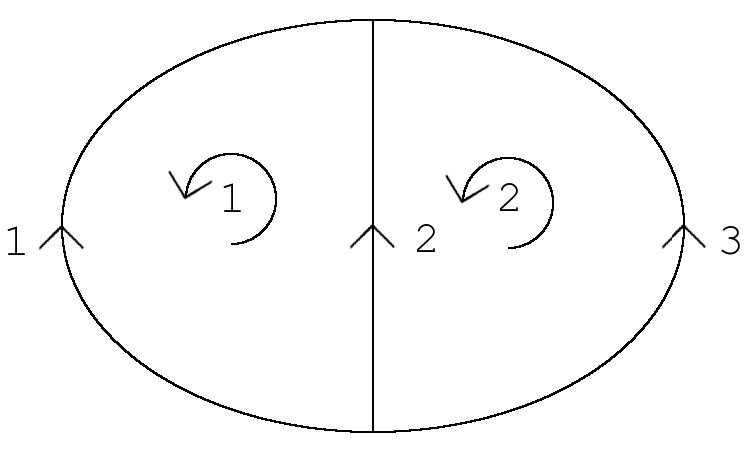}}}}
\caption{The unique one-particle irreducible two-loop skeleton diagram. The amplitude is obtained by attaching vertex operators to points on the lines, which are integrated around the diagram. The circular arrows denote the different $B_I$-cycles. The propagators in the skeleton are numbered from 1 to 3 and the arrows on each line indicate the direction of increasing proper time along the line.}
\label{Figure:twoloop}
\end{figure}

The Green function for the  $N$-point amplitude with $L>1$ loops  was determined in \cite{Dai:2006vj} by using an electric circuit analogue. This involves a basis of $B_I$-cycles defined by the $L$ inequivalent internal counter-clockwise loops. An example at two loops is shown in figure \ref{Figure:twoloop}. Using the $B_I$-cycles one can define the non-trivial one-forms of the skeleton,
\eqnb
\omega_I
	&\equiv&
		{a_{I}}^{i} d\tau_i,
\eqne
where
\eqnb
{a_{I}}^{i}
	&\equiv&
		\left\{
		\begin{array}{ll}
			1  & B_{I}\mathrm{\; in \; the \; same \; direction \; as \;} d\tau_i \\
			-1 & B_{I}\mathrm{\; in \; the \; opposite \; direction \; as \;} d\tau_i \\
			0  & \mathrm{else}
		\end{array}
		\right. .
\eqne
The period matrix of the skeleton is defined in terms of the one-forms $\omega_{I}$ and the $B_I$-cycles by 
\eqnb
\Omega_{IJ}
	&\equiv&
		\oint_{B_I}\omega_{J}.
\eqne
Observe that the period matrix is symmetric, $\Omega_{IJ}=\Omega_{JI}$, as can be seen  from the relation $\oint_{B_I}\omega_{J} = \int_{F_L} (\omega_{I}\omega_{J}/d\tau)$. One can compute the entries of the period matrix in a simple way for any planar skeleton. The diagonal element, $\Omega_{II}$, are equal to the length of the $B_I$-cycles and the off-diagonal entries, $\Omega_{IJ}$, are equal to minus the length of the lines common to the cycles $B_I$ and $B_J$. For non-planar skeletons, the computation of the non-diagonal pieces is more complicated since two different $B_I$-cycles can have the same direction on a line, giving a plus sign. The determinant of the period matrix is important in the computation of the skeleton amplitudes and is denoted by $\Delta \equiv \det{\Omega_{IJ}}$.
These expressions are closely related to those that arise in discussing Riemann surfaces, which are  central to computations of string theory amplitudes \cite{Verlinde:1986kw,D'Hoker:1988ta}. In the point particle case the formalism can be simplified since the integrand is constant on each line and the integral reduce to simple matrix multiplication. 

Consider now the $L>1$ amplitudes. One here glue together the ends of the propagators in momentum space with the vertex $\left<V_B;\tau\right|$. This will introduce a delta function for the momenta,
\eqnb
\int d^Dx\,\left<V_B;\tau\right|\left(\left|p_{f}^1,b_f^i;\tau\right>\left|p_{f}^2,b_f^i;\tau\right>\left|p_{f}^3,b_f^i;\tau\right>\right) 
&=& 
\int d^D x\, e^{i\left(p_f^1+p_f^2+p_f^3\right)x}
\no
&=&
\delta\left(p_f^1+p_f^2+p_f^3\right),
\eqne
here determined when all propagators are in the final state. This we will call an internal vertex.

The components in the construction are $3L-3$ propagators in (\ref{scalarIbuild}), labelled  $j=1, \ldots, 3L-3$, giving a total of  $N$ external particles. Furthermore, there are $2L-2$ internal vertices, labelled $l=1,\ldots,2L-2$. The ends of the propagators can be glued together with vertices in a variety of ways described by  matrices $\{d^f\}^l_j$ and $\{d^i\}^l_j$. The matrix $\{d^f\}^l_j$ is equal to one if the propagator $j$ ends on the vertex $l$ and is zero otherwise. In the same way the matrix $\{d^i\}^l_j$ is defined to be one if the propagator $j$ begins on vertex $l$ and zero otherwise. Consider the picture where all the vertices are integrated. The amplitude with $L$ loops is obtained by integrating over all positions of the external vertices and over the lengths of the $3L-3$ propagators and summing over all different choices of the matrices $\{d^f\}^l_j$ and $\{d^i\}^l_j$.  This results in an expression for the amplitude of the form
\eqnb
A^{(L)}_0\left(s_{ij}\right)
	&=&
		\sum_{\{d^f\}^l_j,\{d^i\}^l_j}\int dT_1\ldots dT_{3L-3} \int_{F_L} \prod_{r=1}^{N}d\tau_r \int \prod_{j=1}^{3L-3}d^Dp^j_f\, d^Dp^j_i\, db_f^j\, db_i^j
	\no
	&\times&
		\prod_{l=1}^{2L-2} \delta\left(\{d^f\}^l_jp^j_f - \{d^i\}^l_jp^j_i\right)\prod_{j=1}^{3L-3}B_j\left(p^j_f,b^j_f,p^j_i,b^j_i;T_j\right).
\label{LamplscalarI}
\eqne 
Here we have suppressed the vertex dependence of $B_j$ for simplicity. In this equation, $\sum_{\{d^f\}^l_j,\{d^i\}^l_j}$ denotes the sum over all non-equivalent matrices. This is equivalent to the sum over all different skeletons with $L$ loops. There is a  total of $3L-3$ insertions of $b$ ghosts since the functions $B_j$ are linear in $b$. 

Let us now evaluate (\ref{LamplscalarI}). The ghost part of the amplitude is trivially integrated out since
\eqnb
\int db_i\, db_f\, \left<b_f\right|b\left|b_i\right>
	=
		\int db_i\, db_f \,  \delta(b_f-b_i)\, b_i
	=
		\int db\,  b
	&=&
		1.
\eqne
Explicitly performing the functional integral (\ref{LamplscalarI}) using (\ref{scalarIbuild}) and (\ref{Greenscontr}) gives
\eqnb
A^{(L)}_0\left(s_{ij}\right)
	&=&
		\sum_{F_L}\int dT_1\ldots dT_{3L-3} \int_{F_L} \prod_{r=1}^{N}d\tau_r 
	\no
	&\times&
		\int \prod_{j=1}^{3L-3}d^Dp^j_f\, d^Dp^j_i d^Dx\, \prod_{l=1}^{2L-2} \delta\left(\{d^f\}^l_jp^j_f - \{d^i\}^l_jp^j_i\right)
	\no
	&\times&
		\prod_{j=1}^{3L-3} e^{-T_j\frac{\left(p^j_i\right)^2}{2}}\delta\left(p_f^j-p_i^j\right)e^{-\sum_{r<s}k_rk_sG\left(\tau_r,\tau_s\right) + i\left(\sum_{r=1}^{N}k_r\right)x}.
\label{LamplscalarII}
\eqne
It is useful to change variables for the momentum integrals in order to make the loop momenta explicit by defining  $p_f^j = \ell^I\, (\omega_I/d\tau_j) + {p'}_f^j$ and $p_i^j = \ell^I\, (\omega_I/d\tau_j) + {p'}_i^j$. Observe that there are $5L-6$ independent variables ${p'}_f^j$ and ${p'}_i^j$. By construction, the delta functions in (\ref{LamplscalarII}) are independent of $\ell^I$. Therefore, the surplus momenta variables ${p'}_i^j$ and ${p'}_f^j$ can be integrated out giving
\eqnb
A^{(L)}_0\left(s_{ij}\right)
	&=&
		\sum_{F_L}\int dT_1\ldots dT_{3L-3} \int_{F_L} \prod_{r=1}^{N}d\tau_r 
		\int d^Dx\, \prod_{I=1}^{L}d^D\ell^I\,
	\no
	&\times&
		e^{-\ell^I\Omega_{IJ}\ell^J/2}e^{-\sum_{r<s}k_rk_sG\left(\tau_r,\tau_s\right) + i\left(\sum_{r=1}^{N}k_r\right)x}
	\no
	&=&
		\delta\left(\sum_{r=1}^{N}k_r\right)\sum_{F_L}\int \frac{dT_1\ldots dT_{3L-3} }{\Delta^{D/2}}\int_{F_L} \prod_{r=1}^{N}d\tau_r e^{-\sum_{r<s}k_rk_sG\left(\tau_r,\tau_s\right)},
\label{LamplscalarIII}
\eqne
where $\Delta$ is the determinant of the period matrix. One essential property of the amplitude is the integration over zero modes, in the first line of (\ref{LamplscalarIII}) (the $x$ and $\ell^I$ integrals). There is one zero mode for each component of the coordinate field $X$, enforcing overall momentum conservation. Such fields with one zero mode for each component are world-line scalars.   The coordinates $\ell^I$  in (\ref{LamplscalarIII}) are the $L$ zero modes of the momentum, $P$ conjugate to $X$, which correspond to the loop momenta of the amplitude. Fields of this type, which have $L$ zero modes, are  world-line vector fields. The vertex is constructed in such a manner that world-line scalar fields have a common value at the vertex, and the components of a  world-line vector are conserved.

%%%%%%%%%%%%%%%%%%%%%%%%%
\section{Pure spinor particle}
\label{sec:purespinor}
%%%%%%%%%%%%%%%%%%%%%%%%%

An action for the pure spinor particle describing ten-dimensional $\cN=1$ supersymmetric Yang--Mills was introduced in \cite{Berkovits:2001rb}. This action was written down in the ``minimal''-formalism and can easily be generalised to the ``non-minimal'' form (introduced for the string in \cite{Berkovits:2005bt}) by adding additional pure spinor fields to the action
\eqnb
S_{YM} &=& \int d\tau \left( \dot{X}P + \dot{\theta} p + \dot{\lambda} w + \bar{w}\dot{\bar{\lambda}}-s\dot{r}-\frac{P^2}{2}\right),
\label{actionN1}
\eqne
written here in the gauge $e=1$. The world-line fields in this action consist of the classical superspace fields together with a set of pure spinor fields. The classical fields are the bosons, $X^m$ and $P_m$ (where $m=1,\ldots ,10$) and the fermions $\theta^{\alpha}$ and $p_{\alpha}$ (where $\alpha = 1,\ldots ,16$). The non-minimal pure spinor fields consists of the bosonic fields\footnote{When computing amplitudes, the non-minimal pure spinor field $\bar{\lambda}_{\alpha}$ should be interpreted as the complex conjugate of the minimal pure spinor field $\lambda^{\alpha}$ \cite{Berkovits:2005bt}.} $\lambda^{\alpha},w_{\alpha},\bar{\lambda}_{\alpha}$ and $\bar{w}^{\alpha}$ and the fermionic field $r_{\alpha}$ and $s^{\alpha}$. The coordinate fields are $X^m,\theta^{\alpha},\lambda^{\alpha},\bar{\lambda}_{\alpha}$ and $r_{\alpha}$, and the corresponding conjugated momentum fields are $P_m,p_{\alpha},w_{\alpha},\bar{w}^{\alpha}$ and $s^{\alpha}$. In the next section we will find that the coordinate fields are world-line scalar fields and the conjugated momentum fields are world-line vector fields.

The commutation relations between the various coordinates and their conjugate momenta follow in the usual manner
\eqnb
\left[P_m,X^n\right] &=& -{\delta_{m}}^{n} \,,\quad
\left[p_\alpha,\theta^{\beta}\right] = -{\delta_\alpha}^\beta \,,\quad
\left[w_\alpha,\lambda^\beta\right] = -{\delta_\alpha}^\beta \no
\left[\bar{w}^\alpha,\bar{\lambda}_\beta\right] &=& -{\delta^\alpha}_\beta \,,\quad
\left[s^\alpha,r_\beta\right] = -{\delta^\alpha}_\beta \,, \quad
\left[d_\alpha,d_\beta\right] = -P^m\left(\gamma_m\right)_{\alpha\beta} \no
\left[d_\alpha,X^m\right] &=& -\frac{1}{2}\left(\gamma^m\theta\right)_{\alpha} \,,\quad
\left[d_\alpha,P_m\right] =0 \,, \quad
\left[d_\alpha,\theta^{\beta}\right] = -{\delta_\alpha}^\beta,
\eqne
where we have defined $d_{\alpha} = p_{\alpha}+\frac{1}{2}P^m\left(\gamma_{m}\theta\right)_{\alpha}$.

The world-line fields $\lambda^{\alpha},\bar{\lambda}_\alpha$ and $r_\alpha$ are pure spinor fields satisfying the constraints
\be
\lambda\gamma_{m}\lambda = 0\,, \qquad 
\bar{\lambda}\gamma_{m}\bar{\lambda} = 0\,,\qquad
\bar{\lambda}\gamma_{m}r = 0,
\label{psconst}
\ee
which imply that $\lambda^{\alpha},\bar{\lambda}_{\alpha}$ and $r_{\alpha}$ have eleven degrees of freedom. The pure spinor constraints generates a gauge invariance of the action which acts non-trivially on the conjugated momenta for the pure spinor fields
\eqnb
\delta w_\alpha
	&=&
		- \epsilon^m_1\left(\gamma_{m}\lambda\right)_{\alpha} \no
\delta \bar{w}^\alpha
	&=&
		- \epsilon^m_2\left(\gamma_{m}\bar{\lambda}\right)^{\alpha} - \epsilon^m_3\left(\gamma_{m}r\right)^{\alpha} \no
\delta s^\alpha
	&=&
		- \epsilon^m_3\left(\gamma_{m}\bar{\lambda}\right)^{\alpha},
\eqne
where $\epsilon^m_i$ are infinitesimal vectors. One can make linear combinations of $w_{\alpha}$, $\bar{w}^{\alpha}$ and $s^{\alpha}$ that are gauge invariant. These gauge invariant combinations are
\eqnb
J &=& \lambda w\,,\phantom{123456789}\; \bar{J} = \bar{w}\bar{\lambda} - sr \,\no
N_{mn} &=& \frac{1}{2} w\gamma_{mn}\lambda\,, \phantom{12}
\bar{N}_{mn} = \frac{1}{2}\bar{w}\gamma_{mn}\bar{\lambda}- \frac{1}{2}s\gamma_{mn}r\no
S &=&s\bar{\lambda} \,, \phantom{12345678} S_{mn} = \frac{1}{2}s\gamma_{mn}\bar{\lambda}\,,
\label{ginvcomb}
\eqne
where $N_{mn},\bar{N}_{mn}$ and $S_{mn}$ has ten linearly independent components. Writing all expressions involving $w_{\alpha}$, $\bar{w}^{\alpha}$ and $s^{\alpha}$ in terms of these combinations, one can strongly impose the pure spinor constraints in computations.

Note that the Hamiltonian of the action in (\ref{actionN1}) is
\eqnb
H &=& \frac{1}{2}P^2\,,
\eqne
so the solutions of the equations of motion of all fields other than $X$ are locally constants. This will be an important feature in the computation of the amplitudes in later sections.

The BRST charge for the action in (\ref{actionN1}) is given by
\eqnb
Q &=& \lambda d + \bar{w}r\,,
\label{BRSTchargeYM}
\eqne
which is nilpotent using the pure spinor constraints. In the non-minimal formalism one can construct a composite $b$ ghost by mimicking the procedure in \cite{Berkovits:2005bt} of the pure spinor superstring. The equation which the $b$ ghost satisfies is
\eqnb
[Q,b] &=& H,
\eqne
as well as $b^2=0$. The solution for the $b$ ghost resembles the string theory expression and is given by\footnote{The condition $b^2=0$ for the zero mode part of the string has been checked by Aisaka and Cederwall \cite{Cederwall:2010} and the general case by Chandia \cite{Chandia:2010ix}.}
\eqnb
b &=& \frac{G^\alpha\bar{\lambda}_{\alpha}}{\left(\lambda\bar{\lambda}\right)} + \frac{\bar{\lambda}_{\alpha}r_\beta H^{[\alpha\beta]}}{\left(\lambda\bar{\lambda}\right)^2} + \frac{\bar{\lambda}_{\alpha}r_\beta r_\gamma K^{[\alpha\beta\gamma]}}{\left(\lambda\bar{\lambda}\right)^3} + \frac{\bar{\lambda}_{\alpha}r_\beta r_\gamma r_\delta L^{[\alpha\beta\gamma\delta]}}{\left(\lambda\bar{\lambda}\right)^4},
\label{bghostN1}
\eqne
where $[\;]$ denote anti-symmetrisation of the indices. Here the $G^{\alpha}, H^{[\alpha\beta]}, K^{[\alpha\beta\gamma]}$ and $L^{[\alpha\beta\gamma\delta]}$ are the zero modes of the expressions that arise for the pure spinor string in equation (3.16) of \cite{Berkovits:2005bt},
\eqnb
G^\alpha
	&\equiv&
		-\frac{1}{2}P^m\left(\gamma_m d\right)^\alpha
	\no
H^{[\alpha\beta]}
	&\equiv&
		-\frac{1}{384}\gamma^{\alpha\beta}_{mnp}\left[\left(d\gamma^{mnp}d\right) - 24 N^{mn}P^p\right] \no
K^{[\alpha\beta\gamma]}
	&\equiv&
		\frac{1}{192}\gamma_{mnp}^{[\alpha\beta}\left(\gamma^md\right)^{\gamma]}N^{np} \no
L^{[\alpha\beta\gamma\delta]}
	&\equiv&
		\frac{1}{12244}\gamma_{mnp}^{[\alpha\beta}{{\gamma^m}_{qr}}^{\gamma\delta]}N^{np} N^{qr},
\label{bfields}
\eqne
which satisfy\footnote{These properties can shown with the help of the software package GAMMA \cite{Gran:2001yh}.}
\eqnb
\left[Q,G^{\alpha}\right] 
	&=&
		\lambda^\alpha H \,, \phantom{1234}\,\, \phantom{12}\,\left[Q,H^{[\alpha\beta]}\right]
	=
		\lambda^{[\alpha} G^{\beta]} \,,\no 
\left[Q,K^{[\alpha\beta\gamma]}\right]	&=&
		\lambda^{[\alpha} H^{\beta\gamma]} \,, \phantom{123}  \left[Q,L^{[\alpha\beta\gamma\delta]}\right]
	=
		\lambda^{[\alpha} K^{\beta\gamma\delta]} \,,\qquad 
		\lambda^{[\alpha}L^{\beta\gamma\delta\epsilon]}
	=
		0.
\eqne
In the derivation above, the following gamma matrix identities has been used
\eqnb
\frac{1}{2}\left(\gamma_{m}\right)^{\beta\delta}\left(\gamma^{m}\right)_{\alpha\gamma}
	&=&
		\delta^{\beta}_{\gamma}\delta^{\delta}_{\alpha} + \frac{1}{4}\delta^{\beta}_{\alpha}\delta^{\delta}_{\gamma} + \frac{1}{8}{\left(\gamma_{mn}\right)^{\beta}}_{\alpha}{\left(\gamma^{mn}\right)^{\delta}}_{\gamma}
	\no
\left(\gamma_{mnp}\right)^{\alpha\beta}{\left(\gamma_{mn}\right)_{\gamma}}^{\delta}
	&=&
		4\left(3{\delta_{\gamma}}^{[\alpha}\left(\gamma_p\right)^{\beta]\delta}+{\left(\gamma_{qp}\right)_{\gamma}}^{[\alpha}\left(\gamma_q\right)^{\beta]\delta}\right)\, .
\eqne 
For future reference, note that the $b$ ghost involves $P$. 

We will now construct the single-particle vertices of the theory (the three-point vertices with one physical particle). This also follows by analogy with the case of the pure spinor string \cite{Berkovits:2000fe}. The  unintegrated and integrated vertices, denoted by $U_{YM}$ and $V_{YM}$ respectively, are obtained from the properties
\eqnb
\left[Q,U_{YM}\right] &=& 0\no
\left[Q,V_{YM}\right] &=& \left[H,U_{YM}\right].
\eqne
One can construct these vertices in the minimal picture where the unintegrated vertices only depend on $X^m,\theta^{\alpha}$ and $\lambda^{\alpha}$. The solution for the unintegrated vertex is
\eqnb
U_{YM}\left(X,\theta,\lambda\right) &=& \lambda^{\alpha} A_\alpha\left(X,\theta\right),
\label{Vertexymui}
\eqne
if the field $A_\alpha\left(X,\theta\right)$ satisfies the linearised field equations of supersymmetric Yang-Mills in ten dimensions.  The field equations and $\theta$-expansions are summarised in  appendix \ref{sec:YM}. The integrated vertex operator is given (after use of the equations of motion) by
\eqnb
V_{YM}\left(P,X,p,\theta,w,\lambda\right)
	&\equiv&
		P^mA_m\left(X,\theta\right) - d_{\alpha}W^{\alpha}\left(X,\theta\right) + \frac{1}{2}N^{mn}\mathcal{F}_{mn}\left(X,\theta\right)\,,
\label{Vertexymi}
\eqne 
if the fields in  (\ref{Vertexymi}) satisfy the linearised field equations. As we are mainly interested in amplitudes with external gluons, it is enlightening to consider where the field strength, $F_{mn}$, arises for the first time in the $\theta$-expansion of the different superfields. For $A_{\alpha}$ it arises as $\theta^3\, F$, for $A_m$ it arises as $\theta^2\, F$ because $A_m \sim D A_{\alpha}$. For the superfield $W^{\alpha}$ it arises in $\theta\, F$ as $W \sim D^2 A_{\alpha} $. Here $D_{\alpha}$ is the fermionic superderivative defined in  (\ref{covariantderiv}) of appendix \ref{sec:YM}. An important observation is that there is a term in the integrated vertex operator, which depends on the momentum. When such insertions arise in pairs they can produce contact terms (vertices with more than three legs) in the amplitude. Since we are considering on-shell  amplitudes we will make a plane wave expansion of the fields and consider vertex operators with fixed momentum, which will be denoted  $U_{YM}\left(k_r,\tau_{r}\right) \equiv U_{YM}\left(e^{ik_rX(\tau_r)},\theta\right)$ and $V_{YM}\left(k_r,\tau_{r}\right) \equiv V_{YM}\left(e^{ik_rX(\tau_r)},\theta\right)$.

The integrated and unintegrated vertex operators are related, up to BRST exact terms, by the $b$ ghost
\eqnb
[b,U_0\left(k,\tau\right)] 
	&=& V_0\left(k,\tau\right).
\eqne
However, the relation is not as simple as for the scalar particle, because the $b$ ghost involves the non-minimal pure spinor fields.

Standard gauge transformations of the fields, which act as $\delta_{\rho} A_{\alpha}=-D_{\alpha}\rho$ and $\delta_{\rho} A_m = -\partial_m\rho$, generalise to BRST transformations of the three-point vertices, 
\eqnb
\delta_{\rho} U_{YM}
	&=&
		\left[Q,\rho\right]\,,
	\no
\delta_{\rho} V_{YM}
	&=&
		\left[Q,\left[\rho,b\right]\right] + \left[H,\rho\right]\,.
\eqne
Note that it is possible for $[\phi,b] \neq 0$ even if $\phi$ does not depend on the pure spinor fields since the $b$ ghost is a composite field. The above considerations apply to the abelian theory but the generalisation to the non-abelian  theory is straightforward by introducing  colour factors. This will be discussed in more detail in later sections.

The above vertices are constructed for the linearised theory. We may also need to include non-linear effects, by considering  how the vertex operators deform the BRST charge and the Hamiltonian. The effect of the perturbation is to deform the free expressions of the BRST charge and the Hamiltonian as 
\eqnb
Q &\rightarrow& Q - \epsilon\, U_{YM}(\epsilon)  \no
H &\rightarrow& H - \epsilon\, V_{YM}(\epsilon),
\eqne
where $\epsilon$ is a small parameter. The nilpotency condition of the BRST charge then implies
\eqnb
[Q,U_{YM}(\epsilon)] &=& \frac{\epsilon}{2}[U_{YM}(\epsilon),U_{YM}(\epsilon)].
\label{nonlYM}
\eqne
By using the form of the vertex operators, one obtain the equations of motion of $\cN = 1$ supersymmetric Yang--Mills in ten dimensions, see appendix \ref{sec:YM}. The BRST invariance of the Hamiltonian follow from the equations of motion and the properties of the $b$ ghost.

All of the preceding can be generalised to describe maximally supersymmetric supergravity in ten dimensions. Mimicking the transition from the open string to the closed string, this involves doubling all the fields in the action (\ref{actionN1}), apart from $X$ and $P$, giving the action
\eqnb
S_{SG} &=& \int d\tau \left( \dot{X}P + \dot{\theta} d + \dot{\lambda} w + \bar{w}\dot{\bar{\lambda}}-s\dot{r}- \hat{d}\dot{\hat{\theta}} + \hat{w}\dot{\hat{\lambda}} + \dot{\hat{\bar{\lambda}}}\hat{\bar{w}} + \dot{\hat{r}}\hat{s}-\frac{P^2}{2}\right).
\eqne
The new fields are the ones with a hat on them. The BRST charge for this theory is
\eqnb
Q_{tot} &=& \lambda d + \bar{w}r + \hat{d}\hat{\lambda} + \hat{r}\hat{\bar{w}}\,,
\eqne
which  decomposes into two separate pieces, $Q$ and $\hat{Q}$, corresponding to the unhatted and hatted fields. There are various different ways to construct a $b$ ghost, $b_{tot}$, for the model, but the most natural way is to construct a $b$ ghost which also decompose into a sum of two terms, $b_{tot}=b+\hat{b}$, one involving  the unhatted  fields and one involving the hatted fields. This  construction also resembles that of the  pure spinor string. The generalisation of  (\ref{bghostN1}) implies that the ghost satisfy
\eqnb
[Q_{tot},b_{tot}] = [Q,b] + [\hat{Q},\hat{b}] = H,
&\;\;&
[Q_{tot},b-\hat{b}] = [Q,b] - [\hat{Q},\hat{b}] = 0,
\eqne 
the second condition corresponds to the level matching condition for the string. Observe that the ghost for one of the sectors of the theory is half of the corresponding expression for Yang--Mills, see (\ref{bghostN1}).

The construction of the unintegrated and integrated vertices are a bit different in the supergravity case since the BRST charge and $b$ ghost decompose into two parts. As for the string, we need to consider integrated and unintegrated vertex operators. These satisfy
\eqnb
[Q,U_{SG}] = [\hat{Q},U_{SG}] = 0 &\phantom{123}&
[Q,V_{SG}] = \frac{1}{2}[H,\hat{K}]
\no
[\hat{Q},V_{SG}] = -\frac{1}{2}[H,K] &\phantom{123}&
[Q,\hat{K}] = [\hat{Q},K] = 0
\no
[\hat{Q},\hat{K}] = \frac{1}{2}[H,U_{SG}] &\phantom{123}&
[\hat{Q},K] = \frac{1}{2}[H,U_{SG}].
\label{eqn:defSGUV}
\eqne
From these conditions one can determine that the unintegrated vertex operator has the form
\eqnb
U_{SG}(X,\theta,\hat{\theta},\lambda,\hat{\lambda}) &=& \lambda^{\alpha}{A_{\alpha}}^{\beta}(X,\theta,\hat{\theta})\hat{\lambda}_{\beta}\,,
\eqne
if the fields satisfy the linearised field equations of type IIA supergravity\footnote{One can also describe IIB supergravity by changing the hatted fields to have the same chirality as the unhatted ones.}\cite{Berkovits:2001ue,Grassi:2004ih}. Similarly, the integrated vertex operator is of the form\footnote{The dependence of the classical superfields has been suppressed on the right-hand side for the economy of space.}
\eqnb
V_{SG}(P,X,p,\theta,\hat{p},\hat{\theta},w,\lambda,\hat{w},\hat{\lambda}) 
	&=&
		P^m G_{mn} P^n + d_{\alpha}{{W}^{\alpha}}_{\beta}\hat{d}^{\beta} - d_{\alpha}{\hat{E}^{\alpha}}_{m}P^m - P^m{E}_{m\alpha}\hat{d}^{\alpha} + 
	\no
	&+&
		\frac{1}{2}N^{mn}\hat{\Omega}_{mn;p}P^p + \frac{1}{2}P^m \Omega_{m;np}\hat{N}^{np} - \frac{1}{2}N^{mn}\hat{C}_{mn;\alpha}\hat{d}^{\alpha}
	\no
	&-&
		\frac{1}{2} d_{\alpha} {C^{\alpha}}_{mn}\hat{N}^{mn} + \frac{1}{4} N^{mn}S_{mn;pq}\hat{N}^{pq}\,,
\label{inVSG}
\eqne
if the fields satisfy the linearised type IIA field equations. We will consider amplitudes with external gravitons, so it is of interest to note the lowest  components of the $\theta$- and $\hat{\theta}$-expansion of the superfields  in which the Riemann tensor arises. In particular, the Riemann tensor arises in the $\theta^3\, \hat{\theta}^3 \, \cR$ component of   ${A_{\alpha}}^{\beta}$ and in the  $\theta^2\, \hat{\theta}^2 \, \cR$ component of $G_{mn}$ as $G \sim D\,\hat{D}\, A$. Furthermore, the bispinor ${W^{\alpha}}_{\beta}$ contains the curvature  in the term $\theta \, \hat{\theta} \, \cR$ because $W \sim D\, \hat{D}\, G $. The fields ${\hat{E}^{\alpha}}_{m}$ and ${E}_{m\alpha}$ contains the curvature in the term $\theta\, \hat{\theta}^2 \, \cR$ and $\theta^2 \, \hat{\theta} \, \cR$, respectively. Here $\hat{D}^{\alpha}$ is the fermionic superderivative for $\hat{\theta}_{\alpha}$ defined in the same way as $D_{\alpha}$. An important observation is that there are momentum factors $P$ in different parts of the vertex operators. When they arise in pairs they can generate contact terms in the amplitude.

We can also determine the ``intermediate'' fields in (\ref{eqn:defSGUV})
\eqnb
\hat{K}(P,X,\theta,\hat{p},\hat{\theta},\lambda,\hat{w},\hat{\lambda})
	&=&
		\lambda^{\alpha}\left(A_{\alpha m}P^m - E_{\alpha; \beta}\hat{d}^{\beta} + \frac{1}{2}\Omega_{\alpha;mn}\hat{N}^{mn}\right)
\no
K(P,X,p,\theta,\hat{\theta},w,\lambda,\hat{\lambda})
	&=&
		\left(P^m{A_{m}}^{\beta} - d_{\alpha} E^{\alpha; \beta} + \frac{1}{2}N^{mn}{\Omega_{mn}}^{\beta}\right)\hat{\lambda}_{\beta},
\eqne 
where the fields involved satisfy the linearised field equations of type IIA supergravity. The vertex operators and intermediate fields satisfy the relations 
\eqnb
[b,U_{SG}] = \hat{K}, &\phantom{12}&
[\hat{b},U_{SG}] = K,
\no
[b,[\hat{b},U_{SG}]] = V_{SG}, &\phantom{12}&
\frac{1}{2}\left([b,\hat{K}]-[\hat{b},K] \right)= V_{SG},
\label{eq:relationsSGV}
\eqne
up to BRST exact terms because the $b$ field involve non-minimal fields.

It is also essential to consider non-linear effects in the above fields. The vertices deform the free theory
\eqnb
Q_{tot} &\rightarrow& Q_{tot} - \frac{\epsilon}{2}\,\left(K(\epsilon)-\hat{K}(\epsilon)\right) \no
H &\rightarrow& H - \epsilon\, V_{SG}(\epsilon),
\eqne
where $\epsilon$ is infinitesimal. From the nilpotency condition of the BRST charge it follows that
\eqnb
[Q_{tot},K(\epsilon)-\hat{K}(\epsilon)] &=& \frac{\epsilon}{4}\,[K(\epsilon)-\hat{K}(\epsilon),K(\epsilon)-\hat{K}(\epsilon)].
\eqne
Using the relation between $K, \hat{K}$ and $U_{SG}$ in (\ref{eq:relationsSGV}) one obtains
\eqnb
[U_{SG}(\epsilon),Q_{tot}] &=& \frac{\epsilon}{4}\, [U_{SG}(\epsilon),[b-\hat{b},U_{SG}(\epsilon)]],
\label{nonlinearSG}
\eqne
up to BRST exact terms. This is the equation, which the vertex operators satisfy if one includes non-linear effects. Observe at this stage one cannot exclude the presence of higher-point, fundamental, vertices. But as we will see in section \ref{sec:BRST}, these are not present within perturbation theory.

In the next section we will define the perturbation theory of the pure spinor particle by extension of the construction of the scalar theory. This makes use of the connection between the $\cN=2$ topological string and the pure spinor string \cite{Berkovits:2005bt}.

%%%%%%%%%%%%%%%%%%%%%%%%%
\section{Amplitude prescription for the pure spinor particle}
\label{sec:defamp}
%%%%%%%%%%%%%%%%%%%%%%%%%

In this section we will discuss the multi-loop four-particle amplitude for the pure spinor particle. A web of relations between different theories motivates the form of the amplitude. First, the non-minimal pure spinor string is related to the $\cN = 2$ topological string \cite{Berkovits:2005bt}. The amplitudes of the latter are modelled on the bosonic string. Furthermore, as we have shown in section~\ref{sec:firstquant}, the amplitude prescription of the scalar particle is closely related to the bosonic string\footnote{In principle, it is the tachyonic part of the bosonic string.}. Therefore, we conjecture that the form of the amplitude of the pure spinor particle is closely related to that of the scalar particle. The consistency will be checked by an analysis of BRST invariance. In the Yang--Mills case, we conjecture a form for the propagator, motivated by the bosonic particle. The internal vertex is constructed by imposing locality, which is enough for the vertex to be BRST invariant. For supergravity, the propagator and the internal vertex are determined by mimicking the relation between the open string and the closed string. In the end, we will reproduce the expressions in \cite{Bjornsson:2010wm} but will learn much about the structure of the amplitudes,  which will make the BRST invariance more transparent.

We will consider the Yang--Mills and supergravity theories separately since there are essential differences between them. These differences  are important in discussing  the BRST invariance of the amplitude,  which will be studied in section \ref{sec:BRST}.

The amplitudes obtained are generalisations of those presented for the tree and one-loop amplitude in \cite{Berkovits:2001rb,Anguelova:2004pg}.

%%%%%%%%%%%%%%%%
 
\subsection{Yang--Mills}

To simplify the expressions we will denote any coordinate fields by $\phi$ and any momentum field by $\Phi$.  The propagator is the operator which takes a particle with momentum $\Phi_{i;b_i}$ at $T=0$ to $\Phi_{f;b_f}$ at time $T_j$. Here $b_i$ and $b_f$ denote the colour factors of the different states.  By analogy with the expression for the scalar particle the obvious expression for the pure spinor particle propagator is 
\eqnb
&&
\left<\Phi_{f;b_f};T_j\left|\mathcal{N}\left(\phi,\Phi\right) b\left(\phi,\Phi\right)\right|\Phi_{i;b_i};0\right>\,
\label{propYM}
\eqne
where we have emphasised that the $b$ ghost is not a fundamental field.  Observe that we also have inserted a regulator, $\cN$, to regularise possible singularities. We impose the condition that the regulator is BRST exact and it has an expansion of the form 
\eqnb
\mathcal{N}\left(\phi,\Phi\right) = 1 + \epsilon\,\int_{0}^{T_j}d\tau\,  [Q,\chi(\tau)]+\mathcal{O}(\epsilon^2).
\label{regulatorcond}
\eqne 
Since there is  an odd number of fermionic fields and one insertion of a $b$ ghost, the operator in (\ref{propYM}) is bosonic. 

\begin{figure}[t]
\begin{center}
\begin{tabular}{cc}
{\rotatebox{0}{\scalebox{0.27}{\includegraphics*{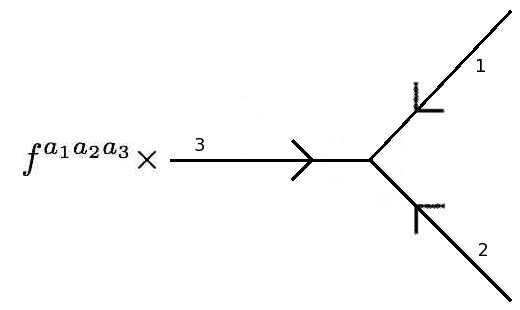}}}}
\hspace{2cm}
&
{\rotatebox{0}{\scalebox{0.27}{\includegraphics*{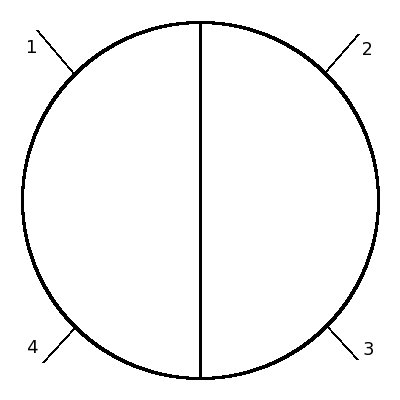}}}}
\\
(a) & (b)
\end{tabular}
\end{center}
\caption{(a)  The three-point vertex with colour factor. (b) A two-loop amplitude}
\label{fig:colour}
\end{figure}

Let us now discuss the colour factors of the amplitude. First, we can consider a particular ordering of the external particles. The complete amplitude is determined by summing over all permutations of the external particles. We will usually consider the ordering in figure~\ref{fig:colour}(b), where the amplitude is a function of $s$ and $t$.  Each vertex contains a  factor of the structure constant, $f^{abc}$,  where one of the indices has been raised by the Killing form making it totally anti-symmetric. The convention we use is that the indices are placed in clockwise order (see figure \ref{fig:colour}(a)). Each propagator has an insertion of the inverse of the Killing form, which contracts the indices in the vertices. One example of a colour factor is $f^{a_1b_1b_2}f^{b_3b_4b_5}f^{a_2b_7b_6}f^{a_3b_9b_8}f^{b_{10}b_{11}b_{12}}f^{a_4b_{14}b_{13}}\kappa_{b_1b_3}\kappa_{b_4b_6}\kappa_{b_7b_8}\kappa_{b_9b_{10}}\kappa_{b_5b_{12}}\kappa_{b_{11}b_{13}}\kappa_{b_2b_{14}}$ for the amplitude in figure \ref{fig:colour}(b).  This can be rewritten  in terms of single- and double-trace contributions
\eqnb
&&N_c^2\left(\Tr[t^{a_1}t^{a_2}t^{a_3}t^{a_4}]+\Tr[t^{a_1}t^{a_4}t^{a_3}t^{a_2}]\right)\left(A_{P}(s,t) + \mathcal{O}(N_c^{-2})\right) 
\no
&&\; + \,N_c \Tr[t^{a_1}t^{a_2}]\Tr[t^{a_3}t^{a_4}]A_{NP}(s,t) + \mathcal{O}(N_c^{0}) \,,
\eqne
where we have expanded the expression for a large number of colours,  $N_c$.
The leading term is the planar single-trace contribution, $A_{P}(s,t)$, which is proportional to  $N_c^{L}$. The next to leading term is the double-trace contribution, $A_{NP}(s,t)$, which is proportional to $N_c^{L-1}$. In this section we will not be concerned about factoring out the colour,  although analysing the relative behaviour of planar and non-planar contributions will be  important later when we discuss the leading low energy behaviour of the single-trace and double-trace contributions\footnote{Although the large-$N_c$ limit is not necessary, it will be used in section \ref{sec:four-point} as it gives an intuitive picture of the difference between the single- and double-trace operators.}.

We will now construct the bolding blocks for the amplitudes. Using the above, one can obtain expressions for the propagator which absorbs $N$ external particles. From section \ref{sec:firstquant} we know that there are two different approaches. In the unintegrated picture the propagator is
\eqnb
B^U_j\left(\phi,\Phi_{f;b_f},\Phi_{i;b_i},\{k_r\},\{\tau_s\}\right)
	&=&
		\left<\Phi_{f;b_f};\sum_{s=1}^{N+1}\tau_s\left|\prod_{r=1}^{N}\left(b U_{YM}^{r}\right)\mathcal{N}\,b\right|\Phi_{i;b_i};0\right>\,.
\label{NpropUYM}
\eqne
The index $r$ of the vertex operator $U_{YM}^{r}$ here denotes the momentum, position and colour.

In the integrated picture there is only a single $b$ ghost insertion. This $b$ commutes with the integrated vertices and can therefore be located anywhere. We will make the choice that the $b$ ghost is inserted to the right of the vertex operators, so that
\eqnb
B^I_j\left(\phi,\Phi_{f;b_f},\Phi_{i;b_i},\{k_r\},\{\tau_r\};T\right)
	&=&
		\left<\Phi_{f,b_f};T\right|\prod_{r=1}^{N} V_{YM}^{r}\mathcal{N}\,b\left|\Phi_i;0\right>\, .
\label{NpropVYM}
\eqne
  
By using the relation between integrated and unintegrated vertex operators one can show that (\ref{NpropUYM}) and (\ref{NpropVYM}) are equivalent up to a BRST exact term\footnote{Note that we have chosen the order  $\tau_{r-1}\geq\tau_{r}$ for the integrated propagator.  If we were to integrate over all orders it would be necessary to include a sum over all permutations in the unintegrated picture.}.   The BRST invariance of the amplitude is easiest to see in the first picture,  but it is easy to determine the low energy properties of the amplitude using the second approach. Since  $B^U_j \sim B^I_j$,  in the following it will not be necessary to keep the superscript $I$ or $U$.

In the previous section we determined the single-particle vertices of the theory i.e.\ the absorption of a single physical state from the propagator. These vertices were not constructed from a general three-point vertex. We also need to consider the general three-point vertices, which arise as internal vertices in loops. For the vertex, denoted by $\left<V_{YM};\tau\right|$, we only need to impose locality -- momentum conservation follows from integration over the interaction point. Furthermore, the colour factor of the vertex is $f^{b_1b_2b_3}$, where the colour indices are distributed in a clockwise direction, see figure~\ref{fig:colour}(a). A consequence of imposing locality at the vertex is that the coordinate fields are  world-line scalars and the momentum fields are  world-line vectors as will be shown later in the section. We will in the construction of the amplitudes be interested in the vertex contracted with three momentum states 
\eqnb
\int d\phi\, \left<V_{YM};\tau\right|\left(\left|\Phi_{b_1}^1;\tau\right>\left|\Phi_{b_3}^3;\tau\right>\left<\Phi_{b_3}^3;\tau\right|\right)
&=&
\int d\phi\, \left<V_{YM};\tau\right|\left(\left|\Phi_{b_1}^1;\tau\right>\left|\Phi_{b_3}^3;\tau\right>\left|-\Phi_{b_3}^3;\tau\right>\right)
\no
&=&
\delta\left(\Phi_{b_1}^1+\Phi_{b_2}^2-\Phi_{b_3}^3\right).
\eqne
Our sign convention assigns a plus sign to an incoming momentum and a minus sign to an outgoing momentum. As the number of fermionic fields involved are odd\footnote{The fields $r_{\alpha}$ and $s^{\alpha}$ each have eleven degrees of freedom.}, the delta function is fermionic and an odd function. This delta function also includes the colour factor of the vertex. No other conditions need to be  imposed on the vertex in order to conserve the  BRST charge (unlike the  situation in the string calculation in  \cite{Green:1982tc} where additional conditions were  imposed to ensure that the vertex  conserves supersymmetry). This follows from the form of the BRST charge given in (\ref{BRSTchargeYM}). 

We will first consider loop amplitudes. We will  here consider the same set of diagrams as those described in section \ref{sec:firstquant} and figure \ref{fig:twoloopexamples}(a)-(c).  We start with the special case of one loop. The amplitude is constructed by again generalising  the procedure for the scalar particle in section \ref{sec:firstquant}. In the integrated picture one use the operator in (\ref{NpropVYM})
\eqnb
A^{(1)}_{YM}(s_{ij})
	&=&
		\int_{0}^{\infty} dT\int_{0}^T \prod_{r=1}^{N-1} d\tau_r \int d\Phi\; \kappa^{cd}\left<\Phi_c;T\right|\mathcal{N}\, \prod_{r=1}^{N-1} V_{YM}^{r}\, b\, U_{YM}^{N}\left|\Phi_d;0\right>\, .
\label{intN1L1}
\eqne
In the unintegrated picture one has to integrate over the length of the propagators connecting the unintegrated vertices as well as summing over all different permutations of the first $N-1$ external particles
\eqnb
A^{(1)}_{YM}(s_{ij})
	&=&
		\sum_{\sigma \in S_{N-1}} \int_0^{\infty} \prod_{r=1}^{N}d\tau_r \int d\Phi\; \kappa^{cd} \left<\Phi_c;\sum_{s=1}^{N}\tau_s\left|\mathcal{N}\, \prod_{r=0}^{N}\left(b\, U_{YM}^{\sigma(r)}\right)\right|\Phi_d;0\right>\,.
\label{unintN1L1}
\eqne
As described earlier, the two different prescriptions are equivalent if the amplitudes are BRST invariant. We introduce both pictures as it is easier to prove BRST invariance in the unintegrated picture (\ref{unintN1L1}),  but the low energy properties are more transparent in the latter picture   (\ref{intN1L1}). We note that the expression  (\ref{intN1L1}) is equal to the functional integral
\eqnb
A^{(1)}_{YM}(s_{ij})
	&=&
		\int_0^\infty dT \int_0^{T} \prod_{r=1}^{N-1} d\tau_{r} K_{YM}\left(\{k_r\},T,\tau_r\right)
	\no
K_{YM}\left(\{k_r\},T,\tau_r\right)
	&=&
		\int \mathcal{D}\Phi\, \mathcal{D}\phi\, \mathcal{N}\, b\, \prod_{r=1}^{N-1} V_{YM}^{r}\left(k_r,\tau_r\right) U_{YM}^{N}(k_{N},0)\, ,
\eqne
where we have separated the integration of the moduli and position of the vertex operators and the functional integration of the fields for fixed values of the moduli and position of the vertices.

We will find in the next section, that the above prescription has to be generalised in order to be consistent with  BRST invariance. This generalisation can be expressed in one of  two ways --  either  by including higher-point external vertices or by the addition of trees  attached to the one-loop diagram. 

For amplitudes with more than one loop, the $N$-point amplitude is constructed by using $3L-3$ propagators with a total of $N$ vertex operators and $2L-2$ internal vertex operators. In the unintegrated picture the result is
\eqnb
A^{(L)}_{YM}(s_{ij})
	&=&
		\sum_{\{d^f\}^l_j,\{d^i\}^l_j}\sum_{N}\int_0^{\infty} \prod_{r=1}^{3L-3+N}d\tau_r \int \prod_{j=1}^{3L-3} d\Phi^j_f\, d\Phi^j_i 
	\no
	&\times&
		\prod_{l=1}^{2L-2} \delta\left(\{d^f\}^l_j\Phi^j_{f;b_f} - \{d^i\}^l_j\Phi^j_{i;b_i}\right) \prod_{j=1}^{3L-3}B_j\left(\phi,\Phi_{f;b_f}^j,\Phi_{i;b_i}^j\right)\, ,
\label{LamplUI}
\eqne
where $\sum_{\{d^f\}^l_j,\{d^i\}^l_j}$ denotes the sum over all different choices of $\{d^f\}^l_j$ and $\{d^i\}^l_j$ and $\sum_{N}$ denotes the sum over all different distributions and permutations of external vertices. This equation will be the basis of the study of BRST invariance of the amplitude.

In the unintegrated picture, the amplitude is
\eqnb
A^{(L)}_{YM}(s_{ij})
	&=&
		\sum_{\{d^f\}^l_j,\{d^i\}^l_j} \int_0^{\infty} dT_1\ldots dT_{3L-3} \int_{F_L} \prod_{r=1}^{N}d\tau_r \int \prod_{j=1}^{3L-3} d\Phi^j_f\, d\Phi^j_i 
	\no
	&\times&
		\prod_{l=1}^{2L-2} \delta\left(\{d^f\}^l_j\Phi^j_{f;b_f} - \{d^i\}^l_j\Phi^j_{i;b_i}\right) \prod_{j=1}^{3L-3}B_j\left(\phi,\Phi_{f;b_f}^j,\Phi_{i;b_i}^j,T_j\right)\, .
\label{LamplI}
\eqne
To match this with the expression of the functional integral in \cite{{Bjornsson:2010wm}}, we observe that the sum over all inequivalent $\{d^f\}^l_j$ and $\{d^i\}^l_j$ matrices is equal to the sum over all different skeleton graphs. Therefore, the amplitude prescription in (\ref{LamplI}) is equal to the functional integral
\eqnb
A^{(L)}_{YM}(s_{ij})
	&=&
		\sum_{F_L} A^{(F_L)}_{YM}(s_{ij})
	\no
A^{(F_L)}_{YM}(s_{ij})
	&=&
		\int_0^{\infty} dT_1\ldots dT_{3L-3} \int_{F_L} \prod_{r=1}^{N}d\tau_r K^{F_L}_{YM}\left(\{k_r\},T_j,\tau_r\right)
	\no
K^{F_L}_{YM}\left(\{k_r\},T_j,\tau_r\right)
	&=&
		\int \mathcal{D}\Phi\,\mathcal{D}\phi\, \mathcal{N}\, \prod_{j=1}^{3L-3} b^j \, \prod_{r=1}^{N} V_{YM}^{r}\left(k_r,\tau_r\right)\, ,
\eqne
where $b^j$ denotes the $b$ ghost insertion on line $j$ and $\sum_{F_L}$ denotes the sum over all possible skeletons.

We will now show that each component of a coordinate field is a world-line  scalar with a single zero mode while each momentum field is a world-line vector and has $L$ zero modes on a skeleton with $L$ loops. This is a generalisation of the discussion in section \ref{sec:firstquant}. Consider the basic bolding block $B_j$ in (\ref{NpropVYM}) add two identity operators in $B_j$
\eqnb
\int\left.\left<\Phi_{f,b_f};T_j\right|\phi_1^j;T_j\right>d\phi_1^j\left<\phi_1^j;T_j\right|\prod_{r=1}^{N} V_{YM}^{r}\mathcal{N}b\left|\phi_2^j;0\right>d\phi_2^j\left.\left<\phi_2^j;0\right|\Phi_i;0\right>.
\label{phiBYM}
\eqne
One can then integrate over a subset of the momenta fields. This subset can be separated by a change of variables for the momenta fields $\Phi^j_{f;b_f} = \Phi^I_{b_f}\, (\omega_I/d\tau_j) + \Phi'^j_{f;b_f}$ and $\Phi^j_{i;b_i} = \Phi^I_{b_i}\, (\omega_I/d\tau_j) + \Phi'^j_{i;b_i}$. There are $L$ $\Phi^I$ and $5L-6$ $\Phi'^{j}_{f}$ and $\Phi'^{j}_{i}$ for each component. This change of variables makes the delta functions independent of the fields $\Phi^I$. By first integrating over $2L-2$ of the fields $\Phi'$ using the delta functions, the $3L-4$ remaining integrals of $\Phi'$ will give $3L-4$ delta functions involving each of the fields $\phi^j_{1}$ and $\phi^j_{2}$. As each propagator involves a delta function of $\phi^j_1$ and $\phi^j_2$ (see (\ref{phiBYM})), one have in total $6L-7$ delta functions. Performing the integrals over the surplus $\phi$'s give
\eqnb
K^{F_L}_{YM}\left(\{k_r\},T_j,\tau_r\right)
	&=&
		\int d\phi \prod_{I=1}^L d\Phi^I \prod_{j=1}^{3L-3}B_j\left(\phi,\Phi^I_{b_f}\left(\frac{\omega_I}{d\tau_j}\right),\Phi^I_{b_i}\left(\frac{\omega_I}{d\tau_j}\right)\right)\,.
\label{proofzero}
\eqne
This shows that the coordinate fields are world-line scalars and the conjugate 
momenta fields are world-line vectors. These properties are a consequence of imposing locality of the interaction and integrating the interaction point over space-time and pure spinor space. Note that the integration over the surplus fields can give contractions between different fields. This will depend crucially on the skeleton considered. We will not consider these in detailed, as we are mainly interested in the qualitative behaviour of the amplitudes.

We conclude the Yang--Mills discussion by obtaining expressions for three- and four-point tree amplitudes. The three-point tree amplitude is 
\eqnb
A^{(Tree)}_{YM}\left(\{k_r\}\right)
	&=&
		\lim_{T_j\rightarrow 0}\left<U_{YM}^{a_1}\left(k_1,T_1\right)U_{YM}^{a_2}\left(k_2,T_2\right)U_{YM}^{a_3}\left(k_1,T_3\right)\right> \, ,
\eqne
where the proper time of the vertex is $\tau=0$. The colour factor of the interaction is $f^{a_1 a_2 a_3}$.

The four-point amplitude is defined using the three-point amplitude by introducing a second vertex and integrating over its position (see figure \ref{fig:fourpointtree}(a)-(c))
\eqnb
&&A^{(Tree)}_{YM}\left(s,t,u\right)
\no
&&\phantom{12}=
		\lim_{T_j\rightarrow \infty} \sum_{j=1}^{3}\int_{0}^{T_j}d\tau \left<0\right| U_{YM}^{a_1}\left(k_1,T_1\right) U_{YM}^{a_2}\left(k_2,T_2\right) U_{YM}^{a_3}\left(k_3,T_3\right) V_{YM}^{a_4}\left(k_4,\tau\right) \left|0\right>\, .
\no
\eqne
As discussed at the beginning of this section, this amplitude can be separated into different parts corresponding to different colour orderings. For the four-point tree amplitude this is fairly simple as there are only single-trace terms. For example, the  $\Tr\left[t^{a_1}t^{a_2}t^{a_3}t^{a_4}\right]$ factor has only $s$- and $t$-channel contributions. Separating this factor from the amplitude gives
\eqnb
A^{(Tree)}_{YM}\left(s,t\right)
	&=&
		\lim_{T_j\rightarrow \infty}
		\left\{
		\int_{0}^{T_1}d\tau \left<0\right| U_{YM}\left(k_1,T_1\right) U_{YM}\left(k_2,T_2\right) U_{YM}\left(k_3,T_3\right) V_{YM}\left(k_4,\tau\right) \left|0\right> 
		\right.
	\no
	&+&
		\left.
		\int_{0}^{T_3}d\tau \left<0\right| U_{YM}\left(k_1,T_1\right) U_{YM}\left(k_2,T_2\right) U_{YM}\left(k_3,T_3\right) V_{YM}\left(k_4,\tau\right) \left|0\right> 
		\right\}\, .
\eqne
We have now defined the amplitudes for Yang--Mills. Before discussing BRST invariance, we will consider the supergravity amplitude.

\subsection{Supergravity}

The definitions of the supergravity amplitudes follow from a series of steps similar to those used in the Yang--Mills case. This mimics the transition from the open string to the closed string by doubling the fields, with the exception $X$ and $P$. Another difference is that supergravity does not have colour. As we will see, these properties of the supergravity construction are essential for the BRST consistency of the amplitude. In this case it is natural to take the propagator  between momenta $\Phi_i$ at proper time 0 and $\Phi_f$ at proper time $T_j$ to be
\eqnb
\left<\Phi_f;T_j\left|\mathcal{N}\, \hat{\mathcal{N}}\, b\, \hat{b}\right|\Phi_i;0\right>\, .
\eqne
The propagator is bosonic because there are two insertions of the $b$ ghosts. 
One argument of the doubling of the $b$ ghosts is that one corresponds to $b_{tot} = b + \hat{b}$ and the other the level matching condition, $b - \hat{b}$. A more indirect argument is that one needs two $b$ ghosts for the amplitudes to be compatible with BRST invariance. We have also inserted a regulator and assumed that it can be written as a product of two functions, where both are assumed to be equal to the Yang--Mills regulator\footnote{One factor is a function of the unhatted fields and the other is a function of the hatted fields.}. Observe that the conjecture is also in line with the KLT-relations \cite{Kawai:1985xq}, which relate amplitudes for the open string and the closed string.

The basic bolding block of the amplitudes is the propagator which absorbs $N$ physical particles evaluated between two general momentum states. From the discussion in section~\ref{sec:firstquant} we know that there are two different pictures,
\eqnb
B^U_j\left(\phi,\Phi_f,\Phi_i,\{k_r\},\{\tau_s\}\right)
	&=&
		\left<\Phi_f;\sum_{s=1}^{N+1}\tau_s\left| \prod_{r=1}^{N} \left(b\, \hat{b}\, U_{SG}^{r}\right)\, \mathcal{N}\, \hat{\mathcal{N}}\,b\, \hat{b}\right|\Phi_i;0\right>
\label{NpropUSG}
	\\
B^I_j\left(\phi,\Phi_f,\Phi_i,\{k_r\},\{\tau_r\};T_j\right)
	&=&
		\left<\Phi_f;T_j\right| \prod_{r=1}^{N} V_{SG}^{r}\,\mathcal{N}\, \hat{\mathcal{N}}\, b\, \hat{b}\left|\Phi_i;0\right>\, .
\label{NpropVSG}
\eqne
As for Yang--Mills, one can show that (\ref{NpropUSG}) and (\ref{NpropVSG}) are equivalent up to a BRST exact term\footnote{Note that one has to restrict the positions of the vertices in the integrated picture ($\tau_{r-1}\geq \tau_{r}$)  or include permutations of the vertices in the unintegrated picture for the relation to be true.}. Since $B_j^I \sim B_j^U$, in the following it will not be necessary to keep the superscript $I$ or $U$.

The three-point vertex in supergravity satisfies the same conditions as the vertex in Yang--Mills. The important properties are that the interaction is local and conserves momentum. If the vertex is contracted with three general momentum states it reduce to a delta function
\eqnb
\int\, d\phi\, \left<V_{SG};\tau\right|\left(\left|\Phi^1_f;\tau\right>\, \left|\Phi^2_f;\tau\right>\, \left<\Phi^3_i;\tau\right|\right)
&=&
\int\, d\phi\, \left<V_{SG};\tau\right|\left(\left|\Phi^1_f;\tau\right>\, \left|\Phi^2_f;\tau\right>\, \left|-\Phi^3_i;\tau\right>\right)
\no
&=&
\delta\left(\Phi^1_f + \Phi^2_f - \Phi^3_i\right)\, .
\eqne
A difference between the Yang--Mills and the supergravity cases is that the delta function is bosonic and therefore is an even function.

Gluing together the two ends of the operator in (\ref{NpropUSG}), which absorbs $N-1$ physical particles, with an unintegrated vertex operator involving the $N$'th physical particle, one obtains the expression for the one-loop amplitude in the unintegrated picture
\eqnb
A^{(1)}_{SG}\left(s_{ij}\right)
	&=&
		\sum_{\sigma \in S_{N-1}} \int_{0}^{\infty} \prod_{r=1}^{N}d\tau_r \int d\Phi \left<\Phi; \sum_{s=1}^{N}\tau_s\left|\prod_{r=1}^{N}\left(b\hat{b}U_{SG}^{\sigma(r)}\right)b\hat{b}U_{SG}^{N}\right|\Phi;0\right>\, . 
\eqne
In the integrated picture, the one-loop amplitude is
\eqnb
A^{(1)}_{SG}(s_{ij})
&=&
\int_0^{\infty} dT \int_0^T \prod_{r=1}^{N-1} d\tau_r \int d\Phi \left<\Phi;T\right| \prod_{r=1}^{N-1} V_{SG}^{r}\, \mathcal{N}\,\hat{\mathcal{N}} \, b\, \hat{b}\,U_{SG}^{N}\left|\Phi;0\right>\, ,
\eqne
which is equal to the functional integral
\eqnb
A^{(1)}_{SG}(s_{ij})
&=&
\int_0^{\infty} dT \int_0^T \prod_{r=1}^{N-1} d\tau_r K_{SG}\left(\{k_r\},T,\tau_r\right)
\no
K_{SG}\left(\{k_r\},T,\tau_r\right)
&=&
\int \cD\phi\, \cD \Phi\, \mathcal{N}\,\hat{\mathcal{N}}\, b\, \hat{b}\, \prod_{r=1}^{N-1} V_{SG}\left(k_r,\tau_r\right)U_{SG}\left(k_N,0\right)\, .
\eqne
In the expression above we have separated the amplitude into two parts. One is the functional integral with fixed length of the moduli and position of the vertices. The second part is the integration over the moduli and position of the vertices.

The multi-loop amplitudes are determined in the same way as Yang--Mills. In the unintegrated picture the amplitude is
\eqnb
A^{(L)}_{SG}(s_{ij})
	&=&
		\sum_{\{d^f\}^l_j,\{d^i\}^l_j}\sum_{N}\int_0^{\infty} \prod_{r=1}^{3L-3+N}d\tau_r \int \prod_{j=1}^{3L-3} d\Phi^j_f\, d\Phi^j_i\, 
	\no
	&\times&
		\prod_{l=1}^{2L-2}\delta\left(\{d^f\}^l_j\Phi^j_f -\{d^i\}^l_j\Phi^j_i\right) \prod_{j=1}^{3L-3} B_j\left(\phi,\Phi^j_f,\Phi^j_i\right)\, .
\eqne
In the integrated picture, the amplitude is
\eqnb
A^{(L)}_{SG}(s_{ij})
	&=&
		\sum_{\{d^f\}^l_j,\{d^i\}^l_j} \int_0^{\infty} dT_1\ldots dT_{3L-3} \int_{F_L} \prod_{r=1}^{N}d\tau_r \int \prod_{j=1}^{3L-3} d\Phi^j_f d\Phi^j_i 
	\no
	&\times&
		\prod_{l=1}^{2L-2}\delta\left(\{d^f\}^l_j\Phi^j_f -\{d^i\}^l_j\Phi^j_i\right) \prod_{j=1}^{3L-3} B_j\left(\phi,\Phi^j_f,\Phi^j_i;T_j\right)\, .
\label{GravityAmpLI}
\eqne
This expression of the amplitude is equal to \cite{Bjornsson:2010wm},
\eqnb
A^{(L)}_{SG}
	&=&
		\sum_{F_L}A^{(F_L)}_{SG}
	\no
A^{(F_L)}_{SG}
	&=&
		\int_{0}^{\infty} dT_1\ldots dT_{3L-3} \int_{F_L} \prod_{r=1}^{L} d\tau_r K^{F_L}_{SG}\left(\{k_r\}, T_j, \tau_r \right)
	\no
K^{F_L}_{SG}\left(\{k_r\}, T_j, \tau_r \right)
	&=&
		\int \cD \phi\, \cD \Phi\, \cN\, \hat{\cN}\, \prod_{j=1}^{3L-3}b^j\, \hat{b}^j\, \prod_{r=1}^{N} V_{SG}\left(k_r,\tau_r\right)\, ,
\eqne 
where $b^j\, \hat{b}^j$ denote the insertion of the term $b\,\hat{b}$ on line $j$. By doing the same analysis as in Yang--Mills, one can show that the field $\Phi$ has $L$ zero modes for each component and $\phi$ has one zero mode for each component on a skeleton with $L$ loops.

As in the Yang--Mills case, we will find that the above prescription has to be generalised to be consistent with BRST invariance. This generalisation can be expressed in one of two ways -- either by including higher-point external vertices or by the addition of trees attached to the skeleton diagram. 

The three- and four-point amplitude is defined in the same way as for the Yang--Mills case. The three-point tree amplitude is
\eqnb
A^{(Tree)}_{SG}\left(k_j\right)
	&=&
		\lim_{T_j\rightarrow\infty}\left<0\right|U_{SG}\left(k_1,T_1\right)U_{SG}\left(k_2,T_2\right)U_{SG}\left(k_3,T_3\right)\left|0\right>\, ,
\eqne
where the proper time of the vertex is $\tau=0$. The four-point tree diagram is \eqnb
A^{(Tree)}_{SG}(s,t,u)
	&=&
		\lim_{T_j\rightarrow \infty} \sum_{j=1}^{3} \int_0^{T_j}d\tau \left<0\right| U_{SG}\left(k_1,T_1\right)U_{SG}\left(k_2,T_2\right)U_{SG}\left(k_3,T_3\right)V_{SG}\left(k_4,\tau\right)\left|0\right>\, ,
	\no
\eqne
where the position of the fourth particle should be integrated over the whole three-point diagram.

%%%%%%%%%%%%%%%%%%%%%%%%%
\section{BRST invariance}
\label{sec:BRST}
%%%%%%%%%%%%%%%%%%%%%%%%%

\begin{figure}[t]
\begin{center}
\begin{tabular}{cc}
{\rotatebox{0}{\scalebox{0.20}{\includegraphics*{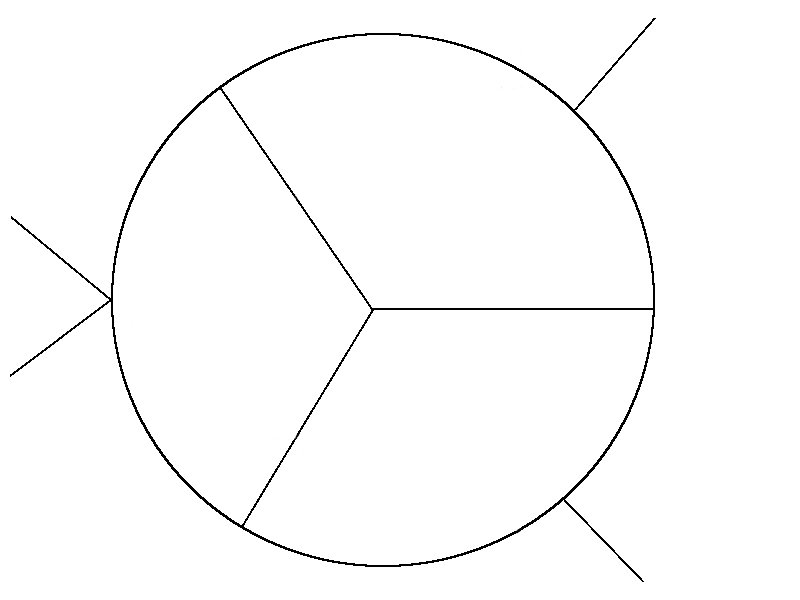}}}}
&
\hspace{0.5cm}{\rotatebox{0}{\scalebox{0.20}{\includegraphics*{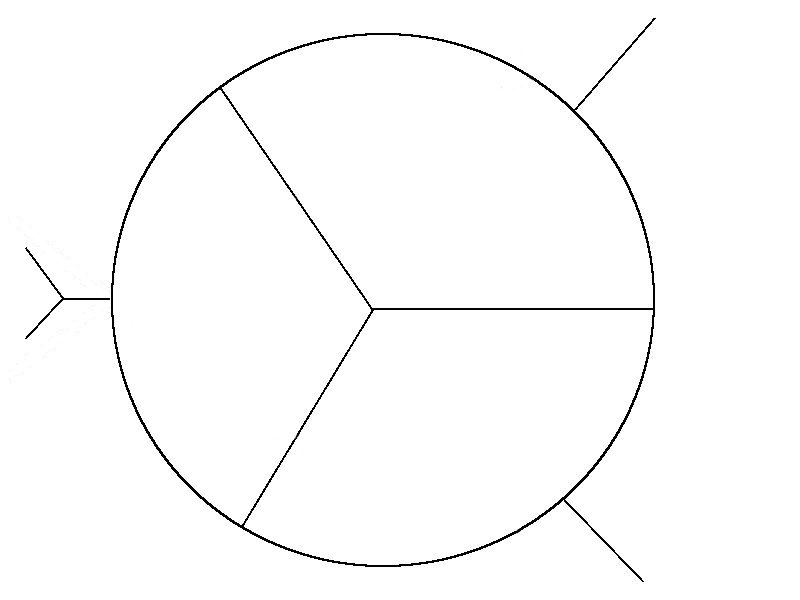}}}}
\\
(a) & (b)
\end{tabular}
\end{center}
\caption{(a) A three-loop skeleton with a four-point contact term attached. (b) A three-loop skeleton with a tree attached.}
\label{fig:contactterms}
\end{figure}

In the previous section presented the propagator and the internal three-point vertex for Yang--Mills and supergravity in terms of the pure spinor fields. This led to the three- and four-point tree as well as loop amplitudes with any number of external states. In this section we will make the first non-trivial test of these expressions. We will see that the amplitudes need to be generalised to be consistent with BRST invariance. This can be expressed in two equivalent ways. Either by including higher-point external vertices (as in figure~\ref{fig:contactterms}(a)) or by addition of trees attaching the skeleton (as in figure~\ref{fig:contactterms}(b)). We will choose the former as it makes the ultraviolet properties of the amplitude more transparent. In addition to these higher-point contact interaction\footnote{A contact term is a vertex that has more than three points.}  there are ``effective'' contact terms generated by contractions of $P$ factors that enters into the $b$ ghost and vertex operators.

The regulator is a BRST trivial term and will be suppressed in the analysis. The only time it could have an effect is when vertices collide. As the regulator is assumed to have the form $\cN = 1 + \epsilon\, \int_0^{T_i}d\tau\, [Q,\chi(\tau)] + \cO\left(\epsilon^2\right)$ and the function $\chi(\tau)$ is a continuous function, the regulator satisfies $\lim_{T_i\rightarrow 0}\cN = 1$. Therefore, the regulator will not contribute when vertices collide.

In this section we will use the unintegrated picture and analyse the two theories separately as there are some important differences. We will first consider loop-amplitudes as the BRST invariance of the trees follows from this.

\subsection{Yang--Mills}

Consider first the one-loop amplitude and make a BRST transformation of the $N$'th vertex, 
\eqnb
U_{YM}^N &\rightarrow& U_{YM}^N + [Q,\rho],
\eqne
where $\rho$ is an arbitrary function of the coordinate fields. The BRST charge can be commuted through the vertices and $b$ ghosts to act on the bra and ket,
\eqnb
\delta_{\rho}A^{(1)}_{YM}(s_{ij})
	&=&
		\sum_{\sigma\in S_{N-1}} \int_0^{\infty}\prod_{r=1}^{N}d\tau_{r}\, \int d\Phi\, \kappa^{cd}\,Q \left<\Phi_c;\sum_{s=1}^{N}\tau_r\left|\prod_{r=1}^{N-1}\left(bU_{YM}^{\sigma(r)}\right)b\rho\right|\Phi_d;0\right>
	\no
	&-&
		\sum_{\sigma\in S_{N-1}} \int_0^{\infty}\prod_{r=1}^{N}d\tau_{r}\, \int d\Phi\,\kappa^{cd} \left<\Phi_c;\sum_{s=1}^{N}\tau_r\left|\prod_{r=1}^{N-1}\left(bU_{YM}^{\sigma(r)}\right)b\rho\right|\Phi_d;0\right>Q
	\no
	&+&
		\sum_{\sigma\in S_{N-1}} \int_0^{\infty}\prod_{r=1}^{N}d\tau_{r}\, \int d\Phi\,\kappa^{cd}\, U^{\sigma(1)} \left<\Phi_c;\sum_{s=2}^{N}\tau_s\left|\prod_{r=2}^{N-1}\left(bU_{YM}^{\sigma(r)}\right)b\rho\right|\Phi_d;0\right>
	\no
	&-&
		\sum_{\sigma\in S_{N-1}} \int_0^{\infty} \prod_{r=1}^{N-1}d\tau_r \int d\Phi\, \kappa^{cd}\,\left<\Phi_c;\sum_{s=1}^{N-1}\tau_s\left|\prod_{r=1}^{N-2}\left(bU_{YM}^{\sigma(r)}\right)bU_{YM}^{\sigma(N-1)}\rho\right|\Phi_d;0\right>
	\no
	&-&
		\sum_{\sigma\in S_{N-1}} \sum_{s=2}^{N-1}\int_0^{\infty}\!\!\! \prod_{r=1\,r\neq s }^{N}d\tau_r \int d\Phi\, \kappa^{cd}\,\left<\Phi_c;\sum_{r=1 r\neq s}^{N}\tau_r\right|\prod_{r=1}^{s-2}\left(bU_{YM}^{\sigma(r)}\right) 
	\no
	&\times&
		\left.\left.b\left(U_{YM}^{\sigma(s-1)}U_{YM}^{\sigma(s)}\right)\prod_{r=s+1}^{N-1}\left(bU_{YM}^{\sigma(r)}\right) b\rho\right|\Phi_d;0\right>\, ,
\label{BRST1loop}
\eqne
where we have used $[Q,b]=H$ and the fact that an insertion of the Hamiltonian is equal to a total derivative w.r.t.\ the length of the propagator. Integration over this modulus gives an overall minus sign and a contact term. This contact term is expressed in the last line of (\ref{BRST1loop}) as two vertex operators adjacent to each other without a $b$ ghost inserted between them\footnote{For loop amplitudes, the upper limit vanishes for fixed nonzero values of the other moduli. For tree amplitudes, the upper limit vanishes after an appropriate analytic continuation of the momenta.}. The two terms involving the BRST charge cancel and one can move the $U^{\sigma(1)}$ vertex in the third term to the right. Using the symmetric group one can show that the amplitude is BRST invariant up to contact terms
\eqnb
\delta_{\rho}A_{YM}(s_{ij})
	&=&
		-\sum_{\sigma\in S_{N-1}} \int_0^{\infty} \prod_{r=1}^{N-1}d\tau_r \int d\Phi\, \kappa^{cd}
	\no
	&\times&
		\left<\Phi_c;\sum_{s=1}^{N-1}\tau_r\left|\prod_{r=1}^{N-2}\left(bU_{YM}^{\sigma(r)}\right)[U_{YM}^{\sigma(N-1)},\rho]\right|\Phi_d;0\right>
	\no
	&-&
		\frac{1}{2}\sum_{\sigma\in S_{N-1}} \sum_{s=2}^{N-1}\int_0^{\infty} \!\!\!\!\prod_{r=1\,r\neq s }^{N} \!\!d\tau_r \int d\Phi\, \kappa^{cd}\,\left<\Phi_c;\sum_{r=1\, s\neq r}^{N}\tau_r\right|\prod_{r=1}^{s-2}\left(bU_{YM}^{\sigma(r)}\right) 
	\no
	&\times&
		\left.\left.
		b[U_{YM}^{\sigma(s-1)},U_{YM}^{\sigma(s)}]\prod_{r=s+1}^{N-1}\left(bU_{YM}^{\sigma(r)}\right) b\rho\right|\Phi_d;0\right>\, .
\eqne
Soon we will see how these can be cancelled by addition of contact vertices. Before that we will analyse the multi-loop amplitude.

\begin{figure}[t]
\begin{center}
\begin{tabular}{ccc}
{\rotatebox{0}{\scalebox{0.5}{\includegraphics*{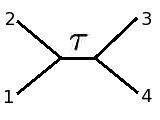}}}}
&
{\rotatebox{0}{\scalebox{0.5}{\includegraphics*{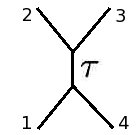}}}}
&
{\rotatebox{0}{\scalebox{0.5}{\includegraphics*{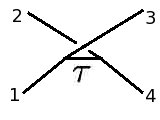}}}}
\\
(a) & (b) & (c)
\end{tabular}
\end{center}
\caption{The three different diagrams describing interaction between four particles involving only three-point vertices.} 
\label{fig:4pointYM}
\end{figure}

A BRST transformation of one of the external vertices has the form $[Q,\rho] = Q\rho - \rho Q$. One can move the BRST charge in the first of these terms to the left around the network using the BRST invariance of the internal vertices and external states so that it cancels the second term. This is a generalisation of the one-loop discussion. The only part in the amplitude which has nonzero commutator with the BRST charge is the $b$ ghost. This will produce one insertion of the Hamiltonian, which can be written as a total derivative w.r.t.\ to the moduli. Integrating over this modulus produces a contact term with four points. Therefore, four-point amplitudes with off-shell states form an essential part of the analysis. 

Pairs of three-point vertices can be glued together to form three distinct contributions to the four-point amplitude (as shown in figure~\ref{fig:4pointYM}(a)-(c)), which we call the $s$-, $t$- and $u$-channel, respectively. Including colour explicitly is essential in the analysis, so we separate it from the kinematics. The contributions of the different channels to the amplitude are
\eqnb
s:&&\kappa_{de}f^{a_1a_2d}f^{a_3a_4e}\int_0^{\infty}d\tau\, \left<U_{YM}^1U_{YM}^2\,b\,U_{YM}^3U_{YM}^4\right> \no
t:&&\kappa_{de}f^{a_4a_1d}f^{a_2a_3e}\int_0^{\infty}d\tau\, \left<U_{YM}^4U_{YM}^1\,b\,U_{YM}^2U_{YM}^3\right> \no
u:&&\kappa_{de}f^{a_1a_3d}f^{a_4a_2e}\int_0^{\infty}d\tau\, \left<U_{YM}^1U_{YM}^3\,b\,U_{YM}^4U_{YM}^2\right>\, ,
\eqne
where $U^a$ in this equation is an unintegrated vertex operator which can be off-shell. This operator only depends on the coordinate fields $X^m, \theta^{\alpha},\lambda^{\alpha},\bar{\lambda}_{\alpha}$ and $r_{\alpha}$. Consider the case when a BRST charge acts only on the $b$ ghost insertion. As described above, one will then get an insertion of the Hamiltonian between the two sets of vertices. This can be exchanged by a total derivative w.r.t.\ the modulus for the propagator. Integrating over this modulus, the two three-point vertices will collide and one has in each case a four-point vertex
\eqnb
s:&&-\kappa_{de}f^{a_1a_2d}f^{a_3a_4e}\left<U_{YM}^1U_{YM}^2U_{YM}^3U_{YM}^4\right> \no
t:&&-\kappa_{de}f^{a_4a_1d}f^{a_2a_3e}\left<U_{YM}^4U_{YM}^1U_{YM}^2U_{YM}^3\right> \no
u:&&-\kappa_{de}f^{a_1a_3d}f^{a_4a_2e}\left<U_{YM}^1U_{YM}^3U_{YM}^4U_{YM}^2\right>\, . 
\eqne
Observe that the operators in the above equation can be reordered to have the same kinetic part. Summing the $s$-, $t$- and $u$-channels one will get
\eqnb
-\left(f^{a_1a_2d}{f^{a_3a_4}}_d - f^{a_4a_1d}{f^{a_2a_3}}_d + f^{a_1a_3d}{f^{a_4a_2}}_d\right) \left<U_{YM}^1U_{YM}^2U_{YM}^3U_{YM}^4\right>
&=&
0\, ,
\label{BRSTrelation}
\eqne
where we have used the Jacobi identity. The consequences of this identity are important. Consider first the case when one of the operators above describes a physical particle. This shows that there is no contribution when an external three-point vertex collides with an internal vertex as long as one has the $s$-, $t$- and $u$-channel contributions. Note that this also applies when one of the colliding vertices is the one, which is BRST exact. The case when all vertex operators are off-shell corresponds to collisions of two internal vertices. The above analysis shows that this is cancelled in the sum of the $s$-, $t$- and $u$-channel contributions.

\begin{figure}[t!]
\begin{center}
\begin{tabular}{cc}
{\rotatebox{0}{\scalebox{0.25}{\includegraphics*{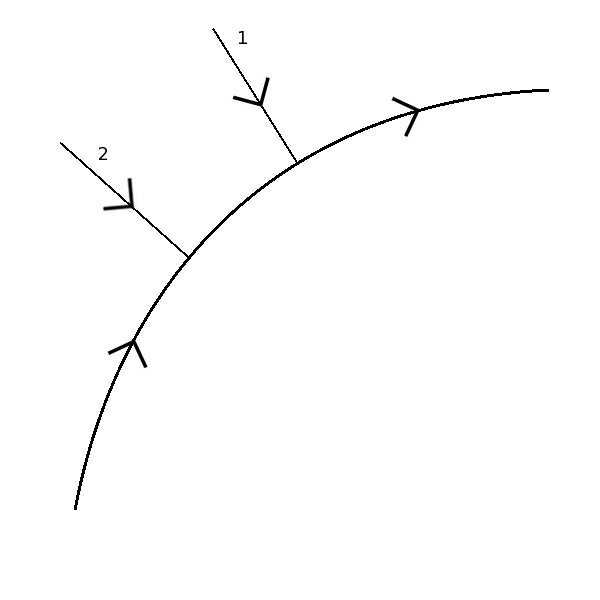}}}}
& {\rotatebox{0}{\scalebox{0.25}{\includegraphics*{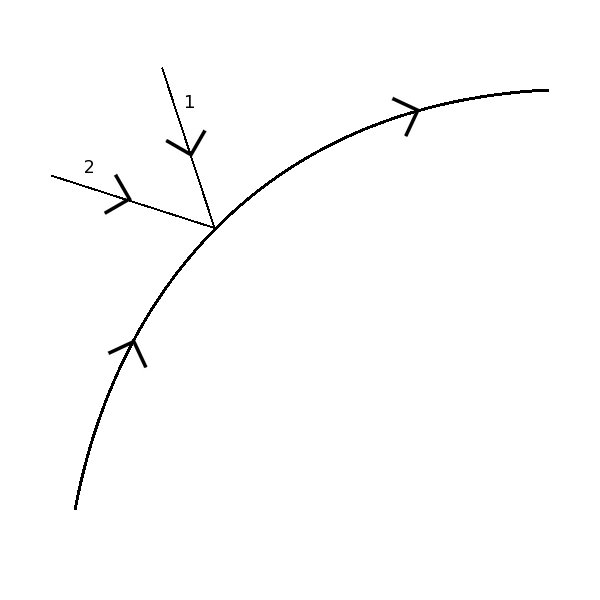}}}} \\
(a) & (b)
\end{tabular}
\caption{The propagation of a particle absorbing two physical states through: (a) Two three point vertices. (b) One four-point vertex.}
\label{Figure:construct}
\end{center}
\end{figure}

The conclusion is that the only case when there is a possibility for the amplitude to not be BRST invariant is when two vertices producing external physical particles collide. This violation of BRST invariance can be cancelled by the addition of an appropriate four-point contact vertex that does not enter the tree amplitude. This vertex can be determined by considering a four-point interaction with two on-shell states. This consists of the sum of two contributions. One of these is formed by gluing together two three-point vertices together with a propagator (figure~\ref{Figure:construct}(a)) while the second is obtained from the four-point vertex under consideration, $U_{YM}^{12}$ (figure~\ref{Figure:construct}(b)),
\eqnb
&&
\sum_{\sigma\in S_2} \int_0^{\infty}\prod_{s=1}^{3} d\tau_{s}\, B_j\left(\phi,\Phi_{f;b^j_f}^j,\Phi_{i;b^j_i}^j,\{k_{\sigma(r)}\},\{\tau_{s}\}\right)
\no
&&+
\int_0^{\infty}\prod_{s=1}^{2}d\tau_s\, B^2_j\left(\phi,\Phi_{f;b^j_f}^j,\Phi_{i;b^j_i}^j,\{k_1,k_2\},\{\tau_s\}\right)
\no
&&=
\sum_{\sigma\in S_2} \int_0^{\infty}\prod_{s=1}^{3} d\tau_{s}\, \left<\Phi^j_{f;b^j_f};T\right|\prod_{r=1}^{2}\left(bU_{YM}^{\sigma(r)}\right)b\left|\Phi^j_{i;b^j_i};0\right>
\no
&&+
\int_0^{\infty}\prod_{s=1}^{2}d\tau_s\, \left<\Phi^j_{f;b^j_f};T\right|bU^{12}_{YM}b\left|\Phi^j_{i;b^j_i};0\right>\, .
\eqne
Multiplying this equation by $Q^j_i$ from the right and rewriting it as $Q^j_f$ on the left together with terms where the BRST charge has commuted with vertex operators and $b$ ghosts gives
\eqnb
&&
\sum_{\sigma\in S_2} \int_0^{\infty}\prod_{s=1}^{3} d\tau_{s}\, B_j\left(\phi,\Phi_{f;b^j_f}^j,\Phi_{i;b^j_i}^j,\{k_{\sigma(r)}\},\{\tau_{\sigma(r)}\};T\right)Q^j_i 
\no
&&+
\int_0^{\infty}\prod_{s=1}^{2}d\tau_s\, B^2_j\left(\phi,\Phi_{f;b^j_f}^j,\Phi_{i;b^j_i}^j,\{k_1,k_2\},\tau;T\right)Q^j_i
\no
&&=
\sum_{\sigma\in S_2} \int_0^{\infty}\prod_{s=1}^{3} d\tau_{s}\, Q^j_f\, B_j\left(\phi,\Phi_{f;b^j_f}^j,\Phi_{i;b^j_i}^j,\{k_{\sigma(r)}\},\{\tau_{\sigma(r)}\};T\right)
\no
&&+
\int_0^{\infty}\prod_{s=1}^{2}d\tau_s\, Q^j_f\, B^2_j\left(\phi,\Phi_{f;b^j_f}^j,\Phi_{i;b^j_i}^j,\{k_1,k_2\},\tau;T\right)
\no
&&-
\sum_{\sigma\in S_2} \int_0^{\infty} d\tau_{1}\, d\tau_{3}\, \left<\Phi^j_{f;b^j_f};T\right|b[U_{YM}^{\sigma(r)},U_{YM}^{\sigma(2)}]b\left|\Phi^j_{i;b^j_i};0\right>
\no
&&+
\int_0^{\infty} d\tau_1\, d\tau_2\left<\Phi^j_{f;b^j_f};T\right|b[U_{YM}^{12},Q]b\left|\Phi^j_{i;b^j_i};0\right>\, + \ldots\, ,
\eqne
where the dots denote terms where the BRST charge commutes with the $b$ ghosts adjacent to the bra or the ket. From the derivation above one obtains the equation, which the four-point vertex satisfies
\eqnb
[Q,U_{YM}^{12}] &=& \frac{1}{2}\left([U_{YM}^1,U_{YM}^2]+[U_{YM}^2,U_{YM}^1]\right)\, ,
\label{defeqnfourpointYM}
\eqne
which is similar to (\ref{nonlYM}). As expected, the four-point vertex is connected to non-linear effects. Let us go through the solution in some detail. We restrict to the case with external bosons (the generalisation to fermions is straightforward). Consider a solution of the equations of motion for Yang--Mills consisting of two plane waves. The solution for the $\theta^0$-component of the superfield $A_m$ is
\eqnb
\left.A_m\left(k_1,k_2\right)\right|_{\theta=0} &=& a^1_m e^{ik_1X} + a^2_m e^{ik_2X}\, .
\eqne
As the equations of motion in Yang--Mills are non-linear, there are terms mixing the two plain waves. Consider the superfield $A_{\alpha}$, which is involved in the unintegrated vertex operator. The solution to the equations of motion for this field can be written as a sum of three terms\footnote{We have here suppressed the $\theta$ dependence of the superfields.}
\eqnb
A_{\alpha}\left(k_1,k_2\right)
	&=&
		A_{\alpha}^{1;l}\left(k_1\right) + A_{\alpha}^{2;l}\left(k_2\right) + A_{\alpha}^{12;nl}\left(k_1,k_2\right)\, .
\eqne
The first two terms satisfy the linearised equations of motion for Yang--Mills. The third term includes the non-linear effects. Inserting this expression into the equation of motion and contracting the two free spinor indices with pure spinor fields
\eqnb
&&
\lambda^{\alpha}\lambda^{\beta}\left(D_{\alpha}A_{\beta} + \frac{1}{2}[A_\alpha,A_{\beta}]\right)
\no
&&
=
		\lambda^{\alpha}\lambda^{\beta}\left(D_{\alpha}A_{\beta}^{12;nl} \left(k_1,k_2\right) + \frac{1}{2}[A_{\alpha}^{1;l}\left(k_1\right) + A_{\alpha}^{2;l}\left(k_2\right),A_{\alpha}^{1;l}\left(k_1\right) + A_{\alpha}^{2;l}\left(k_2\right)]\right)\, ,
\no
&&
=0
\eqne
where we have used the fact that the fields $A_{\alpha}^{i;l}$ satisfy the linearised equations of motion and that the non-linear term commutes with the other fields. Therefore, the non-linear term satisfies
\eqnb
[Q,\lambda^{\alpha} A_{\alpha}^{12,nl}\left(k_1,k_2\right)]
	&=&
		\frac{1}{2}\left([\lambda^{\alpha} A_{\alpha}^{1,l}\left(k_1\right),\lambda^{\alpha} A_{\alpha}^{2,l}\left(k_2\right)] + [\lambda^{\alpha} A_{\alpha}^{2,l}\left(k_2\right),\lambda^{\alpha} A_{\alpha}^{1,l}\left(k_1\right)]\right)\, .\no
\eqne
Apart from $\epsilon$, this is the same equation as (\ref{nonlYM}). Therefore, the four-point vertex has the form
\eqnb
U_{YM}^{12} &\equiv& \lambda^{\alpha} A_{\alpha}^{12,nl}\, ,
\label{UInonlYM}
\eqne
where the colour factor of $U^{12}_{YM}$ is $f^{a_1a_2c}f^{def}\kappa_{cf}$. The unintegrated contact vertex can be defined as $V_{YM}^{12} = [U^{12}_{YM},b]$ just as the three-point vertex (up to BRST trivial terms). It follows that (c.f.\  (\ref{Vertexymi})) that
\eqnb
V^{12}_{YM}(k_1,k_2,\tau) &=& :P^mA^{nl}_m: - :d_{\alpha} W^{\alpha}_{nl}: + \frac{1}{2}N^{mn}\cF^{nl}_{mn}\, ,
\label{InonlYM}
\eqne
where $:\;:$ denotes normal ordering. 

We have now shown that adding one four-point vertex can compensate the non-BRST invariance of the amplitude with only external three-point vertices. However, the BRST transformation of the amplitude with one four-point vertex will generate yet more BRST non-invariant terms. Such terms are cancelled by adding additional contact terms. In the end one will have amplitudes with three-, four- and higher-point external vertices\footnote{The analysis above can be generalised to cases involving one, or more, higher-point vertices in straightforward way.}. The higher-point vertices will all satisfy equations of the form (\ref{defeqnfourpointYM}). To simplify the analysis, one can introduce a parameter, $\epsilon$, and consider a solution consisting of the $N$ external particles,
\eqnb
\left.A_m\left(k_1,\ldots,k_N\right)\right|_{\theta=0}
&=&
\epsilon \sum_{j=1}^{N}  a^j_m e^{ik^jX}\, .
\eqne
The solution $A_{\alpha}$ of the Yang--Mills equations can be written as a linear combination
\eqnb
A_{\alpha}\left(k_1,\ldots,k_N\right)
&=&
\epsilon \sum_{j=1}^{N} A^{j;l}_{\alpha} + \sum_{l=2}^{N}\epsilon^l\sum_{j_1<\ldots<j_l=1}^{N} A^{j_1\ldots j_l;nl}_{\alpha}\, .
\eqne 
We now insert this solution into the equations of motion contracted with two pure spinor fields,
\eqnb
\lambda^{\alpha}\lambda^{\beta}\left(D_{\alpha}A_{\beta} + \frac{1}{2}[A_\alpha,A_{\beta}]\right)
	&=&
		0\, ,
\eqne
which gives an equation that can be separated into different powers of $\epsilon$. The solution to order $\epsilon^2$ is a generalisation of the simplest case that give $U^{12}_{YM}$ (\ref{UInonlYM}). The equation to order $\epsilon^3$ describes the collision of a three- and a four-point vertex. This generalises to all higher orders in $\epsilon$. 

The conclusion is that the Yang--Mills loop amplitudes can be made compatible with BRST invariance by adding multi-point vertices. The form of the higher-point vertices is completely determined by the form of the three-point vertices.

The BRST invariance of the tree amplitudes follows from the above analysis. The three-point amplitude is trivially invariant and the invariance of the four-point amplitude follows from (\ref{BRSTrelation}) with four physical particles.

It is possible to describe the restoration of BRST invariance without introducing  contact interactions. This alternative description involves attaching trees to the skeleton as in Figure~\ref{fig:contactterms}(b). Roughly speaking, the propagators in the tree are cancelled by momenta in the tree vertices, leading to the contact term discussed above. This description corresponds to the discussion in the context of string theory \cite{Berkovits:2009aw} and in the context of three-loop field theory in \cite{Bern:2010ue}, which manifestly incorporates the duality between colour and kinematics \cite{Bern:2008qj}.

\subsection{Supergravity}

The discussion of BRST invariance in supergravity is analogous to the Yang--Mills case with some important differences. Consider first the one-loop amplitude and make a BRST transformation of the $N$'th vertex, $U_{SG}^N \rightarrow U_{SG}^N + [Q_{tot},\rho]$. Commuting the BRST charge through the vertices and $b$ ghosts to act on the bra and the ket one obtains
\eqnb
\delta_{\rho}A_{SG}^{(1)}\left(s_{ij}\right)
	&=&
		\sum_{\sigma \in S_{N-1}} \int_0^{\infty} \prod_{r=1}^{N}d\tau_r \int d\Phi \; Q_{tot}\,\left<\Phi;\sum_{r=1}^{N}\tau_r\left|\prod_{r=1}^{N-1}\left(b\hat{b}\,U_{SG}^{\sigma(r)}\right)b\hat{b}\,\rho\right|\Phi;0\right>
	\no
	&+&
		\sum_{\sigma \in S_{N-1}} \int_0^{\infty} \prod_{r=1}^{N}d\tau_r \int d\Phi \,\left<\Phi;\sum_{r=1}^{N}\tau_r\left|\prod_{r=1}^{N-1}\left(b\hat{b}\,U_{SG}^{\sigma(r)}\right)b\hat{b}\,\rho\right|\Phi;0\right>\,Q_{tot}
	\no
	&+&
		\frac{1}{2}\sum_{\sigma \in S_{N-1}} \int_0^{\infty} \prod_{r=2}^{N}d\tau_r \int d\Phi 
	\no
	&\times&
		\left<\Phi;\sum_{r=1}^{N-1}\tau_r\left|[{U}_{SG}^{\sigma(1)},b-\hat{b}]\prod_{r=2}^{N-1}\left(b\hat{b}\,U_{SG}^{\sigma(r)}\right)b\hat{b}\,\rho\right|\Phi;0\right>
	\no
	&-&
		\frac{1}{2}\sum_{s=2}^{N-1}\sum_{\sigma \in S_{N-1}} \int_0^{\infty} \prod_{r=1\,r\neq s}^{N}d\tau_r \int d\Phi \left<\Phi;\sum_{r=1\,r\neq s}^{N}\tau_r\right|\prod_{r=1}^{s-2}\left(b\hat{b}\,U_{SG}^{\sigma(r)}\right)
	\no
	&\times&
		\left.\left.b\hat{b}\left(U_{SG}^{\sigma(s-1)}\left(b-\hat{b}\right)U_{SG}^{\sigma(s)}\right)\prod_{r=s+1}^{N-1}b\hat{b}\,U_{SG}^{\sigma(r)}b\hat{b}\,\rho \right|\Phi;0\right>
	\no
	&-&
		\frac{1}{2}\sum_{\sigma \in S_{N-1}} \int_{0}^{\infty} \prod_{r=1}^{N-1}d\tau_r \int d\Phi
	\no
	&\times&
		\left<\Phi;\sum_{r=1}^{N-1}\tau_r\left|\prod_{r=1}^{N-2}\left(b\hat{b}\,U_{SG}^{\sigma(r)}\right)b\hat{b}\,[U_{SG}^{\sigma(N-1)},b-\hat{b}]\,\rho \right|\Phi;0\right>\, .
\eqne
The first two terms cancel. Using the symmetric group and the fact that the $b$ ghosts anti-commute, one obtains
\eqnb
A^{(1)}_{SG}\left(s_{ij}\right)
	&=&
		-\frac{1}{4}\sum_{s=2}^{N-1}\sum_{\sigma \in S_{N-1}} \int_0^{\infty} \prod_{r=1\,r\neq s}^{N}d\tau_r \int d\Phi \left<\Phi;\sum_{r=1\,r\neq s}^{N}\tau_r\right|\prod_{r=1}^{s-2}\left(b\hat{b}\,U_{SG}^{\sigma(r)}\right)
	\no
	&\times&
		\left.\left.b\hat{b}\,\left[U_{SG}^{\sigma(s-1)},\left[b-\hat{b},U_{SG}^{\sigma(s)}\right]\right]\prod_{r=s+1}^{N-1}\left(b\hat{b}\,U_{SG}^{\sigma(r)}\right)b\hat{b}\,\rho \right|\Phi;0\right>
	\no
	&-&
		\frac{1}{2}\sum_{\sigma \in S_{N-1}} \int_0^{\infty} \prod_{r=1}^{N-1}d\tau_r \int d\Phi 
	\no
	&\times&
		\left<\Phi; \sum_{r=1}^{N-1}\tau_r\left|\prod_{r=1}^{N-2}\left(b\hat{b}\,U_{SG}^{\sigma(r)}\right)b\hat{b}\,\left[\left[U_{SG}^{\sigma(N-1)},b-\hat{b}\right],\rho\right] \right|\Phi;0\right>\, .
\eqne
Therefore, the amplitude is BRST invariant up to contact terms. Before we consider the form of these contact terms, we will consider the BRST transformation of multi-loop amplitudes where such terms also arise. 

Analysing BRST invariance of the multi-loop amplitude follow the same steps as Yang--Mills. Therefore, one needs to analyse four-point amplitudes where particles can be off-shell. Consider the three channels of the four-point function depicted in figure~\ref{fig:fourpointtree}(a)-(c),
\eqnb
s: && \delta(\Phi_i^1+\Phi_i^2+\Phi_i^3)\, \left<\Phi^1_f;T_1\right|\ldots \left|\Phi^1_i;0\right>
\left<\Phi^2_f;T^2\right|\ldots \left|\Phi^2_i;0\right>
\left<\Phi^3_f;T^3\right|\ldots U_{SG}^4\,b\hat{b}\left|\Phi^3_i;0\right>
\no
t: && \delta(\Phi_i^1+\Phi_i^2+\Phi_i^3)\, \left<\Phi^1_f;T_1\right|\ldots U_{SG}^4\,b\hat{b}\left|\Phi^1_i;0\right>
\left<\Phi^2_f;T_2\right|\ldots \left|\Phi^2_i;0\right>
\left<\Phi^3_f;T_3\right|\ldots \left|\Phi^3_i;0\right>
\no
u: && \delta(\Phi_i^1+\Phi_i^2+\Phi_i^3)\, \left<\Phi^1_f;T_1\right|\ldots \left|\Phi^1_i;0\right>
\left<\Phi^2_f;T_2\right|\ldots U_{SG}^4\,b\hat{b}\left|\Phi^2_i;0\right>
\left<\Phi^3_f;T_3\right|\ldots \left|\Phi^3_i;0\right>.
\no
\eqne
Assume now that the BRST charge is first located at line 1 and move it to act on the $b\hat{b}$ insertion between the two vertices. This will make the two vertices collide since $[Q_{tot},b\hat{b}] = -\frac{1}{2}H\left(b-\hat{b}\right)$. Adding together the three channels one obtains
\eqnb
\sum_{s,t,u} \delta{A} &=&	
\frac{1}{2}\left(\left[U_{SG}^{4},b-\hat{b}\right]+\left[U_{SG}^{4},b-\hat{b}\right]+\left[U_{SG}^{4},b-\hat{b}\right]\right)\,\delta(\Phi_i^1+\Phi_i^2+\Phi_i^3)
\no
&\times&
\left<\Phi^1_f;T_1\right|\ldots \left|\Phi^1_i;0\right>
\left<\Phi^2_f;T_2\right|\ldots \left|\Phi^2_i;0\right>
\left<\Phi^3_f;T_3\right|\ldots \left|\Phi^3_i;0\right>
\no
&=&
0\, .
\eqne
The vanishing comes about because the operator $[U^{a},b-\hat{b}]$ is a world-line vector. If the fourth particle is a physical state one concludes that one has no contribution when an external vertex collides with an internal vertex as long as all channels of the four-point interaction are included. The case when the fourth particle is off-shell describes the collision of two internal vertices and it follows that the amplitude is BRST invariant as long as  the three channels are included.

The analysis above shows that the amplitude is BRST invariant up to terms when external vertices collide. We again consider the four-point interaction with two on-shell states and two off-shell states. This has contributions from the product of two vertex operators (one for each on-shell state) joined by a propagator and from the new four-particle vertex with two on-shell states, $U^{12}_{SG}$,
\eqnb
&&
\sum_{\sigma \in S_2} \int_0^{\infty} \prod_{r=1}^{3}d\tau_r\, B_j(\phi,\Phi_{f}^j,\Phi_{i}^j,\{k_{\sigma(r)}\},\{\tau_s\})
+
\int_0^{\infty} d\tau_1\, d\tau_2\, B^2_j(\phi,\Phi_{f}^j,\Phi_{i}^j,\{k_1,k_2\},\{\tau_s\}) 
\no
&&=
\sum_{\sigma \in S_2} \int_0^{\infty} \prod_{r=1}^{3}d\tau_r \left<\Phi_{f}^j;\sum_{r=1}^{3}\tau_r\left|\prod_{r=1}^{2}\left(b\hat{b}\,U_{SG}^{\sigma(r)}\right)b\hat{b}\right|\Phi_{i}^j;0\right>
\no
&&+
\int_0^{\infty} d\tau_1\, d\tau_2 \left.\left.\left<\Phi_{f}^j;\sum_{r=1}^{2}\tau_r\right|b\hat{b}\,U_{SG}^{12}\,b\hat{b}\right|\Phi_{i}^j;0\right>\, .
\eqne

Multiplying this expression with $Q_i^j$ from the right and commuting it through the vertices and $b$ ghost insertions one obtains the same term multiplied with $Q_f^j$ from the left and terms where $Q_{tot}$ are commuted with the $b$ insertions,
\eqnb
&&
\sum_{\sigma \in S_2}\int_0^{\infty} \prod_{r=1}^{3}d\tau_r\, B_j(\phi,\Phi_{f}^j,\Phi_{i}^j,\{k_{\sigma(r)}\},\{\tau_s\})\,Q_i^j
\no
&&+
\int_0^{\infty} d\tau_1\, d\tau_2\, B^2_j(\phi,\Phi_{f}^j,\Phi_{i}^j,\{k_1,k_2\},\{\tau_s\})\,Q_i^j \;\; =
\no
\nonumber
\eqne
\eqnb
&&=
\sum_{\sigma \in S_2}\int_0^{\infty} \prod_{r=1}^{3}d\tau_r\, Q_f^j\,B_j(\phi,\Phi_{f}^j,\Phi_{i}^j,\{k_{\sigma(r)}\},\{\tau_s\})
\no
&&+
\int_0^{\infty} d\tau_1\, d\tau_2\, Q_f^j\, B^2_j(\phi,\Phi_{f}^j,\Phi_{i}^j,\{k_1,k_2\},\{\tau_s\})
\no
&&
-\frac{1}{2}\sum_{\sigma \in S_2} \int_0^{\infty} d\tau_1\, d\tau_3\left<\Phi_{f}^j;\tau_1+\tau_3\right|b\hat{b}\,U_{SG}^{\sigma(1)}\,(b-\hat{b})\,U_{SG}^{\sigma(2)}\,b\hat{b}\left|\Phi_{i}^j;0\right>
\no
&&
+\int_0^{\infty} d\tau_1\, d\tau_2\left<\Phi_{f}^j,\tau_1+\tau_2\right|b\hat{b}\,[U_{SG}^{12},Q_{tot}]\,b\hat{b}\left|\Phi_{i}^j;0\right> + \ldots
\no
&&
=
\sum_{\sigma \in S_2}\int_0^{\infty} \prod_{r=1}^{3}d\tau_r\, Q_f^j\,B_j(\phi,\Phi_{f}^j,\Phi_{i}^j,\{k_{\sigma(r)}\},\{\tau_s\})
\no
&&+
\int_0^{\infty} d\tau_1\, d\tau_2\, Q_f^j\, B^2_j(\phi,\Phi_{f}^j,\Phi_{i}^j,\{k_1,k_2\},\{\tau_s\})
\no
&&
-\frac{1}{4}\sum_{\sigma \in S_2} \int_0^{\infty} d\tau_1\, d\tau_3\left<\Phi_{f}^j;\tau_1+\tau_3\right|b\hat{b}\,\left[U_{SG}^{\sigma(1)},\left[(b-\hat{b}),U_{SG}^{\sigma(2)}\right]\right]\,b\hat{b}\left|\Phi_{i}^j;0\right>
\no
&&
+\int_0^{\infty} d\tau_1\, d\tau_2\left<\Phi_{f}^j,\tau_1+\tau_2\right|b\hat{b}\,[U_{SG}^{12},Q_{tot}]\,b\hat{b}\left|\Phi_{i}^j;0\right>+ \ldots \; ,
\eqne
where the dots denote terms when $Q_{tot}$ is commuted with the $b$ ghosts adjacent to the bra or the ket. From the expression above one obtains the equations, which the four-point vertex satisfies
\eqnb
[U^{12}_{SG},Q_{tot}] &=& \frac{1}{4}\left(\left[U_{SG}^{1},\left[(b-\hat{b}),U_{SG}^{2}\right]\right] + \left[U_{SG}^{2},\left[(b-\hat{b}),U_{SG}^{1}\right]\right]\right)\, .
\eqne
The similarities to (\ref{nonlinearSG}) show that the solution is connected to non-linear effects in supergravity. The interesting superfield is ${A_{\alpha}}^{\beta}$ as it is involved in the unintegrated vertex operator. Consider a superposition of two plane waves. As the theory is non-linear, there are non-linear terms in the ${A_{\alpha}}^{\beta}$ superfield. The solution can be separated into the sum of three terms as before,
\eqnb
{A_{\alpha}}^{\beta}\left(k_1,k_2\right)
	&=&
		{A_{\alpha}}^{\beta; l}\left(k_1\right) + {A_{\alpha}}^{\beta; l}\left(k_2\right) + {A_{\alpha}}^{\beta; nl}\left(k_1,k_2\right)\, ,
\eqne
where the two first terms satisfy the linearised equations of motion. Inserting this into (\ref{nonlinearSG}) shows that the non-linear piece must satisfy
\eqnb
[\lambda^{\alpha}{A_{\alpha}}^{\beta; nl}\left(k_1,k_2\right)\hat{\lambda}_{\beta},Q_{tot}] &=& \frac{1}{4}\left(\left[U_{SG}^{1},\left[(b-\hat{b}),U_{SG}^{2}\right]\right] + \left[U_{SG}^{2},\left[(b-\hat{b}),U_{SG}^{1}\right]\right]\right)\, .\phantom{123}
\eqne
We have now improved BRST invariance at lowest order by introducing a single four-point contact vertex. As we saw in the Yang--Mills case, inserting this vertex leads to further BRST violation, which is restored by the insertion of more four-point vertices as well as higher-point vertices.

The higher $n$-point vertices are obtained in the same way. Expanding the $\theta^0$-component of $G_{mn}$ in terms of $N$ plane waves,
\eqnb
\left.G_{mn}\right|_{\theta =0}
	&=&
		\epsilon \sum_{j=1}^{N}h^j_{mn}e^{ik_jX},
\eqne
The solution of the equations of motion for ${A_{\alpha}}^{\beta}$ can be written as an expansion in terms of $\epsilon$
\eqnb
{A_{\alpha}}^{\beta} &=& \epsilon \sum_{j=1}^{N}{A_{\alpha}}^{\beta; l}\left(k_j\right) + \sum_{m=2}^{N} \epsilon^{l} \sum_{j_1<\ldots< j_m = 1}^{n}{A_{\alpha}}^{\beta; nl}\left(k_{j_1},\ldots,k_{j_m}\right)\, .
\eqne
Using (\ref{nonlinearSG}), one finds that the term $\lambda^{\alpha}{A_{\alpha}}^{\beta; nl}\left(k_{j_1},\ldots,k_{j_m}\right)\hat{\lambda}_{\beta}$ corresponds to the $m+2$ vertex where the particles with momentum $k_{j_1},\ldots,k_{j_m}$ are involved.

As in the Yang--Mills case, an alternative description of these contact vertices can be given in terms of attaching tress to the skeleton. This makes contact with observations in the context of $D=11$ supergravity \cite{Cederwall:2009ez,Cederwall:2010tn}. 

\begin{figure}[t!]
\begin{center}
\begin{tabular}{cc}
{\rotatebox{0}{\scalebox{0.22}{\includegraphics*{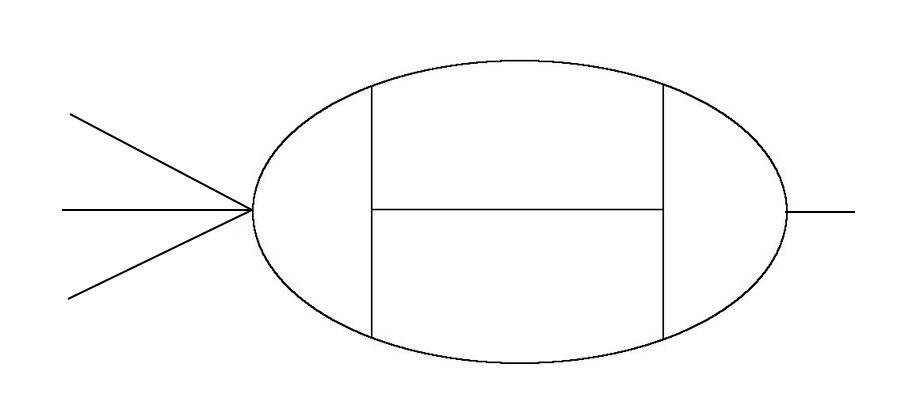}}}}
& {\rotatebox{0}{\scalebox{0.22}{\includegraphics*{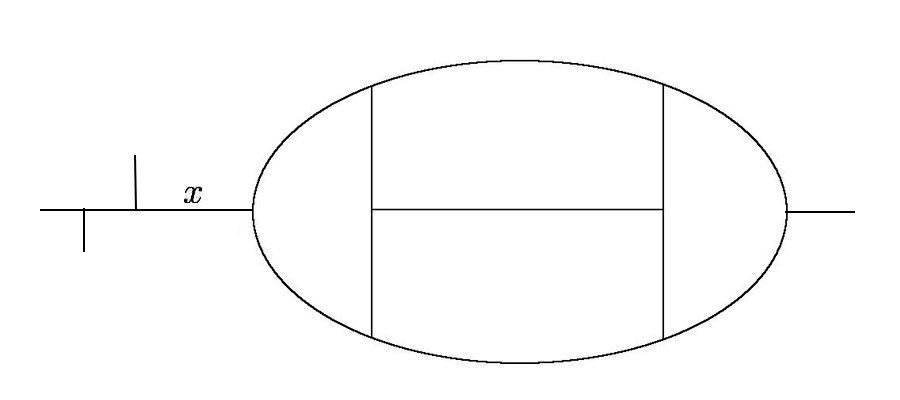}}}} \\
\;\;\;\;\;\;\;\;\;\;(a) & \;\;\;\;\;\;\;\;\;\;(b)
\end{tabular}
\caption{(a) A four-point amplitude with a five-point contact term attached. (b) The alternative solution with a tree attached where the propagator connecting the tree with the skeleton is marked with $x$.}
\label{Figure:fourloop5contact}
\end{center}
\end{figure}

One advantage of the description of contact terms as trees attached to skeletons is that it makes clear that $n$-point vertices with $n>4$ do not contribute to the four-point amplitude in either Yang--Mills or supergravity.  For example, consider the four-loop amplitude with one $n=5$ contact term shown in figure~\ref{Figure:fourloop5contact}(a). This can be expressed as a tree attached to the skeleton as in figure~\ref{Figure:fourloop5contact}(b). However the leg marked $x$ is necessarily on-shell and the diagram would give a mass renormalisation to the Yang--Mills state (or supergravity state) which cannot happen because the two-point function vanishes on-shell (as will be reiterated in section~\ref{sec:four-point}).

%%%%%%%%%%%%%%%%%%%%%%%%%
\section{Properties of the amplitude}
\label{sec:propamp}
%%%%%%%%%%%%%%%%%%%%%%%%%

In the two previous sections we have defined the amplitude and shown that it is compatible with BRST invariance by including multi-point external vertices. 

In this section we will obtain some general properties of these multi-loop amplitudes that follow from saturation of fermionic zero modes. These properties are deeply related to supersymmetry and strongly constrain the pattern of allowed diagrams, their low energy behaviour and their leading ultraviolet divergences. We will review and somewhat extend the discussion given in \cite{Bjornsson:2010wm}, which is closely related to the discussion in \cite{Berkovits:2005bt} for the pure spinor string.

In the following, we do not need to distinguish between Yang--Mills and supergravity as the hatted fields work in the same way as the unhatted. Furthermore, we will use the integrated picture.

Before we consider the structure of the amplitudes, we derive the form of the regulator needed for amplitudes with few enough loops. This is the so-called large-$\lambda$ regulator \cite{Berkovits:2005bt}. The regulator needed to regulate amplitudes with many loops, the so-called ``small-$\lambda$ regulator'' \cite{Berkovits:2006vi,Aisaka:2009yp} (see also \cite{Grassi:2009fe} for an alternative regulator) will not be discussed in detail although some properties of this regulator will be needed in the next section.

\subsection{The regulator and integration over zero modes}

The first step in obtaining the properties of the amplitude is to discuss the measure of the zero modes. This is non-trivial as there are pure spinor fields and was determined in \cite{Berkovits:2005bt,Berkovits:2004px,Berkovits:2004bw}. Consider first the part of the measure involving world-line scalar fields $X^m,\theta^{\alpha},\lambda^{\alpha},\bar{\lambda}_{\alpha}$ and $r_{\alpha}$, which has the form
\eqnb
d\phi &=& d^{D}x\, d^{16}\theta \, d^{11}\lambda\, d^{11}\bar{\lambda}\, d^{11}r\, .
\label{measurecoordf}
\eqne
In this expression the dimension is generic, as we will dimensionally reduce the theory in a way that preserves supersymmetry. This is done by reducing the superfields to lower dimensions and choosing special kinematic configurations for the external particles. The measure for the pure spinor fields in (\ref{measurecoordf}) is
\eqnb
d^{11}\lambda \, \lambda^{\alpha}\lambda^{\beta}\lambda^{\gamma}
	&=&
		\epsilon_{\rho_1\ldots\rho_{11}\kappa_{1}\ldots\kappa_5}\,\left(T^{-1}\right)^{(\alpha\beta\gamma)[\kappa_{1}\ldots\kappa_5]}\,d\lambda^{\rho_1}\ldots d\lambda^{\rho_{11}} \,,
\no
d^{11}\bar{\lambda} \, \bar{\lambda}_{\alpha}\bar{\lambda}_{\beta}\bar{\lambda}_{\gamma}
	&=&
		\epsilon^{\rho_1\ldots\rho_{11}\kappa_{1}\ldots\kappa_5}\,T_{(\alpha\beta\gamma)[\kappa_{1}\ldots\kappa_5]}\,d\bar{\lambda}_{\rho_1}\ldots \bar{d}\lambda_{\rho_{11}} \,,
\no
d^{11}r
	&=&
		\epsilon_{\rho_1\ldots\rho_{11}\kappa_{1}\ldots\kappa_5}\,\left(T^{-1}\right)^{(\alpha\beta\gamma)[\kappa_{1}\ldots\kappa_5]}\,\bar{\lambda}_{\alpha}\bar{\lambda}_{\beta}\bar{\lambda}_{\gamma}dr^{\rho_1}\ldots dr^{\rho_{11}}\, ,
\label{measurecoord2}
\eqne
where $(\;)$ and $[\;]$ denote symmetrisation and anti-symmetrisation of the indices, respectively. The tensor $T^{(\alpha\beta\gamma)[\kappa_{1}\ldots\kappa_5]}$ is given by \cite{Berkovits:2000fe,Berkovits:2004px}
\eqnb
\epsilon^{\rho_1\ldots\rho_{11}\kappa_{1}\ldots\kappa_5}T_{(\alpha\beta\gamma)[\kappa_{1}\ldots\kappa_5]}
	&=&
		\epsilon^{\rho_1\ldots\rho_{16}}\gamma^m_{\rho_{12}\delta}\gamma^n_{\rho_{13}\sigma}\gamma^p_{\rho_{14}\tau}
	\no
	&\times&
		\left(\left(\gamma_{mnp}\right)_{\rho_{15}\rho_{16}}\left(\delta^{\delta}_{(\alpha}\delta^{\sigma}_{\beta}\delta^{\tau}_{\gamma)}-\frac{1}{40}\gamma^q_{(\alpha\beta}\delta^{\delta}_{\gamma)}\gamma^{\sigma\tau}_q\right)\right)\, .
\eqne
It follows from the equations above that one has to extract sixteen $\theta$'s as well as eleven $r$'s from the $b$ ghost, vertices and the regulator. Furthermore, one also needs three additional $\lambda$'s from the measure of the world-line vector fields, $b$ ghost, vertices and regulator compared to the number of $\bar{\lambda}$'s.

The measure for the zero modes of the world-line vector fields is
\eqnb
d\Phi^I &=& d^{D}\ell^I\, d^{16}d^I \, d^{10}N^I \, dJ^I\, d^{10}\bar{N}^I \, d\bar{J}^I\, d^{10}S^I \, dS^I\, ,
\eqne
where we have used gauge invariant combinations as well as $d_{\alpha}$ instead of $p_{\alpha}$. Furthermore, $\ell^I$ is the loop momenta. In this expression the measure for the momenta for the pure spinors is\cite{Berkovits:2005bt}
\eqnb
d^{10}N^I\,dJ^I\, \lambda^{\alpha_1}\ldots\lambda^{\alpha_8}
	&=&
		M^{\alpha_{1}\ldots\alpha_{8}}_{m_1n_1\ldots m_{10}n_{10}}\,dN^{I\,m_1n_1}\ldots dN^{I\,m_{10}n_{10}}dJ^I
	\no
d^{10}\bar{N}^I\,d\bar{J}^I\, \bar{\lambda}_{\alpha_1}\ldots\bar{\lambda}_{\alpha_8}
	&=&
		\left(M^{-1}\right)_{\alpha_{1}\ldots\alpha_{8}}^{m_1n_1\ldots m_{10}n_{10}}\,d\bar{N}^I_{m_1n_1}\ldots d\bar{N}^I_{m_{10}n_{10}}d\bar{J}^I
	\no
d^{10}S^I\,dS^I  
	&=&
		M^{\alpha_{1}\ldots\alpha_{8}}_{m_1n_1\ldots m_{10}n_{10}}\,\bar{\lambda}_{\alpha_1}\ldots\bar{\lambda}_{\alpha_8}\,dS^I_{m_1n_1}\ldots dS^I_{m_{10}n_{10}}dS^I\, ,
\eqne
where $M^{\alpha_{1}\ldots\alpha_{8}}_{m_1n_1\ldots m_{10}n_{10}}$ satisfies \cite{Berkovits:2005bt}
\eqnb
M^{\alpha_{1}\ldots\alpha_{8}}_{m_1n_1\ldots m_{10}n_{10}}\bar{\lambda}_{\alpha_1}\ldots\bar{\lambda}_{\alpha_8}\psi^{m_1n_1}\ldots\psi^{m_{10}n_{10}}
	&=&
		\left(\bar{\lambda}\gamma_{m_1n_1m_2m_3m_4}\bar{\lambda}\right)\left(\bar{\lambda}\gamma_{m_5n_5n_2m_6m_7}\bar{\lambda}\right)
	\no
	&\times&
		\left(\bar{\lambda}\gamma_{m_8n_8n_3n_6m_9}\bar{\lambda}\right)\left(\bar{\lambda}\gamma_{m_{10}n_{10}n_4n_7n_9}\bar{\lambda}\right)
	\no
	&\times&
		\psi^{m_1n_1}\ldots\psi^{m_{10}n_{10}}\, ,
\eqne
and $\psi^{mn}$ is a fermionic, anti-symmetric, two-form. The measure for the world-line vector fields behaves as $\left(\bar{\lambda}/(\lambda\bar{\lambda})\right)^{8L}$. To simplify the equations, we make a change of variables for the non-minimal fields; $\bar{N}_I^{mn} \equiv \Omega_{IJ}\bar{N}^{J;\,mn}, \bar{J}_I \equiv \Omega_{IJ} \bar{J}^J, S_I^{mn} \equiv \Omega_{IJ}S^{J;\,mn}$ and $S_{I} \equiv \Omega_{IJ}S^J$; which has a trivial Jacobian. From the measure one finds that one has to extract $16L$ $d$ zero modes and $11L$ $s$ zero modes from the $b$ ghost insertions, vertices and the regulator. 

Before we can consider the constraints on the amplitude one has to determine a regulator. One cannot set this equal to one, as this would give a $0/0$ singularity, which is apparent from the form of the vertices and $b$ ghost. The numerator zero arise because there are insufficiently many fermionic zero modes to saturate the integrals, while the denominator zero arise at large values of $\lambda,N_{mn}$ and $J$. One can write an expansion for the regulator in (\ref{regulatorcond}) using
\eqnb
\chi
	&=&
		q_1\theta^{\alpha} \bar{\lambda}_{\alpha} + q_2\left(N_{mn}S^{mn} + JS\right)\, ,
\label{chireg}
\eqne
where $q_1$ and $q_2$ are generic complex numbers. The regulator which follows this is
\eqnb
\cN 
	&=&
		e^{\int d\tau [Q,\chi]}
	\no
	&=&
		e^{-q_1 \int d\tau \left(\lambda^{\alpha}\bar{\lambda}_{\alpha}-\theta^{\alpha}r_{\alpha}\right) - q_2\int d\tau\left(N_{mn}\bar{N}^{mn} + J\bar{J} - \frac{1}{2}\left(d\gamma_{mn}\lambda\right) S^{mn} - \left(\lambda d\right)S \right)}\, .
\label{reglargeL1}
\eqne
For consistency, one has to assume that the real part of $q_1$ and $q_2$ is positive otherwise the bosonic integrals do not converge.

Consider the regulator in (\ref{reglargeL1}) for a generic skeleton and assume that only zero modes of the fields contribute. One can then calculate the product of the regulators on the different propagators of the skeleton to get
\eqnb
\cN_{tot}
&=&
\prod_{j=1}^{\max(3L-3,1)} \cN^j
\no
&=&
\prod_{j=1}^{\max(3L-3,1)} e^{-q_1 \left(\lambda^{\alpha}\bar{\lambda}_{\alpha}-\theta^{\alpha}r_{\alpha}\right)\int_{0}^{T_j} d\tau}
\no
&\times&
e^{- q_2\left(N^I_{mn}\bar{N}_J^{mn} + J^I\bar{J}_J - \frac{1}{2}\left(d^I\gamma_{mn}\lambda\right) S_J^{mn} - \left(\lambda d^I\right)S_J \right)\left(\Omega^{-1}\right)^{JK}\int_{0}^{T_j} d\tau\, (\omega_I/d\tau)\, (\omega_K/d\tau)}
\no
&=&
e^{-q_1 \left(\lambda^{\alpha}\bar{\lambda}_{\alpha}-\theta^{\alpha}r_{\alpha}\right)\int_{F_L} d\tau}
\no
&\times&
e^{- q_2\left(N^I_{mn}\bar{N}_J^{mn} + J^I\bar{J}_J - \frac{1}{2}\left(d^I\gamma_{mn}\lambda\right) S_J^{mn} - \left(\lambda d^I\right)S_J \right)\left(\Omega^{-1}\right)^{JK} \int_{F_L} d\tau\, (\omega_I/d\tau)\, (\omega_K/d\tau)}
\no
&=&
e^{-q_1 \left(\lambda^{\alpha}\bar{\lambda}_{\alpha}-\theta^{\alpha}r_{\alpha}\right)\sum_{j=1}^{\max(3L-3,1)}T_j - q_2\left(N^I_{mn}\bar{N}_I^{mn} + J^I\bar{J}_I - \frac{1}{2}\left(d^I\gamma_{mn}\lambda\right) S_I^{mn} - \left(\lambda d^I\right)S_I \right)}\, ,
\eqne
where we have used $\int_{F_L}d\tau\, (\omega_I/d\tau)\, (\omega_K/d\tau) = \Omega_{IK}$. Redefining $q_1$ to absorb the total length of the skeleton one obtains
\eqnb
\cN_{tot}
&=&
e^{-q_1 \left(\lambda^{\alpha}\bar{\lambda}_{\alpha}-\theta^{\alpha}r_{\alpha}\right) - q_2\left(N^I_{mn}\bar{N}_I^{mn} + J^I\bar{J}_I - \frac{1}{2}\left(d^I\gamma_{mn}\lambda\right) S_I^{mn} - \left(\lambda d^I\right)S_I \right)}\, ,
\label{reglargeL2}
\eqne
which is the regulator in \cite{Berkovits:2005bt}. As a remark, the regulator for the tree amplitudes is obtained by setting the world-line vector fields to zero in (\ref{reglargeL2}).

We can now make a detailed analysis of the integration over zero modes. The analysis is the same as the one for the string presented in \cite{Berkovits:2005bt}. This will constrain the allowed skeletons and the position of the vertex operators attached to them.

First consider the integration over the world-line vector fields. A key observation is that the vertices and $b$ ghost are independent of $\bar{N}^{mn}$. Therefore, the factors of $N_{mn}$ in the vertices and $b$ ghost cannot contribute by their zero modes. One can therefore integrate over ${N}^I_{mn},{J}^I,\bar{N}_I^{mn}$ and $\bar{J}_I$ giving the schematic form
\eqnb
\int \prod_{I=1}^{L}d^{10L}N^I\, d^{L}J^I\, d^{10L}\bar{N}_I\, d^{L}\bar{J}_I\, \cN_{tot}
	&\sim&
		q_2^{-11L}\,\frac{1}{\left(\lambda\bar{\lambda}\right)^{8L}} \left. \cN_{tot}\right|_{N_{mn}=J=0}\, .
\label{intoNJ}
\eqne
Furthermore, the vertices and $b$ ghosts are also independent of $S_{mn}$ and $S$. Performing the integration over these fields, using the regulator (\ref{intoNJ}), gives schematically
\eqnb
q_2^{-11L}\int \prod_{I=1}^{L} d^{11L}S_I\, \frac{1}{\left(\lambda\bar{\lambda}\right)^{8L}}\left. \cN_{tot}\right|_{N_{mn}=J=0}
	&\sim&
		q_2^{-11L}\, q_2^{11L}\, \frac{{\bar{\lambda}}^{8L}}{\left(\lambda\bar{\lambda}\right)^{8L}}\lambda^{11L} \prod_{I=1}^{L}\left(d^I\right)^{11} \left. \cN_{tot}\right|_{\Phi^I =0}
	\no
	&\sim&
		 \lambda^{3L} \prod_{I=1}^{L}\left(d^I\right)^{11}\left. \cN_{tot}\right|_{\Phi^I =0}\, ,
\label{intoNJS}
\eqne
where $\left. \cN_{tot}\right|_{\Phi^I =0}$ denotes the regulator for the tree. The result of the integration is independent of $q_2$, as expected because the regulator is BRST exact.

An important observation in (\ref{intoNJS}) is that, after integrating over the zero modes of the pure spinor momenta fields, one will have eleven insertions of $d$ zero modes for each loop in the skeleton. Therefore, the $b$ ghost insertions and vertices has to contribute five $d$ zero modes for each loop in the diagram. This will constrain the skeletons, the positions of the vertices and the ultraviolet behaviour of the amplitude.

Assume we have integrated over the zero modes of $d$, picking out possible combinations of the terms in the $b$ ghost insertions as well as the vertices. Consider now the integration over the world-line scalar fields. Whereas the $b$ ghost depends on the non-minimal fields the vertices do not. Consider the case when the $b$ ghost contribution is $b^{3L-3} \sim (\bar{\lambda}/(\lambda\bar{\lambda}))^{3L-3}(r/(\lambda\bar{\lambda}))^n$ with $n<11$ \footnote{ The restriction to $n < 11$ is crucial in order to avoid small-$\lambda$ singularities.}. The integral over the pure spinor fields and $\theta$ is of the form
\eqnb
&&
\int d^{11}r\, d^{11}\lambda\,d^{11}\bar{\lambda}\,d^{16}\theta\,\left\{\lambda^{3L}\, \left(\frac{\bar{\lambda}}{\left(\lambda\bar{\lambda}\right)}\right)^{3L-3}\,\left(\frac{r}{\left(\lambda\bar{\lambda}\right)}\right)^{n}\,\cO\left(P,X,\theta\right)\,\left. \cN_{tot}\right|_{\Phi^I =0}\right\}
\no
&&
\sim
\int d^{11}r\, d^{11}\lambda\,d^{11}\bar{\lambda}\,d^{16}\theta\,\left\{\lambda^{3}\, \left(\frac{1}{q_1\left(\lambda\bar{\lambda}\right)}\right)^{n}\,D^n\cO\left(P,X,\theta\right)\,\left. \cN_{tot}\right|_{\Phi^I =0}\right\}
\no
&&
\sim
\int d^{11}r\, d^{11}\lambda\,d^{11}\bar{\lambda}\,d^{16}\theta\,\left\{\lambda^{3}\, D^n\cO\left(P,X,\theta\right)\,\left. \cN_{tot}\right|_{\Phi^I =0}\right\}\, ,
\label{Int:trl2}
\eqne
where $\cO\left(P,X,\theta\right)$ denotes the combined part of the vertices and $b$ ghosts insertions which depends only on $P_m,X^m$ and $\theta^{\alpha}$. Above we have rewritten $r_{\alpha}$ as $D_{\alpha}/q_1$ acting on the regulator using momentum conservation. Here $D_{\alpha}$ is the fermionic superderivative defined in appendix \ref{sec:YM} in (\ref{covariantderiv}). After  partial integration, the operator  $D_{\alpha}$ acts on the operator $\cO\left(P,X,\theta\right)$. Furthermore, we have replaced $1/\left(q_1\left(\lambda\bar{\lambda}\right)\right)$ by $1$ as this factor is independent of $q_1$ in the integral.

Using the measure in (\ref{measurecoord2}), it follows that the integration in (\ref{Int:trl2})  extracts a specific component of $\lambda^3 D^n \cO\left(P,X,\theta\right)$ involving three $\lambda$ and five $\theta$. This motivates the introduction of the matrix element
\eqnb
\left.\left<\lambda^3 \cO\left(P,X,\theta\right)\right>\right|_{\theta^5} 
	&\equiv&
		\int d^{11}r\, d^{11}\lambda\,d^{11}\bar{\lambda}\,d^{16}\theta\,\lambda^{3}\, \cO\left(P,X,\theta\right)\,\left. \cN_{tot}\right|_{\Phi^I =0}
	\no
	&\sim&
		\left.D^5\cO\left(P,X,\theta\right)\right|_{\theta=0}\, .
\label{MYM}
\eqne
For economy of notation we will suppress the notation $\left.\right|_{\theta=0}$  in the following.
 
Up to now we have considered the Yang--Mills case. The generalisation to supergravity is straightforward, involving a doubling of all the fields with the exception of $X$ and $P$. The regulator involving the hatted fields has the same form as (\ref{reglargeL2}). Integrating over the zero modes of the momenta for the pure spinor fields, one obtains a symmetric factor of $\lambda^{3L}\prod_{I=1}^{L}\left(d^I\right)^{11}\hat{\lambda}^{3L}\prod_{I=1}^{L}\left(\hat{d}^I\right)^{11}$. Integration over the zero modes of the world-line scalar fields picks out a certain $\theta^5\hat{\theta}^5$ component of the operator $\cO(P,X,\theta,\hat{\theta})$. Therefore, we define the matrix element
\eqnb
\left.\left<\lambda^3 \hat{\lambda}^3 \cO(P,X,\theta,\hat{\theta})\right>\right|_{\theta^5\hat{\theta}^5} 
	&\equiv&
		\int \left(d^{11}r\right)^2\,  \left(d^{11}\lambda\right)^2\,  \left(d^{11}\bar{\lambda}\right)^2\,  \left(d^{16}\theta\right)^2\, \lambda^{3}\, \hat{\lambda}^{3}
	\no
	&\times&
		\cO(P,X,\theta,\hat{\theta})\,\left.\left( \cN_{tot}\hat{\cN}_{tot}\right)\right|_{\Phi^I =0}
	\no
	&\sim&
		\left.D^5\hat{D}^5\cO(P,X,\theta,\hat{\theta})\right|_{\theta=\hat{\theta}=0}\, ,
\label{MSG}
\eqne
where $\left(d^{11}r\right)^2 = d^{11}r\, d^{11}\hat{r}$, and similarly for the other fields. As before, we will suppress the notation $\left.\right|_{\theta=\hat{\theta}=0}$ in the following.

\subsection{Properties of the $b$ insertions}

The analysis of the consequences of the $b$ ghost insertions is of importance for the properties of the amplitudes.   We here  assume for simplicity that the fields in the $b$ ghost only contributes through their zero modes. As described in \cite{Bjornsson:2010wm},  the $b$  ghost consists of the product of two world-line vector fields and world-line scalar fields and can be expanded using the world-line one-forms
\eqnb
\left.b^j\right|_{zero}
	&=&
		b^{IJ} \left(\omega_I/d\tau_j\right)\left(\omega_J/d\tau_j\right)
	\no
	&=&
		b^{IJ} \frac{\partial\Omega_{IJ}}{\partial T_j}\, ,
\eqne
where we have used the definition of the period matrix. This demonstrates the manner in which the insertion of $b$ ghosts is connected to the period matrix. It follows that the product of $3L-3$ insertions of $b$ for a specific skeleton has the form   
\eqnb
\prod_{j=1}^{3L-3}\left.b^j\right|_{zero}
	&=&
		b^{I_1J_1}\ldots b^{I_{3L-3}J_{3L-3}}\frac{\partial\Omega_{I_1J_1}}{\partial T_1}\ldots \frac{\partial\Omega_{I_{3L-3}J_{3L-3}}}{\partial T_{3L-3}}\, .
\label{bins}
\eqne
In considering the ultraviolet behaviour of the amplitude it is important to see whether (\ref{bins}) is nonzero if every factor of $b$ contributes with  the maximal number of two $d$ zero modes.  The  term in the $b$ ghost which has two $d$'s is the second term in  (\ref{bghostN1}) (see also (\ref{bfields})), which will be denoted by $b_{H}^{IJ}$. Since $\left(b_H^{IJ}\right)^2=0$, it is only possible for  (\ref{bins}) to be nonzero when $b$ is replaced by $b_H$  if the period matrix has $3L-3$ nonzero linearly independent components. This condition can be written as
\eqnb
\sum_{j=1}^{3L-3}c_j \frac{\partial\Omega_{IJ}}{\partial T_j}
	&\neq&
		0\, ,
\label{bins2}
\eqne
for any nonzero $c_j$. If this condition does not hold it is not possible for all  $b$ ghosts to contribute via the second term in  (\ref{bghostN1}), which leads to two possibilities. Either at least one of the $b$ ghosts contributes via the first term or at least one of the $b$ ghosts contributes with a nonzero mode. The discussion here again generalises to supergravity in a straightforward manner.

%%%%%%%%%%%%%%%%%%%%%%%%%
\section{Summary of structure of multi-loop amplitudes}
\label{sec:four-point}
%%%%%%%%%%%%%%%%%%%%%%%%%

This section will largely review the computations of four-point amplitudes presented in  \cite{Bjornsson:2010wm}. We will somewhat enlarge the discussion by including the  determination of the four-point one-loop and two-loop amplitudes up to an overall constant using the results in \cite{Berkovits:2005ng,Mafra:2008ar}. We also consider the multi-point one-loop amplitudes and give an alternative proof of the ``no-triangle hypothesis'' \cite{BjerrumBohr:2006yw}. Furthermore, we also show that loop amplitudes with fewer than four external physical particles vanish.

\subsection{Preliminaries}

We will  focus on the ultraviolet behaviour of four-point amplitudes and not on the  infrared behaviour, which needs a more detailed investigation.  More precisely, we will consider the $L$-loop amplitude in sufficiently high dimensions for it to be ultraviolet divergent.  This divergence may be regulated with an ultraviolet momentum cutoff, $\Lambda$. The contributions to the $L$-loop amplitude that come from a particular skeleton, $F_L$, will diverge as a positive power of $\Lambda$ when $D>D_c^{(F_L)}$, where $D_c^{(F_L)}$ is the $L$-loop `critical dimension' for that skeleton.   When $D=D_c^{(F_L)}$ the amplitude  diverges as $\log \Lambda$.

By simple dimensional counting it follows that the term with the leading ultraviolet behaviour of the amplitude is associated with the lowest power of external momenta, $k_r$, in the low momentum limit, $k_r\to 0$. In taking this low momentum limit in a term with an even number of momentum operator insertions arising from the vertices and the $b$ ghost insertions it is possible to set  $e^{ik_rx} = 1$ and make use of the contraction between the momentum insertions 
\eqnb
\left<P^{m}\left(\tau\right)P^{n}\left(\tau'\right)\right>_{F_L}
	&=&
		-\delta^{mn}\left(\omega_I(\tau)/d\tau\right)\left(\Omega^{-1}\right)^{IJ}\left(\omega_J(\tau')/d\tau'\right)\, ,
\eqne
which can be obtained from explicit computation of the loop momenta integrals. If there is an odd number of momenta insertions, then one of these internal momenta has to be replaced  by a linear combination of the external momenta before taking $k_r\to 0$.

In previous sections we saw that the $X$ and $P$ part of the amplitude is described by scalar field theory diagrams with cubic internal vertices, together with certain $n$-point external vertices and numerator momenta insertions. In computing the  low energy limit of a term with $2q$ momentum insertions it is simple to see from dimensional counting  that 
\eqnb
\left<P^{m_1}\left(\tau_1\right)\ldots P^{m_{2q}}\left(\tau_{2q}\right)\right>_{F_L}
	&\sim&
		\Lambda^{L(D-6) + 6 + 2q + 2m}\, ,
\label{Pins}
\eqne
and $m$ denotes the number of external vertices attached to the skeleton (and $D$ must be large enough for the power of $\Lambda$ to be non-negative). 

The critical dimension is determined by the value of $D$ that makes the power of $\Lambda$ in (\ref{Pins}) vanish\eqnb
D_{c}^{\left(F_L\right)}
	&=&
		4 + \frac{2\left(L+m-q-3\right)}{L}\, .
\eqne
In particular, the condition for the expression to be ultraviolet finite in four dimensions is $q<L+m-3$, which limits the number of momentum insertions for a given value of $L$.   Restricting to the case of four-point amplitudes, this condition can be interpreted as a limit on the number of derivatives acting on $F^4$ and $\cR^4$ at any value of $L$.  For $\cN=4$ Yang--Mills in four dimensions one does not need any derivatives for the theory to be ultraviolet finite \cite{{Mandelstam:1982cb},Brink:1982wv} while for supergravity one needs $2L$ derivatives acting on $\cR^4$ for $L > 1$ \cite{Green:2006gt}. 

The amplitudes for Yang--Mills will be considered in the large-$N_c$ limit, where one can use the double line notation to indicate the colour contractions on propagating particles. The group theory is described in terms of external vertices  attached to boundaries of these propagators, as if they were attached to the boundaries of open strips. For single-trace operators, the four external particles are attached to a single boundary and the amplitude behaves as $N_c^L$ on the number of colours. For double-trace operators, the four external particles are attached in pairs to two different boundaries and the amplitude depends as $N_c^{L-1}$ on the number of colours. Non-planar skeletons are suppressed by $1/N_c^2$ compared to the planar ones.   

Before summarising the loop amplitude results, we give the results for tree amplitudes with three and four external particles for Yang-Mills and supergravity
\eqnb
A^{(Tree)}_{YM}(\{k_r\})
	&=&
		f^{a_1a_2a_3}a^1_{m_1}a^2_{m_2}a^3_{m_3}V^{m_1m_2m_3}\, ,
	\no
A^{(Tree)}_{SG}(\{k_r\})
	&=&
		h^1_{m_1n_1}h^2_{m_2n_2}h^3_{m_3n_3}V^{m_1m_2m_3}V^{n_1n_2n_3}\, ,
	\no
A^{(Tree)}_{YM}(s,t)
	&=&
		\frac{\Tr F^4}{st}\, ,
	\no
A^{(Tree)}_{SG}(s,t,u)
	&=&
		\frac{\cR^4}{stu}\, .
\eqne
where 
\eqnb
V^{m_1m_2m_3}
	&=&
		\delta^{m_1m_2}\left(k^{m_3}_1-k^{m_3}_2\right) + \delta^{m_2m_3}\left(k^{m_1}_2-k^{m_1}_3\right) + \delta^{m_3m_1}\left(k^{m_2}_3-k^{m_2}_1\right)\, .
\eqne
These expressions can be obtained using methods presented in section~\ref{sec:firstquant} and the results of the pure spinor integrals for the string \cite{Mafra:2008ar}.

The guiding principle for extracting the leading low energy term is to first consider the term which contributes the maximal number of $d$ zero modes in the $b$ ghost insertions that the skeleton allows (using the first and second term in (\ref{bghostN1})),  The external particles in the four-point amplitude are attached to the skeleton through three- and four-point vertices. For these vertices, one uses the term with the maximal number of $d$ zero modes in (\ref{Vertexymi}) and (\ref{inVSG}). In general, it is not necessary to use the maximal number of $d$ zero modes from the vertex insertions. For the Yang--Mills case, one uses the first term in (\ref{Vertexymi}) for the surplus vertices. In supergravity, one uses the third and fourth terms in (\ref{inVSG}) in pairs  for the surplus vertices.

Before  considering general amplitudes we note that vacuum amplitudes vanish.  This follows since there are five more zero modes of $\theta$ than $r$ and the $b$ ghost is independent of $\theta$, so one cannot get a non-zero result for the $r$ and $\theta$ integrals  at the same time. Thus, zero-point amplitudes vanish.
 
\subsection{One-loop amplitude} 

From the discussion in section \ref{sec:propamp}, five $d$ zero modes need to be obtained from the single $b$ ghost insertion and the integrated vertices. Since the $b$ ghost can contribute two $d$'s, at least three need to come from the integrated vertex operators. Therefore, amplitudes with fewer than four external particles vanish since one vertex is unintegrated. For the four-point amplitude there is a single insertion of $r$, which can be turned into one $D$. The discussion  generalises straightforwardly to the supergravity amplitude by the doubling procedure. Since only one configuration of terms in the four vertices contributes  the leading ultraviolet dependence of the amplitudes in the Yang--Mills and supergravity cases are given by 
\eqnb
A^{(1)}_{YM}(s,t)
	&\sim&
		\left.\left<\lambda^2 D W^3 A_{\alpha}\right>\right|_{\theta^5}\, \Lambda^{D-8}
	\sim
		F^4\, \Lambda^{D-8}\, ,
\no
A^{(1)}_{SG}(s,t,u)
	&\sim&
		\left.\left<\lambda^3 \hat{\lambda}^3 D \hat{D} W^3 A\right>\right|_{\theta^5\hat{\theta}^5}\, \Lambda^{D-8}
	\sim
		\cR^4\, \Lambda^{D-8}\, ,
\label{1Lsugra}
\eqne
where the matrix element are defined by  (\ref{MYM}) and (\ref{MSG})\footnote{We have here also integrated over $X$ and $P$.}. The results are the same for both the single-trace, $\Tr F^4$, and double-trace operators, $(\Tr F^2)^2$, in Yang--Mills. The pure spinor and fermionic integrals were studied in detail in \cite{Mafra:2008ar} in the context of the pure spinor superstring, and using their results one obtains
\eqnb
&&
\hspace{-0.5cm}
A^{(1)}_{YM}(s,t)
\no
&&
\sim F^4 \int_{0}^{\infty} \frac{dT}{T^{D/2}} \int_{F_1'}\prod_{r=2}^{4}d\tau_r\, e^{s\left(G\left(0,\tau_2\right)+G\left(\tau_3,\tau_4\right)\right)+t\left(G\left(0,\tau_4\right)+G\left(\tau_2,\tau_3\right)\right)+u\left(G\left(0,\tau_3\right)+G\left(\tau_2,\tau_4\right)\right)}\, ,\;
\no
&&
\hspace{-0.5cm}
A^{(1)}_{SG}(s,t,u)
\no
&&
\sim \cR^4 \int_{0}^{\infty} \frac{dT}{T^{D/2}} \int_{F_1}\prod_{r=2}^{4}d\tau_r\, e^{s\left(G\left(0,\tau_2\right)+G\left(\tau_3,\tau_4\right)\right)+t\left(G\left(0,\tau_4\right)+G\left(\tau_2,\tau_3\right)\right)+u\left(G\left(0,\tau_3\right)+G\left(\tau_2,\tau_4\right)\right)}\,, \,\,\,\,\;
\label{1LoopSGComp}
\eqne
where $F_1'$ is the integration region which respects the ordering of the external particles. These amplitudes match with \cite{Green:1982sw}.  The corresponding one-loop pure spinor string calculation has been carried out in great detail, including a determination of the overall normalisation in  \cite{Gomez:2009qd}.

We now briefly consider amplitudes with more than four points in order to demonstrate an alternative proof of the ``no-triangle hypothesis'' of $\cN = 8$ supergravity, \cite{BjerrumBohr:2008vc,BjerrumBohr:2008ji,ArkaniHamed:2008gz}. One has to extract five $d$ zero modes from the $b$ ghost and the vertices, which can involve any number of external particles, a priori. To get the contribution which involves the largest number of insertions of $P$ in the diagram one can consider the case in which the $b$ ghost contributes two $d$ zero modes. In this case, the integrated vertices have to contribute three additional $d$'s. The same also holds for the hatted fields. Therefore, one needs at least three integrated vertices and one unintegrated. This contribution has no insertions of numerator momenta and proves that there are no bubble or triangle functions.

\subsection{Two-loop amplitude}

There are two two-loop skeletons. One is the one-particle irreducible  skeleton in figure \ref{Figure:twoloop}  and the other is a diagram consisting of two one-loop diagrams connected by one propagator. From the rules in section~\ref{sec:propamp}, one sees  that the three $b$ ghost insertions and the vertex insertions have to contribute a total of  five $d$ zero modes for each loop. Therefore, the one-particle reducible diagram vanishes unless at least six vertices are attached -- three for each loop.  In the case of the one-particle irreducible diagram, the $b$ ghosts can contribute with at most six $d$ zero modes and needs at least four integrated vertices to be attached in order to provide the remaining $d$ zero modes. Therefore, two-loop amplitudes with less than four particles vanish.

We now consider now the four-point amplitude in more detail. The contribution of six $d$ zero modes from the three $b$ ghosts enter through the term
\eqnb
\prod_{j=1}^{3} b_H^{I_jJ_j}\frac{\partial \Omega_{I_jJ_j}}{\partial T_j}
	&=&
		b_H^{11}\left(b_H^{11} - 2b_H^{12} + b_H^{22}\right)b_H^{22}
	\no
	&=&
		-2\, b_H^{11}\, b_H^{12}\, b_H^{22}\, .
\eqne
The form of this expression constrains the vertices to be attached in pairs to the two different loops. This shows that there are  at most two vertices on each line, which coincides with the distribution found in \cite{Bern:1997nh}. This generalises straightforwardly to the supergravity case where the distribution of the external vertices coincides with \cite{Bern:1998ug}. Using the rules presented in section \ref{sec:propamp}, the low energy behaviour of the amplitudes is 
\eqnb
A^{(2)}_{YM}(s,t)
	&\sim&
		\left.\left<\lambda^3 D^3 W^4\right>\right|_{\theta^5}\, \Lambda^{2D-14}
	\sim
		\partial^2 F^4 \, \Lambda^{2D-14}\, ,
\no
A_{SG}^{(2)}\left(s,t,u\right)
	&\sim&
		\left.\left<\lambda^3\hat{\lambda}^3 D^3 W^4\right>\right|_{\theta^5\hat{\theta}^5}\, \Lambda^{2D-14}
	\sim
		\partial^4 \cR^4 \, \Lambda^{2D-14}\, .
\eqne
where $F^4$ denotes the single- and double-trace term depending on where the vertices are attached. 

In this case one can go further and evaluate the precise form for the amplitude, not just its leading ultraviolet behaviour using the evaluation of the  fermionic and pure spinor integrals, given in the context of the pure spinor  superstring in  \cite{Berkovits:2005ng}.  This leads to 
\eqnb
A_{YM}^{(2)}\left(s,t\right)
	&\sim&
		F^4 \int_0^{\infty} \frac{dT_1\,dT_2\,dT_3}{\Delta^{D/2}}\int_{F'_2} \prod_{r=1}^{4}d\tau_r\,\mathcal{Y}(s,t,u,\tau_s)
	\no
	&\times&
		e^{s\left(G\left(\tau_1,\tau_2\right)+G\left(\tau_3,\tau_4\right)\right)+t\left(G\left(\tau_1,\tau_4\right)+G\left(\tau_2,\tau_3\right)\right)+u\left(G\left(\tau_1,\tau_3\right)+G\left(\tau_2,\tau_4\right)\right)}\, ,
	\no
A_{SG}^{(2)}\left(s,t,u\right)
	&\sim&
		\cR^4 \int_0^{\infty} \frac{dT_1\,dT_2\,dT_3}{\Delta^{D/2}}\int_{F_2} \prod_{r=1}^{4}d\tau_r\,\mathcal{Y}(s,t,u,\tau_s)^2
	\no
	&\times&
		e^{s\left(G\left(\tau_1,\tau_2\right)+G\left(\tau_3,\tau_4\right)\right)+t\left(G\left(\tau_1,\tau_4\right)+G\left(\tau_2,\tau_3\right)\right)+u\left(G\left(\tau_1,\tau_3\right)+G\left(\tau_2,\tau_4\right)\right)}\, ,
\label{eq:twoloop}
\eqne
where $F'_2$ is the integration region which respects the ordering of the external particles and 
\eqnb
\mathcal{Y}(s,t,u,\tau_r)
	&=&
		\left[\left(u-t\right)\Delta\left(1,2\right)\Delta\left(3,4\right)+\left(s-t\right)\Delta\left(1,3\right)\Delta\left(2,4\right)
	\right.
	\no
	&+&
	\left.
	\left(s-u\right)\Delta\left(1,4\right)\Delta\left(2,3\right)\right]\, ,
\eqne
where
\eqnb
\Delta(i,j)
	&=&
		\epsilon^{IJ}\left(\frac{\omega_I}{d\tau_i}\right)\left(\frac{\omega_J}{d\tau_j}\right)\, .
\eqne
The  supergravity amplitude is of the same form as presented in \cite{Green:2008bf}.  Although we have not attempted to determine the overall normalisation of these amplitudes, these coefficients were determined in the case of the pure spinor string in string  in \cite{Gomez:2010ad} (whereas the overall coefficient has not been evaluated directly in the RNS approach although it was fixed by considering a degeneration limit of the amplitude \cite{D'Hoker:2005ht}). 

\subsection{Three-loop amplitude}

\begin{figure}[t!]
\begin{center}
\begin{tabular}{cc}
{\rotatebox{0}{\scalebox{0.28}{\includegraphics*{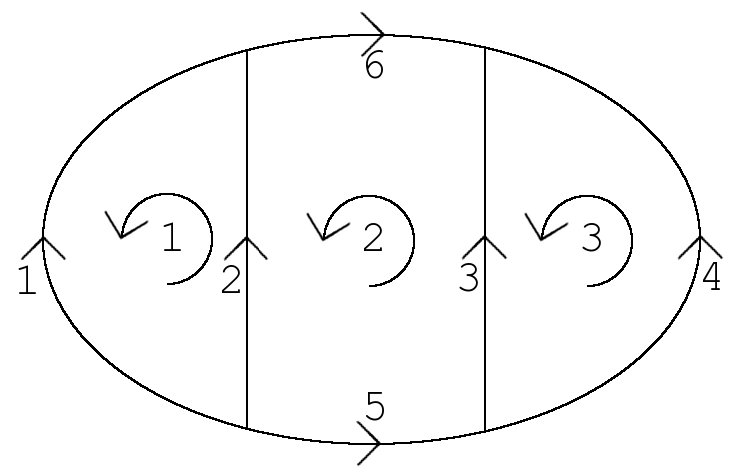}}}}
& {\rotatebox{0}{\scalebox{0.25}{\includegraphics*{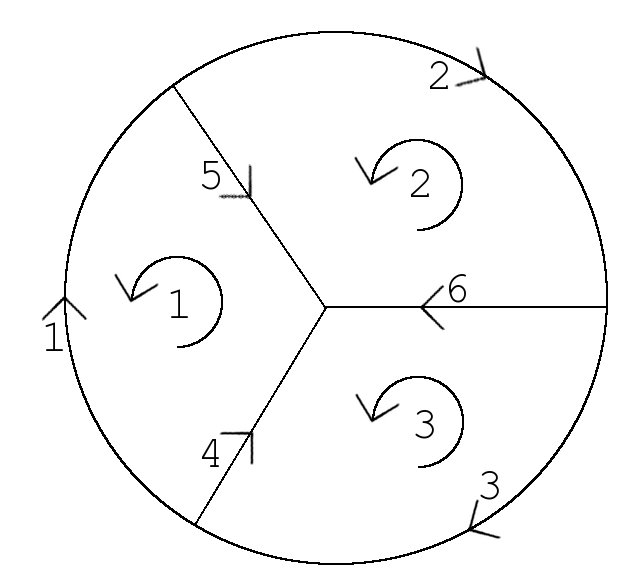}}}} \\
(a) & (b)
\end{tabular}
\caption{The two three-loop one-particle irreducible skeletons. (a) The three-loop ladder skeleton. (b)  The ``Mercedes'' skeleton.}
\label{Figure:threeloop}
\end{center}
\end{figure}

There are five three-loop skeletons, three of which are one-particle reducible. The results in section \ref{sec:propamp} show that one needs to extract fifteen $d$ zero modes, five for each loop, from the vertices and $b$ ghosts. This means that at least six vertices must be attached to any one-particle reducible  skeleton in order to obtain  a nonzero contribution,  so once again only the one-particle irreducible diagrams contribute to the four-point function. 

In particular, twelve $d$ zero modes can be obtained from the $b$-ghost insertions in irreducible diagrams. Therefore, one has to attach at least three vertices to get a nonzero result, so amplitudes with less than three points vanish. Furthermore, certain skeletons do not support twelve $d$ zero modes so they cannot contribute to three-point functions. The twelve $d$ zero modes arising from the $b$ ghost insertions in the remaining ones are accompanied by six $r$'s, which can be converted into six $D$'s. As a result, it follows that   the three-point amplitude involves at least six powers of external momenta and vanishes after imposing the mass-shell condition and conservation of momentum.

The two one-particle irreducible skeletons are depicted in figure \ref{Figure:threeloop}(a) and (b). The first skeleton is the (three-loop) ``ladder diagram'' and the second the ``Mercedes diagram''. The period matrix for the ladder is
\eqnb
\Omega^{(a)}_{IJ}
	&=&
		\left(
		\begin{array}{ccc}
			T_1 + T_2 & -T_2 & 0 \\
			-T_2 & T_2 + T_3 + T_5 + T_6 & -T_3 \\
			0 & -T_3 & T_3 + T_4
		\end{array}
		\right)\, ,
\label{ladderthreePM}
\eqne
while for the Mercedes it is
\eqnb
\Omega^{(b)}_{IJ}
	&=&
		\left(
		\begin{array}{ccc}
			T_1 + T_4 + T_5 & -T_5 & -T_4 \\
			-T_5 & T_2 + T_5 + T_6 & -T_6 \\
			-T_4 & -T_6 & T_3 + T_4 + T_6
		\end{array}
		\right)\, .
\label{mercedesPM}
\eqne
The number of nonzero linearly independent components of the period matrix is five for the ladder diagram and six for the Mercedes. 

As the period matrix for the ladder has only five nonzero linearly independent components, it does not support the maximal number of $d$ zero mode insertions from the $b$ ghost. The maximal number of $d$ zero modes it supports is three for loops 1 and 3 and five for loop 2. The generalisation to the supergravity amplitude is straightforward. The four vertices must be attached in pairs to loops 1 and 3 and the low energy limit of the amplitude  is \cite{Bjornsson:2010wm}
\eqnb
A^{(a)}_{YM}\left(s,t\right)
	&\sim&
		k_m\left.\left<\lambda^3 D^5W^4 \right>\right|_{\theta^5}\, \Lambda^{3D-20}
	\sim
		\partial^4 F^4 \, \Lambda^{3D-20}\, ,
\no
A^{(a)}_{SG}\left(s,t,u\right)
	&\sim&
		k^2_m\left.\left<\lambda^3 \hat{\lambda}^3 D^5 \hat{D}^5 W^4 \right>\right|_{\theta^5\hat{\theta}^5}\, \Lambda^{3D-20}
	\sim
		\partial^8 \cR^4 \, \Lambda^{3D-20}\, .
\eqne
Here $F^4$ contains both the single- and double-trace contribution. 

Consider now the Mercedes skeleton. As the period matrix has six linearly independent components, it supports the maximal number of $d$ zero modes from the $b$ ghosts
\eqnb
\prod_{j=1}^{6} b^{I_jJ_j}_{H} \frac{\partial \Omega_{I_jJ_j}}{\partial T_{j}}
	&=&
		\left(-2\right)^3b_H^{11}\, b_H^{22}\, b_H^{33}\, b_H^{13}\, b_H^{12}\, b_H^{23}\, ,
\eqne
giving four insertions of $d$ zero modes for each loop as well as six insertions of $r$'s, which convert into six $D$'s. Therefore, the diagram has support from three attached vertices (one being a contact term).

\begin{figure}[t!]
\begin{center}
{\rotatebox{0}{\scalebox{0.2}{\includegraphics*{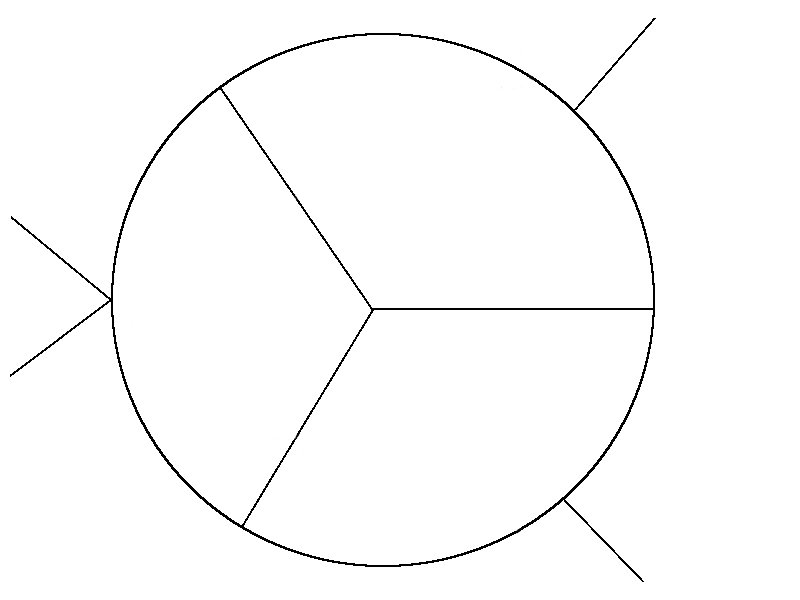}}}}
\caption{The three-loop diagram with one contact term that arises in maximal Yang--Mills and supergravity.  While its contribution makes no qualitative change to the leading behaviour of the supergravity amplitude, in the Yang--Mills case its presence is responsible for the leading behaviour,  $\partial^2\, \Tr F^4$.}
\label{Figure:threecontact}
\end{center}
\end{figure}

Consider first the case with four single-particle vertices, which was obtained in \cite{Bjornsson:2010wm}
\eqnb
A^{(b)}_{YM}\left(s,t\right)
	&\sim&
		k_m\left.\left<\lambda^3 D^6 A W^3 \right>\right|_{\theta^5}\, \Lambda^{3D-20}
	\sim
		\partial^4 F^4 \, \Lambda^{3D-20}\, ,
\no
A^{(b)}_{SG}\left(s,t,u\right)
	&\sim&
		\left.\left<\lambda^3 \hat{\lambda}^3 D^6 \hat{D}^6 \hat{E}E W^2 \right>\right|_{\theta^5\hat{\theta}^5}\, \Lambda^{3D-18}
	\sim
		\partial^6 \cR^4 \, \Lambda^{3D-18}\, ,
\eqne
where $F^4$ denotes both the single- and double-trace contribution. The Mercedes diagram also gets a contribution  from three attached vertices where one of them is a contact term.  The double-trace part of the Yang--Mills amplitude   does not get a contribution from such contact vertices since  this would require  a single vertex attached  to one boundary. This vanishes  because the generators of the algebra are traceless\footnote{In the finite $N_c$ case, the contact term is anti-symmetric in the colour indices but the trace is symmetric, and therefore vanishes.}. The contribution to the single-trace operator is nonzero and the amplitude is shown in figure \ref{Figure:threecontact}. At low energies, it is proportional to
\eqnb
A^{(b)}_{YM}\left(s,t\right)
	&\sim&
		\left.\left<\lambda^3 D^6 \,\Tr(W^{nl} W^2) \right>\right|_{\theta^5}\, \Lambda^{3D-18}
	\sim
		\partial^2 \Tr F^4 \, \Lambda^{3D-18}\, ,
\no
A^{(b)}_{SG}\left(s,t,u\right)
	&\sim&
		\left.\left<\lambda^3 \hat{\lambda}^3 D^6 \hat{D}^6 W^{nl} W^2 \right>\right|_{\theta^5\hat{\theta}^5}\, \Lambda^{3D-18}
	\sim
		\partial^6 \cR^4 \, \Lambda^{3D-18}\, .
\label{singletrace3L}
\eqne
In (\ref{singletrace3L}) we have used the $\theta$-expansion of the non-linear solution of Yang--Mills obtained in the appendix \ref{sec:YM}. Note that the single-trace and double-trace operators have different ultraviolet behaviour. They behave different because the single-trace operators involve contact terms. In supergravity, the contact terms contain two extra momentum factors and  do not change the low energy behaviour of the amplitude. 

The ladder and Mercedes amplitudes match the explicit computations at three loops in \cite{Bern:2007hh}.  The difference between the single-trace and double-trace terms in the Yang--Mills case (first noted in \cite{Dixon:2009tk}) was  partially explained using pure spinor open string theory in \cite{Berkovits:2009aw} and the more complete argument using the pure spinor world-line formalism, reviewed here,  was presented in \cite{Bjornsson:2010wm}). There are also arguments based on supersymmetry that suggest that double-trace operators are more protected than single-trace \cite{Bossard:2009mn}.

\subsection{Four-loop amplitude}

There are seventeen different four-loop skeletons, of which five are one-particle irreducible. From the discussion in section \ref{sec:propamp} it follows that it is necessary to extract twenty $d$ zero modes, five for each loop, from the $b$ ghosts and vertices. The one-particle reducible diagrams vanish unless at least five vertices are attached to the skeletons.  So, as before, the one-particle reducible skeletons do not contribute to the four-point function.

For the one-particle irreducible diagrams, the maximal number of $d$ zero modes the nine $b$ ghosts can contribute with is eighteen. Therefore, the one-point amplitudes vanish as one cannot saturate the $d$ zero mode integrations. The term with the maximal number of $d$ zero modes also has a factor of nine $r$'s, which can be converted into nine $D$'s. For the two-point amplitude, the nine $D$'s give at least seven insertions of external momenta, which vanish by momentum conservation. The vanishing of the three-point amplitude follows from the same arguments as three loops.

\begin{figure}[t!]
\begin{center}
\begin{tabular}{ccc}
{\rotatebox{0}{\scalebox{0.15}{\includegraphics*{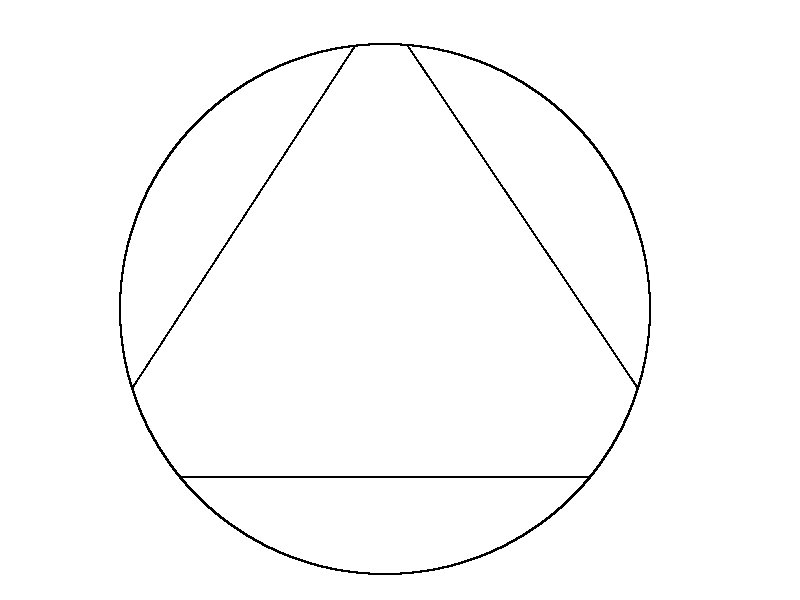}}}}
& {\rotatebox{0}{\scalebox{0.17}{\includegraphics*{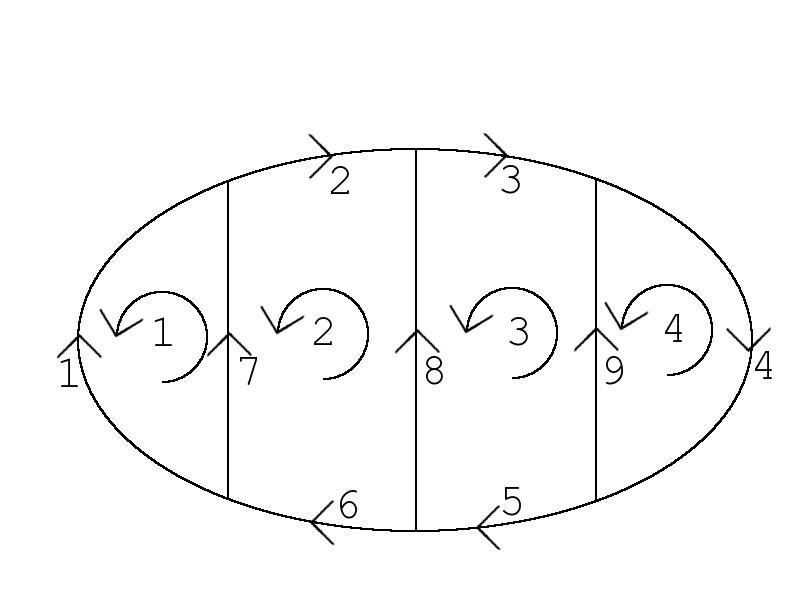}}}} 
& {\rotatebox{0}{\scalebox{0.17}{\includegraphics*{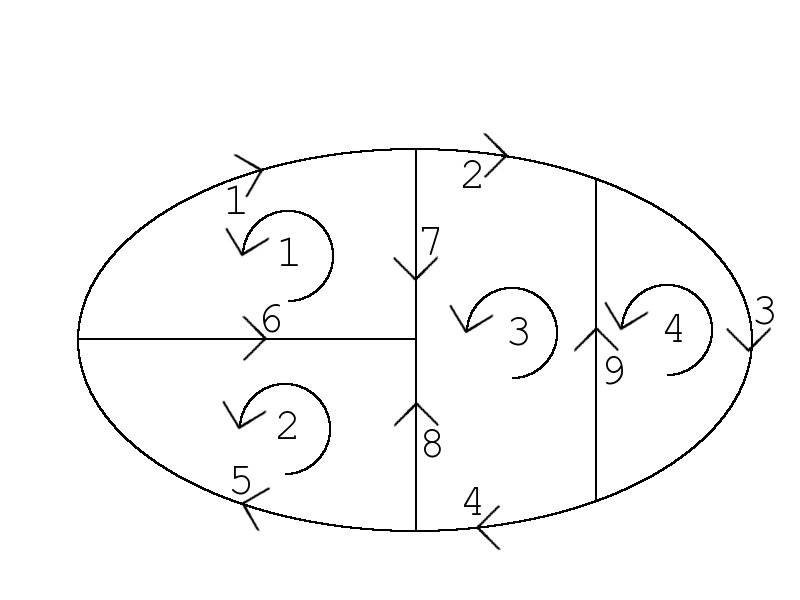}}}} \\
(a) & (b) & (c) 
\end{tabular}
\begin{tabular}{cc}
{\rotatebox{0}{\scalebox{0.17}{\includegraphics*{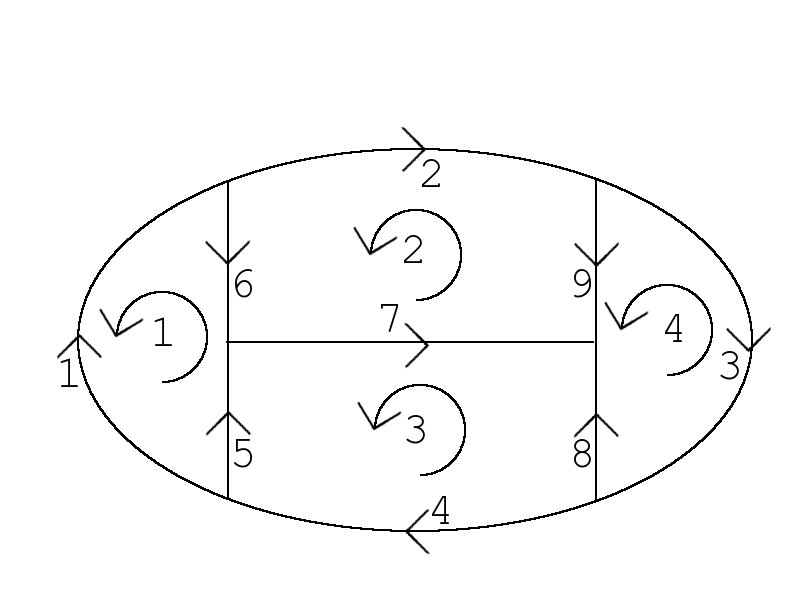}}}} 
& {\rotatebox{0}{\scalebox{0.17}{\includegraphics*{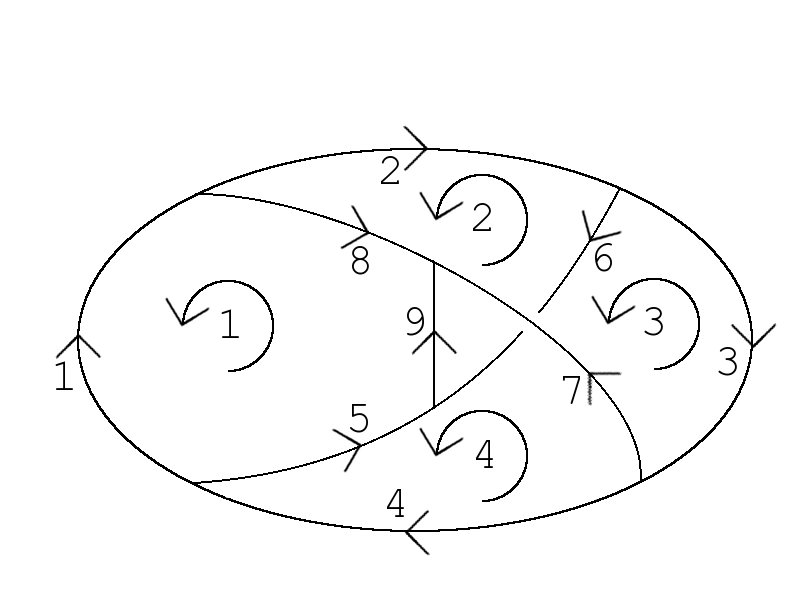}}}} \\
(d) & (e)
\end{tabular}
\caption{The five one-particle irreducible four-loop skeletons. Figure (a) does not contribute to four-particle amplitudes.}
\label{Figure:fourloop}
\end{center}
\end{figure}

It is easy to see that the one-particle irreducible diagram in figure \ref{Figure:fourloop}(a) vanishes unless one attaches at least six vertices. The diagrams contributing to the four-point function are the one-particle irreducible skeletons depicted in figure \ref{Figure:fourloop}(b)-(e). 

In order to analyse the four-point amplitude, we need to determine the period matrices of the different skeletons. The period matrix for the ladder skeleton in figure \ref{Figure:fourloop}(b) is
\eqnb
\Omega^{(b)}_{IJ}
	&=&
		\left(
		\begin{array}{cccc}
			T_1 + T_7 & -T_7 & 0 & 0 \\
			-T_7 & T_2 + T_6 + T_7 + T_8 & -T_8 & 0 \\
			0 & -T_8 & T_3 + T_5 + T_8 + T_9 & -T_9 \\
			0 & 0 & -T_9 & T_4 + T_9
		\end{array}
		\right)\, ,
\label{ladfour}
\eqne
which has seven nonzero linearly independent components. This is two less than the number of moduli. The skeleton diagram in figure \ref{Figure:fourloop}(c) has the period matrix
\eqnb
\Omega^{(c)}_{IJ}
	&=&
		\left(
		\begin{array}{cccc}
			T_1 + T_6 + T_7 & -T_6 & -T_7 & 0 \\
			-T_6 & T_5 + T_6 + T_8 & -T_8 & 0 \\
			-T_7 & -T_8 & T_2 + T_4 + T_7 + T_8 + T_9 & -T_9 \\
			0 & 0 & -T_9 & T_3 + T_9
		\end{array}
		\right)\, ,
\eqne
which has eight nonzero linearly independent components -- one less than the number of moduli. The period matrix for the diagram in figure \ref{Figure:fourloop}(d) is
\eqnb
\Omega^{(d)}_{IJ}
	&=&
		\left(
		\begin{array}{cccc}
			T_1 + T_5 + T_6 & -T_6 & -T_5 & 0 \\
			-T_6 & T_2 + T_6 + T_7 + T_9 & -T_7 & -T_9 \\
			-T_5 & -T_7 & T_4 + T_5 + T_7 + T_8 & -T_8 \\
			0 & -T_9 & -T_8 & T_3 + T_8 + T_9
		\end{array}
		\right)\, ,
\eqne
which has as many nonzero linearly independent components as the number of moduli. The skeleton in figure \ref{Figure:fourloop}(e) has the period matrix
\eqnb
\Omega^{(e)}_{IJ}
	&=&
		{\footnotesize
		\left(
		\begin{array}{cccc}
			T_1 + T_5 + T_8 + T_9 & -T_8 - T_9 & T_9 & -T_5 - T_9 \\
			-T_8 - T_9 & T_2 + T_6 + T_8 + T_9 & -T_6 - T_9 & T_9 \\
			T_9 & -T_6 - T_9 & T_3 + T_6 + T_7 + T_9 & -T_7 - T_9 \\
			-T_5 - T_9 & T_9 & -T_7 - T_9 & T_4 + T_5 + T_7 + T_9
		\end{array}
		\right)\, ,
		}
\eqne
which also has as many nonzero linearly independent components as the number of moduli.

For the ladder, the structure of the period matrix in (\ref{ladfour}) shows that the $b$ ghost  can contribute with at most sixteen factors of $d$ zero modes, which is less than the maximum of eighteen. There can be at most five $d$ zero modes on loops 2 and 3 and three on loops 1 and 4. The $b$ ghost also contribute with seven $r$'s, which are turned into seven $D$'s. Therefore, the amplitude is only non-zero when four single-particle vertices are attached in pairs to loops 1 and 4, and the low energy limit of the amplitude is
\eqnb
A^{(b)}_{YM}(s,t)
	&\sim&
		k_m^2\left.\left<\lambda^3 D^7 W^4 \right>\right|_{\theta^5}\, \Lambda^{4D-26}
	\sim
		\partial^6 F^4 \, \Lambda^{4D-26}\, ,
	\no
A^{(b)}_{SG}(s,t,u)
	&\sim&
		k_m^4\left.\left<\lambda^3 \hat{\lambda}^3 D^7 \hat{D}^7 W^4 \right>\right|_{\theta^5\hat{\theta}^5}\, \Lambda^{4D-26}
	\sim
		\partial^{12} \cR^4 \, \Lambda^{4D-26}\, .
\eqne 
 
The diagrams in figure \ref{Figure:fourloop}(c) also only allow the insertion of  sixteen $d$ zero modes. Naively one would expect seventeen but this cannot occur since the third loop is  a pentagon. As there are four missing $d$ zero modes, the skeleton only contributes to the amplitude with four single-particle vertices attached. The maximal number of $d$ zero modes the $b$ ghost insertions can contribute  is four on loops 1 and 2, five on loop 3 and three on loop 4. Therefore, one has to attach two vertices on loop 4 and one each on loops 1 and 2. The terms in the product of nine $b$ ghost insertions that contribute with sixteen $d$ zero modes also contain one factor of the external momentum flowing across the diagram (i.e.\ in the $s$-, $t$- and $u$-channel) and one internal momentum factor. In the Yang--Mills case, this internal momentum is equivalent to a linear combination of external momenta after functional integration. In supergravity on the other hand, there are two internal momenta insertions, which can contract with each other, producing a contact term in the low energy limit. The resulting low-energy amplitude has the form \cite{Bjornsson:2010wm},
\eqnb
A^{(c)}_{YM}(s,t)
	&\sim&
		k_m^2\left.\left<\lambda^3  D^7 W^4 \right>\right|_{\theta^5}\, \Lambda^{4D-26}
	\sim
		\partial^6 F^4 \, \Lambda^{4D-26}\, ,
	\no
A^{(c)}_{SG}(s,t,u)
	&\sim&
		k_m^2\left.\left<\lambda^3 \hat{\lambda}^3 D^7 \hat{D}^7 W^4 \right>\right|_{\theta^5\hat{\theta}^5}\, \Lambda^{4D-24}
	\sim
		\partial^{10} \cR^4 \, \Lambda^{4D-24}\, .
\eqne

\begin{figure}[t!]
\begin{center}
{\rotatebox{0}{\scalebox{0.2}{\includegraphics*{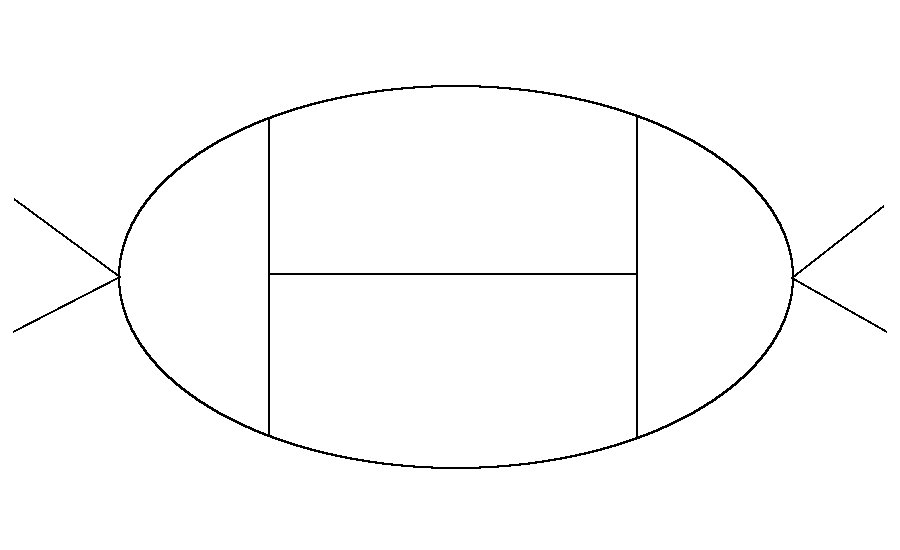}}}}
\caption{The planar four-loop diagram with two contact terms that contributes to $\partial^2\, \Tr F^4$.}
\label{Figure:fourcontact}
\end{center}
\end{figure}

The number of nonzero linearly independent components of the period matrix for the skeleton depicted in figure \ref{Figure:fourloop}(d) is nine, so each of the $b$ ghost insertions  can contribute with the maximal number of two $d$ zero modes,
\eqnb
\prod_{j=1}^{9} b^{I_jJ_j}_{H} \frac{\partial \Omega_{I_jJ_j}}{\partial T_{j}}
	&=&
		\left(-2\right)^5b_H^{11}\, b_H^{22}\, b_H^{44}\, b_H^{33}\, b_H^{13}\, b_H^{12}\, b_H^{23}\, b_H^{34}\, b_H^{24}\, .
\eqne
This shows that the $b$ ghosts contribute with four $d$ zero mode insertions on loops 1 and 4 and five on loops 2 and 3. These insertions lead to nine $r$'s which can be converted into nine $D$'s. As the $b$ ghost insertions contribute  all the $d$ zero modes apart from two,  the amplitude support two attached vertices. The leading low energy contribution for the single-trace amplitude arises from two contact terms, both producing two external particles. As discussed earlier, contact terms do not contribute to the double-trace term, so the leading contribution arises from four single-particle vertices. In the supergravity case, the contact terms do not change the behaviour in the low energy limit. The leading low energy contribution  (obtained in \cite{Bjornsson:2010wm}) is 
\eqnb
A^{(d)}_{YM}(s,t)
	&\sim&
		\left.\left<\lambda^3  D^9 A^2W^2 \right>\right|_{\theta^5}\, \Lambda^{4D-24}
	\sim
		\partial^4 (\Tr F^2)^2 \, \Lambda^{4D-24}\, ,
	\no
A^{(d)}_{YM}(s,t)
	&\sim&
		\left.\left<\lambda^3  D^9\, \Tr\left(W^{nl}\right)^2 \right>\right|_{\theta^5}\, \Lambda^{4D-22}
	\sim
		\partial^2\, \Tr F^4\, \Lambda^{4D-22}\, ,
	\no
A^{(d)}_{SG}(s,t,u)
	&\sim&
		\left.\left<\lambda^3 \hat{\lambda}^3 D^9 \hat{D}^9 E^2\hat{E}^2 \right>\right|_{\theta^5\hat{\theta}^5}\, \Lambda^{4D-22}
	\sim
		\partial^8 \cR^4 \, \Lambda^{4D-22}\, .
\label{eq:fourloop(d)}
\eqne
The four-loop amplitude with two contact terms is shown in figure \ref{Figure:fourcontact}.

Now consider  the non-planar diagram in figure \ref{Figure:fourloop}(e). In Yang--Mills the contribution is suppressed by $1/N_c^2$ and therefore vanishes in the large-$N_c$ limit. For supergravity, there is no planar limit and the non-planar skeleton contributes at the same order as the planar skeletons. One interesting observation for non-planar diagrams is that simple application of the no-bubble and no-triangle rule of sub-diagrams does not work. The reason is that at least one line in the diagram in any basis is involved in more than two loops. In the basis in figure \ref{Figure:fourloop}(e), it is  line 9, which is involved in all four loops. A naive application of the ``no-triangle hypothesis'' shows that one would not need to attach any vertices to figure \ref{Figure:fourloop}(e), which is not correct.

As the number of nonzero linearly independent components of the period matrix for this skeleton is the same as the number of moduli, the $b$ ghost can contribute with the maximal number of $d$ zero modes
\eqnb
\prod_{j=1}^{9} b^{I_jJ_j}_{H} \frac{\partial \Omega_{I_jJ_j}}{\partial T_{j}}
	&=&
		\left(2\right)^5b_H^{11}\, b_H^{22}\, b_H^{33}\, b_H^{44}\, b_H^{14}\, b_H^{23}\, b_H^{34}\, b_H^{12}\left( b_H^{13} + b_H^{24}\right)\, .
\label{4Loosbinsnonplanar}
\eqne
This shows that the insertions of the $b$ ghosts can contribute in two different ways. Either they contributes five $d$ zero modes to loops 1 and 3 and four $d$ zero modes to loops 2 and 4,  or the contributions are the other way round. As both are equivalent, up to a relabeling of the loops in the diagram, we consider the former case.  The leading ultraviolet behaviour was obtained in \cite{Bjornsson:2010wm} where it  was shown to be (\ref{eq:fourloop(d)}). It is also worth stressing that in the Yang--Mills case the contribution of the non-planar skeleton is suppressed by $1/N_c^2$ compared to the planar contributions.

The four-loop results match the known explicit results obtained in  \cite{Bern:2009kd}, and   \cite{Dixon:2009tk,Bern:2010tq}. It is notable  that in the pure spinor framework the low energy behaviour is obtained by a mode-counting argument that does not require  cancellations between different diagrams, in contrast to calculations that do not make supersymmetry manifest \cite{Bern:2009kd}.  

It is worth stressing that according to a variety of arguments \cite{Green:1997tv,Green:1998by,Berkovits:2004px,Berkovits:2006vc,Green:2005ba,Elvang:2010jv,Elvang:2010kc,Drummond:2010fp,Beisert:2010jx,Bossard:2010bd,Vanhove:2010aa,BerkovitsICTP,Howe:1980th} the operators $\cR^4$, $\partial^4 \cR^4$ and $\partial^6 \cR^4$   are fractional BPS operators, or ``F-terms'' that  can be written as integrals over a fraction  of the full superspace and are  protected from getting higher-loop contributions. They should therefore be protected from obtaining contributions beyond one loop, two loops and three loops, respectively.  
   
However, the interaction $\partial^8\cR^4$, which is the leading ultraviolet contribution at four loops, is a D-term that can be written as an integral over the whole superspace. Such terms are not protected from having perturbative corrections from all loops.  In other words, it should come as no surprise if the leading order contribution to the  five-loop amplitude is $\partial^8\cR^4$ as was stressed in a corollary of \cite{Green:2010sp}.

\begin{figure}[b!]
\begin{center}
\begin{tabular}{cccc}
{\rotatebox{0}{\scalebox{0.12}{\includegraphics*{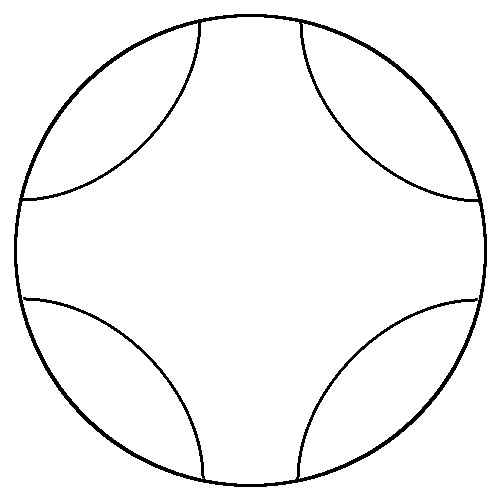}}}}
& {\rotatebox{0}{\scalebox{0.12}{\includegraphics*{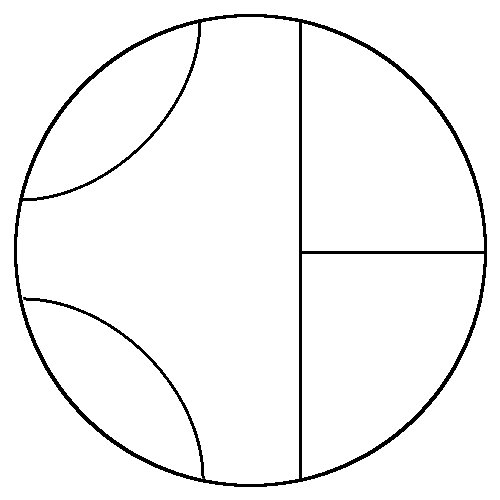}}}}
& {\rotatebox{0}{\scalebox{0.12}{\includegraphics*{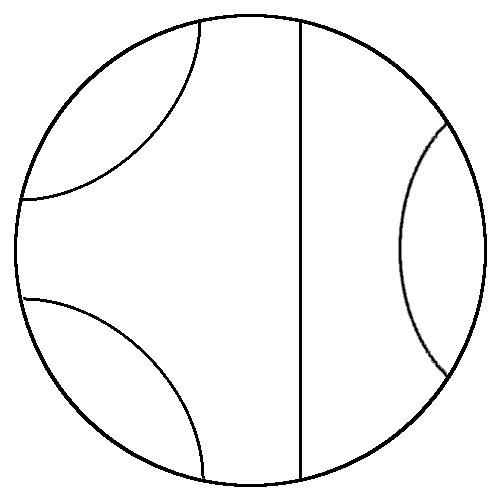}}}}
& {\rotatebox{0}{\scalebox{0.12}{\includegraphics*{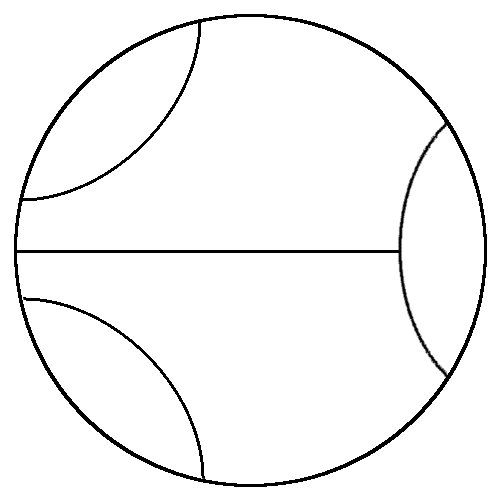}}}} \\
(1) & (2) & (3) & (4) \vspace{0.5cm}\\
{\rotatebox{0}{\scalebox{0.12}{\includegraphics*{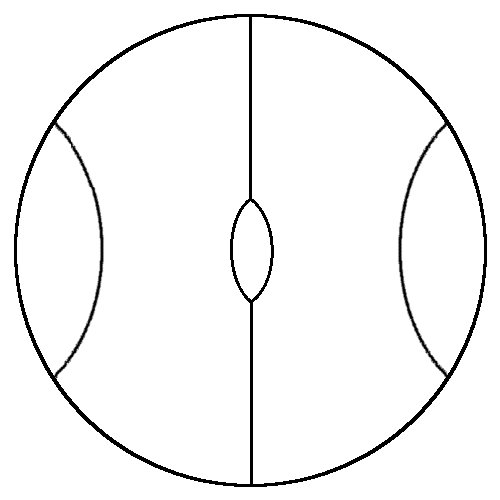}}}}
& {\rotatebox{0}{\scalebox{0.12}{\includegraphics*{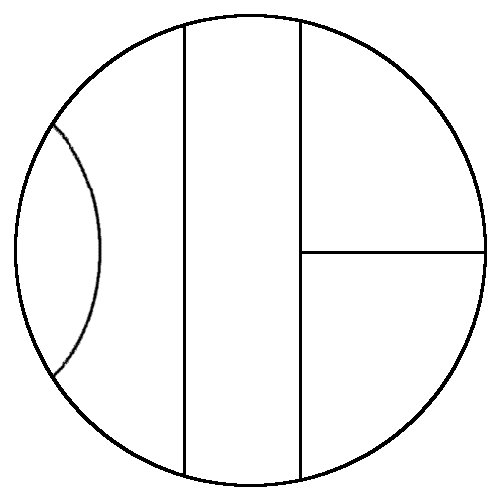}}}} 
& {\rotatebox{0}{\scalebox{0.12}{\includegraphics*{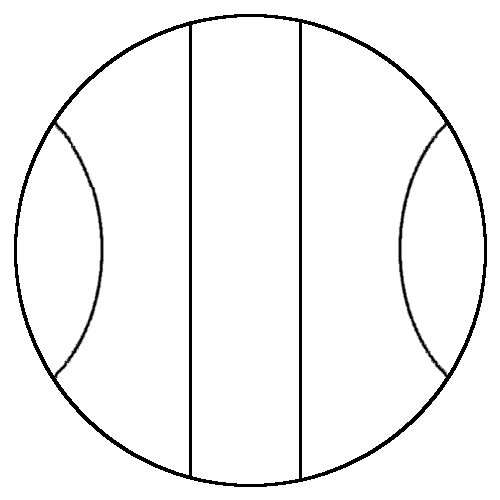}}}} 
& {\rotatebox{0}{\scalebox{0.12}{\includegraphics*{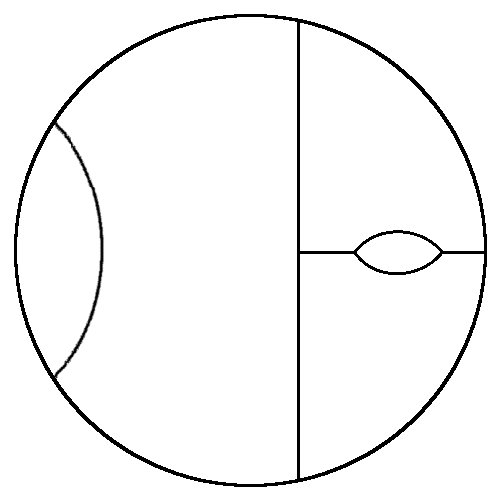}}}} \\
(5) & (6) & (7) & (8) \vspace{0.5cm}\\
{\rotatebox{0}{\scalebox{0.12}{\includegraphics*{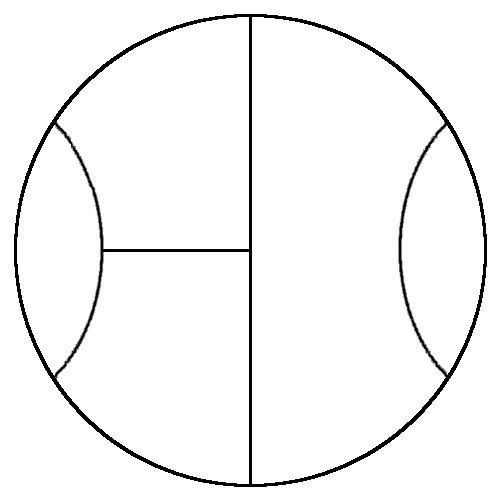}}}}
& {\rotatebox{0}{\scalebox{0.12}{\includegraphics*{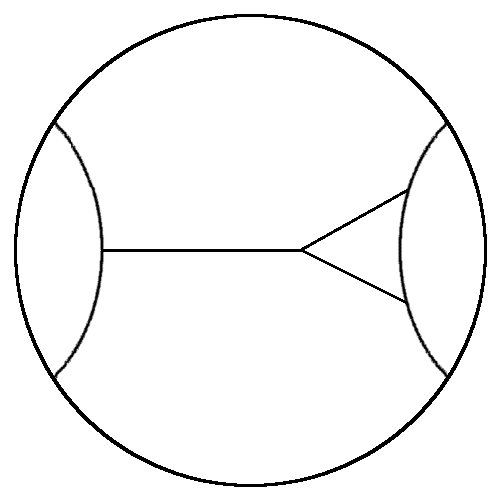}}}} 
& {\rotatebox{0}{\scalebox{0.12}{\includegraphics*{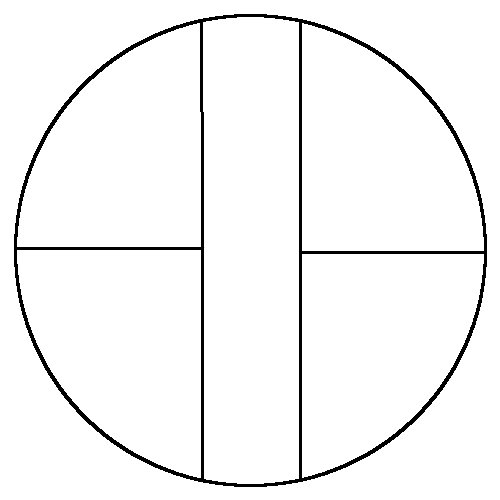}}}} 
& {\rotatebox{0}{\scalebox{0.12}{\includegraphics*{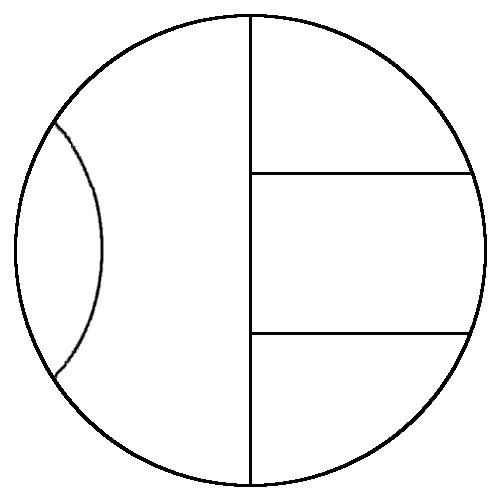}}}} \\
(9) & (10) & (11) & (12) \vspace{0.5cm}\\
{\rotatebox{0}{\scalebox{0.12}{\includegraphics*{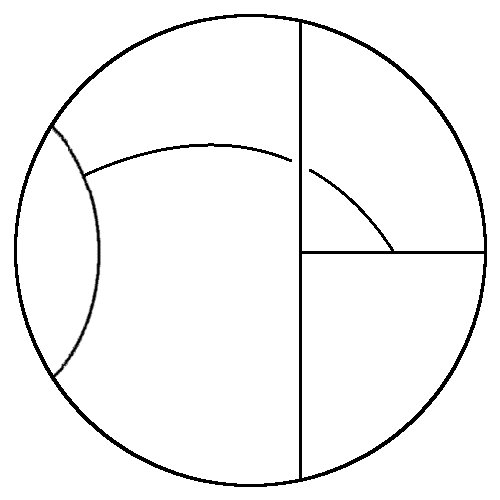}}}}
& {\rotatebox{0}{\scalebox{0.12}{\includegraphics*{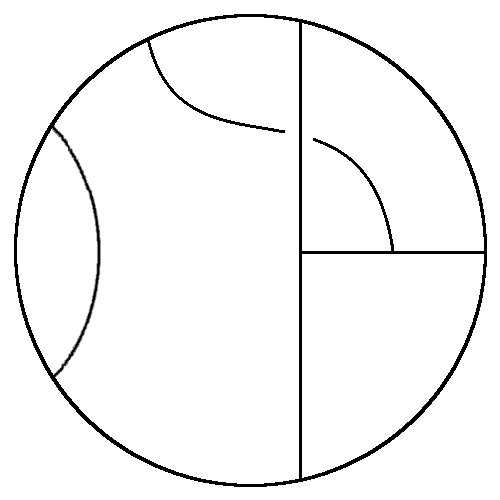}}}} 
& {\rotatebox{0}{\scalebox{0.12}{\includegraphics*{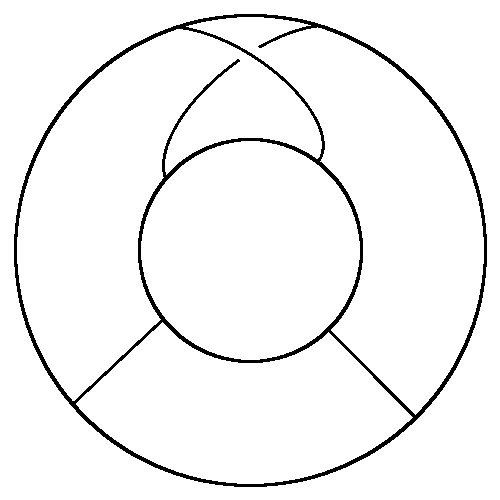}}}} 
& {\rotatebox{0}{\scalebox{0.12}{\includegraphics*{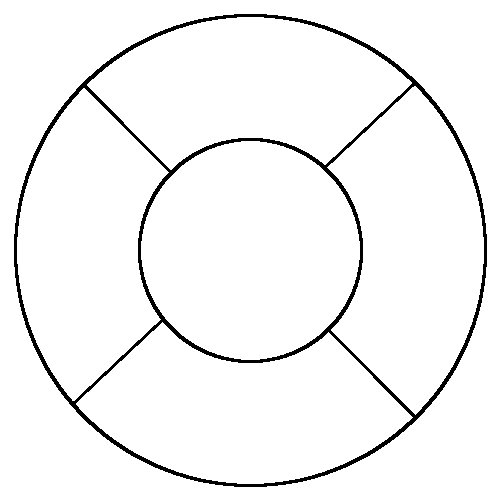}}}} \\
(13) & (14) & (15) & (16) \\
\end{tabular}
\caption{The sixteen one-particle irreducible skeletons with five loops.}
\label{Figure:fiveloop}
\end{center}
\end{figure}

\subsection{Five-loop amplitude}

As expected, at five loops there is a radical change in the pattern of the four-point amplitudes \cite{Bjornsson:2010wm}. A naive computation of the maximal number of $d$ zero modes which the $b$ ghosts can contribute with is twenty-four. The problem with this naive computation is that this term involves twelve insertions of $r$. As there are only eleven zero modes of $r$, one of the $r$'s has to contribute by a nonzero mode. The nonzero mode of $r$ has to contract with one $s$, but there are no factors of $s$ in the amplitude if the regulator, $\cN$, is ignored, so the amplitude  apparently vanishes. However, since this contribution also has more than eleven inverse powers of $(\lambda\bar{\lambda})$ the integral over  $\lambda$ and $\bar{\lambda}$ diverges. In other words, there is a new $0/0$ ambiguity, this time at small values of $\lambda$ and $\bar{\lambda}$.

The regularisation of $r$ can be understood from the form of the large-$\lambda$ regulator in (\ref{reglargeL1}). If the $r$ contributing with a non-zero mode is contracted with an $s$ in the regulator, it produces a factor of $\lambda \bar{\lambda}\, d$, thereby reducing the power of $r/(\lambda\bar{\lambda})$ from twelve to eleven. This resolves the problem with too many $r$'s but it is important to stress that the regulator does not regularise the resulting logarithmic divergence in the $\lambda$ and $\bar{\lambda}$ integrals. To do this one needs the full structure of the regulator \cite{Berkovits:2006vi,Aisaka:2009yp} (for an alternative regulator, see \cite{Grassi:2009fe}), which still has the property that each $r/(\lambda\bar{\lambda})$ is traded for a further factor of $d$. Therefore, the conclusion is that the regularised insertions of $b$ ghosts can contribute all the twenty-five $d$'s and eleven $r$'s needed to saturate the integral.

The only five-loop  skeletons  which are non-zero when the $b$-mode insertions contribute the maximum number of   $d$ zero modes are those numbered  (15) and (16) in figure \ref{Figure:fiveloop} (which is reproduced from \cite{Bjornsson:2010wm}). As the $b$ ghosts contribute the necessary number of $d$ zero modes, the first term in each of the four integrated vertices (see  (\ref{Vertexymi}) and (\ref{inVSG})) can contribute. For the single-trace contribution, the leading term in the ultraviolet arises when there are two contact terms, each producing two external particles. As we have seen, the double-trace part does not get contributions from contact terms and the leading order contribution arises from four single-particle vertices. In supergravity, the presence of contact terms does not alter the ultraviolet behaviour. The leading term at low energies obtained in  \cite{Bjornsson:2010wm} is 
\eqnb
A^{(15) \& (16)}_{YM}(s,t)
	&\sim&
		\left.\left<\lambda^3 D^{11}A^4\right>\right|_{\theta^5}\, \Lambda^{5D-28}
	\sim
		\partial^4(\Tr F^2)^2\, \Lambda^{5D-28}\, ,
	\no
A^{(15) \& (16)}_{YM}(s,t)
	&\sim&
		\left.\left<\lambda^3 D^{11}\Tr \left(A^{nl}\right)^2\right>\right|_{\theta^5}\, \Lambda^{5D-26}
	\sim
		\partial^2 \Tr F^4\, \Lambda^{5D-26}\, ,
	\no
A^{(15) \& (16)}_{SG}(s,t,u)
	&\sim&
		\left.\left<\lambda^3\hat{\lambda}^3D^{11}\hat{D}^{11}G^4\right>\right|_{\theta^5\hat{\theta}^{5}}\, \Lambda^{5D-24}
	\sim
		\partial^8\cR^4\, \Lambda^{5D-24}\, .
\label{5Loop}
\eqne
These have the same form as the four-loop terms. This also highlights the difference between the ultraviolet properties of Yang--Mills and supergravity. Yang--Mills is logarithmically ultraviolet divergent in $26/5$ dimensions whereas supergravity has a logarithmic ultraviolet divergence in $24/5$ dimensions.

As discussed in \cite{Bjornsson:2010wm},  the two skeletons which contribute to $\partial^8\cR^4$ (numbers (15) and (16) in figure \ref{Figure:fiveloop}) lead to amplitudes with a very simple structure.  In the low energy limit the integration over the positions of the external vertices gives a  contribution to the $\partial^8\cR^4$ term that is proportional to the vacuum amplitude of the scalar particle theory. As there are two diagrams, there is a possibility that the two contributions cancel.

We now consider amplitudes with fewer than four points. At five loops there are skeletons that, a priori, do not need attached vertices to give non-vanishing contributions. Such zero point amplitudes were shown to vanish in the preliminary part of this section. The one-point amplitude vanishes because of momentum conservation and the fact that there are insertions of momenta. The two-point amplitude vanishes by the same arguments as at four loops, since there are at least seven insertions of momenta in the amplitude. For three external particles  there are at least eight powers of momenta in the amplitude so it vanishes after using  momentum conservation and the mass-shell condition. This can be generalised for amplitudes with more loops, which shows that all amplitudes with fewer than four points vanish.

\subsection{Beyond five loops}

We will now review  the argument in \cite{Bjornsson:2010wm} that $\partial^2 \Tr F^4$,  $\partial^4 \left(\Tr F^2\right)^2$ and $\partial^8\cR^4$ has contributions from all loops. We will not consider the skeletons that contribute to these terms but we will assume that there are skeletons with no bubble or triangle sub-diagrams has such contributions. For simplicity we assume that the relevant piece of the $b$ ghost is the second term in (\ref{bghostN1}). 

The $b$ ghost contribution for a $L$-loop skeleton arising from the second term in (\ref{bghostN1}) is proportional to $r^{3L-3}d^{6L-6}$. Regulating the surplus $r$'s gives $r^{11}d^{9L-20}$. Only $5L$ $d$'s are needed and the surplus fields can contract pairwise to give $r^{11}d^{5L}P^{2L-10}$. Therefore, the first terms of each vertex (see (\ref{Vertexymi}) and (\ref{inVSG}))   can contribute. The leading contribution for the single-trace amplitude arises from two contact terms. The double-trace part does not get contributions from multi-point vertices and its leading behaviour  arises by attaching four single-particle vertices. In the supergravity amplitude the presence of  contact terms does not change the behaviour of the amplitude in the ultraviolet. The results is
\eqnb
A^{(L)}_{F_L}(s,t)
	&\sim&
		\left.\left<\lambda^3D^{11}A^4\right>\right|_{\theta^5}\, \Lambda^{L\left(D-4\right)-8}
	\sim
		\partial^4 (\Tr F^2)^2\, \Lambda^{L\left(D-4\right)-8}\, ,
\\
A^{(L)}_{F_L}(s,t)
	&\sim&
		\left.\left<\lambda^3D^{11}\Tr\left(A^{nl}\right)^2\right>\right|_{\theta^5\hat{\theta}^{5}}\, \Lambda^{L\left(D-4\right)-6}
	\sim
		\partial^2 \Tr F^4\, \Lambda^{L\left(D-4\right)-6}\, ,
		\label{fiveplanar}
\\
A^{(L)}_{F_L}(s,t,u)
	&\sim&
		\left.\left<\lambda^3\hat{\lambda}^3D^{11}\hat{D}^{11}G^4\right>\right|_{\theta^5\hat{\theta}^{5}}\, \Lambda^{L\left(D-2\right)-14}
	\sim
		\partial^8\cR^4\, \Lambda^{L\left(D-2\right)-14}\, .
		 \label{fivegrav}
\eqne
Observe that the ultraviolet behaviour of the   Yang--Mills and supergravity cases are different.  For example, (\ref{fiveplanar}) shows that $D=4$, ${\cal N}=4$  Yang--Mills is ultraviolet finite to all orders in four dimensions.  However, from (\ref{fivegrav}) we see that $D=4$, ${\cal N}=8$ supergravity receives ultraviolet divergences at more than six loops, with the first such  divergence being  a logarithmic divergence at seven loops.

%%%%%%%%%%%%%%%%%%%%%%%%%
\section{Summary and discussion}
\label{sec:disc}
%%%%%%%%%%%%%%%%%%%%%%%%%

In this paper we have given details of, and  justification for,  the pure spinor world-line formalism  presented in  \cite{Bjornsson:2010wm}.  This provides a framework for constructing multi-loop amplitudes in supersymmetric Yang--Mills theory and supergravity that is manifestly supersymmetric.  The focus of this paper has been on the BRST consistency of the loop amplitudes.  

The $L$-loop amplitude naturally decomposes into a sum of contributions arising from distinct skeleton diagrams, which are $L$-loop vacuum diagrams of scalar field theory with cubic vertices and so have $3L-3$ internal propagators with lengths (or moduli) $T_j$ ($j=1, \dots 3L-3$). The amplitude for each skeleton is given by the expectation value of the product of vertices describing the scattering particles, which are attached to the lines of the skeleton and integrated over all positions around the diagram. In the first instance each vertex operator was modelled on the zero mode content of the string vertex operator, which describes the emission of a single on-shell gauge particle in Yang--Mills  or a graviton in supergravity.  

A key  feature that ensures BRST invariance is the  BRST covariance of  four-particle sub-diagrams formed by trees with four off-shell legs.  We found that this is guaranteed if all such sub-diagrams include all three channels (i.e. the $s$, $t$ and $u$ channel pole contributions).   Although this is the case for sub-diagrams containing purely internal vertices it is not the case for sub-diagrams that contain two external vertices.  One way to remedy this is to modify the rules to allow for external tree diagrams to be attached to the skeleton.  For example, a pair of external particles may couple each other via a three-particle vertex where the third leg is off-shell and is attached to the loop (see figure (\ref{fig:contactterms})).  We saw that this is equivalent to introducing a contact interaction vertex that couples the two external states directly to a leg of the skeleton.    BRST invariance uniquely determined the form of these contact vertices in a manner that is consistent with the nonlinear classical equations of motion.    Further higher-point contact terms are needed to ensure the higher-loop BRST invariance of amplitudes with arbitrary numbers of external legs, but such contact terms do not contribute to the four-point function for $L <3$ (but for general $n$-point function they contribute for $L\geq1$).  However, they  play an important r\^ole in determining the leading ultraviolet divergences in Yang--Mills amplitudes for $L\ge 3$.  They also contribute for $L\ge 3$ in the supergravity case although their presence does not affect the leading ultraviolet properties of the loop.

Interestingly,  the proof of BRST  invariance makes use of the colour factors and the kinematic factors in a related manner in  the Yang-Mills and supergravity cases, respectively. This is reminiscent of the conjectured duality between colour and kinematics in \cite{Bern:2008qj} (further elaborated in \cite{Bern:2010ue,Bern:2010yg}).

For completeness, we also  revisited and reviewed the analysis of the four-point amplitudes considered in \cite{Bjornsson:2010wm}.  We somewhat extended that discussion to demonstrate the vanishing of amplitudes with fewer than four on-shell particles and derive explicit expressions for the one- and two-loop amplitude with four external particles. For the one-loop case, we also gave an alternative proof of the ``no-triangle hypothesis'' of supergravity. 
One important result of the analysis of four-point amplitudes in \cite{Bjornsson:2010wm} is that the $\partial^8\cR^4$ term is not protected from receiving perturbative corrections from all loops. Indeed,  specific five-loop skeletons are identified as giving rise to this interaction.  This is in line with several  arguments \cite{Vanhove:2010aa,Green:2010sp,BerkovitsICTP,Bossard:2009mn,Elvang:2010jv,Elvang:2010kc,Drummond:2010fp,Beisert:2010jx,Bossard:2010bd,Howe:1980th} and indicates that there is a logarithmic divergence of the form $\partial^8\cR^4$ at seven loops in four dimensions.  

There are several avenues of interest to explore.  For example, it would be interesting to derive the results presented here and in \cite{Bjornsson:2010wm}, which were derived from a first quantised approach, from a second quantised field theory lagrangian formulation.   The rules for Yang--Mills can indeed be obtained from the second-quantised action in \cite{Berkovits:2001rb}. This has the form  $S_{YM} \sim \int \left(\frac{1}{2}\Phi Q \Phi + \frac{1}{3}\Phi^3\right)$ where the integration is over all superspace as well as pure spinor space.  From this one obtains a generating function $Z_{YM}[J] \sim e^{\frac{1}{3}\int \left(\frac{\delta}{\delta J}\right)^3}e^{ \frac{1}{2}\int J \frac{\cN b}{P^2}J}$, where $\cN$ is a regulator for the $b$ ghost. This partition function formally gives rise to the same diagrammatic rules and properties obtained for Yang--Mills in the first-quantised approach.  It is more difficult to obtain a second-quantised version of supergravity in the pure spinor formalism but the results of this paper suggest that the supergravity partition function has the form  $Z_{SG}[J] \sim e^{\frac{1}{3}\int \left(\frac{\delta}{\delta J}\right)^3}e^{ \frac{1}{2}\int J \frac{\cN\hat{\cN} b\hat{b}}{P^2}J}$.
This should make contact  with the recent progress in formulating a second-quantised formulation of pure spinor field theory for eleven-dimensional supergravity  \cite{Cederwall:2009ez,Cederwall:2010tn} where an action involving only cubic interactions has been obtained.  This could clarify the origin and r\^ole of the  composite $b$ ghost  \cite{Cederwall:2010,Cederwall:2010tk}.  

\acknowledgments
The author would like to thank Michael Green for many stimulating discussions, comments on the manuscript as well as collaborations on related subjects. Furthermore, the author would like to thank Martin Cederwall for stimulating discussions. Support from the Swedish Research Council under project no.\ 623-2008-7048 is also acknowledged.

%%%%%%%%%%%%%%%%%%%%%%%%%
\appendix

%%%%%%%%%%%%%%%%%%%%%%%%%
\section{Maximally supersymmetric Yang--Mills}
\label{sec:YM}
%%%%%%%%%%%%%%%%%%%%%%%%%

In this appendix we will summarise the field equations and theta expansions which follow from the on-shell constraint $F_{\alpha\beta}=0$ of supersymmetric Yang--Mills in ten dimensions. First we define the covariant derivatives
\eqnb
\nabla_{m} &=& \partial_m + A_m \\
\nabla_{\alpha} &=& D_{\alpha} + A_\alpha \\
D_{\alpha} &=& \partial_{\alpha} +\frac{1}{2}\left(\gamma^m\theta\right)_{\alpha}\partial_m\, ,
\label{covariantderiv}
\eqne
and the field strengths
\eqnb
F_{\alpha\beta} &=& [\nabla_{\alpha},\nabla_{\beta}] - \left(\gamma^m\right)_{\alpha\beta}\nabla_m 
\label{Fab}
\\
F_{\alpha m} &=& [\nabla_{\alpha},\nabla_{m}]
\label{Fam}
\\
F_{mn} &=& [\nabla_{m},\nabla_{n}]\, .
\label{Fmn}
\eqne
Observe that the fermionic super-derivatives satisfy
\eqnb
[D_{\alpha},D_{\beta}] &=& \left(\gamma^m\right)_{\alpha\beta}\partial_m\, .
\eqne
Furthermore, the field strength, $F_{mn}$ is related to the potential by
\eqnb
F_{mn} &=& \nabla_{m}A_n - \nabla_{n}A_m
\label{eqnFA}
\eqne

Important equations follow from the on-shell constraint $F_{\alpha\beta}=0$ and the various Jacobi identities. Combining  $F_{\alpha\beta}=0$ with (\ref{Fab}) gives
\eqnb
\nabla_{\alpha}A_\beta + \nabla_{\beta}A_\alpha &=& \left(\gamma^m\right)_{\alpha\beta}A_m\, .
\label{eqnAA}
\eqne
The Jacobi identity involving three fermionic covariant derivatives, $\nabla_{\alpha}$, gives $F_{\alpha m} \equiv \left(\gamma_mW\right)_{\alpha}$. From (\ref{Fam}) it follows that $\left(\gamma_m W\right)_{\alpha}$ satisfies the equation
\eqnb
\left(\gamma_mW\right)_{\alpha} &=& \nabla_{\alpha}A_m -\nabla_{m}A_{\alpha}\, .
\label{eqnAW}
\eqne
The Jacobi identity involving two $\nabla_m$ and one $\nabla_{\alpha}$ gives the equation
\eqnb
D_{\alpha}F_{mn} + [A_{\alpha},F_{mn}] = \left(\gamma_{m}\left(\partial_n W +[A_n,W]\right)\right)_{\alpha}-\left(\gamma_{n}\left(\partial_m W +[A_m,W]\right)\right)_{\alpha}\, .
\eqne
The Jacobi identity involving one $\nabla_m$ and two $\nabla_{\alpha}$'s results in the equation
\eqnb
\left(\gamma^{n}\right)_{\alpha\beta}F_{nm} 
	&=& 
		\left(\gamma_{m}\right)_{\alpha\gamma}\left(D_{\beta}W^{\gamma} + [A_\beta,W^{\gamma}]\right) 
	\no
	&+&
		\left(\gamma_{m}\right)_{\beta\gamma}\left(D_{\alpha}W^{\gamma} + [A_\alpha,W^{\gamma}]\right)\, .
\label{Feqn1}
\eqne

From (\ref{Feqn1}) one can extract equations for $W^{\alpha}$. Multiplying (\ref{Feqn1}) by  $\left(\gamma^{m}\right)^{\alpha\beta}$ gives
\eqnb
\left(D_{\alpha}W^{\alpha} + [A_\alpha,W^{\alpha}]\right) &=& 0\, .
\label{cond.W}
\eqne
Multiplying (\ref{Feqn1}) by $\left(\gamma^m\right)^{\beta\sigma}$ gives
\eqnb
{\left(\gamma^{mn}\right)_{\alpha}}^{\sigma}F_{mn} 
	&=& 
		10\left(D_{\beta}W^{\sigma} + [A_\beta,W^{\sigma}]\right) 
	\no
	&+&
		\left(\gamma^m\right)^{\beta\sigma}\left(\gamma_{m}\right)_{\beta\gamma}\left(D_{\alpha}W^{\gamma} + [A_\alpha,W^{\gamma}]\right)\, .
\label{Feqn2}
\eqne
Multiplying (\ref{Feqn2}) by $\left(\gamma^p\right)_{\sigma\kappa}\left(\gamma_p\right)^{\alpha\rho}$ and using various gamma matrix identities and (\ref{cond.W}) gives
\eqnb
-6{\left(\gamma^{mn}\right)_{\kappa}}^{\rho}F_{mn} 
	&=& 
		12\left(D_{\kappa}W^{\rho} + [A_\kappa,W^{\rho}]\right) 
	\no
	&+&
		6\left(\gamma^m\right)^{\alpha\rho}\left(\gamma_{m}\right)_{\sigma\rho}\left(D_{\alpha}W^{\sigma} + [A_\alpha,W^{\sigma}]\right)\, .
\label{Feqn3}
\eqne
Using (\ref{Feqn2}) and (\ref{Feqn3}) to cancel out the term involving $(\gamma^{m})^{\beta\sigma}(\gamma^{m})_{\beta\gamma}$ gives the equation
\eqnb
\left(D_{\alpha}W^{\beta} + [A_\alpha,W^{\beta}]\right) 
	&=&
		\frac{1}{4}{\left(\gamma^{mn}\right)_{\alpha}}^{\beta}F_{mn}\, .
\label{eqnWF}
\eqne

One can use the equations following from the on-shell constraint and the Jacobi identities to find the $\theta$-expansions of the various superfields.  Thus, from  (\ref{eqnFA}), (\ref{eqnAA}), (\ref{eqnAW}) and (\ref{eqnWF}) it follows that 
\eqnb
A_{\alpha} &=& \frac{1}{2}\left(\gamma^m\theta\right)_{\alpha}a_m + \frac{1}{3}\left(\gamma^m\theta\right)_{\alpha}\left(\theta\gamma_mw\right) - \frac{1}{32}\left(\gamma_p\theta\right)_{\alpha}\left(\theta\gamma^{mnp}\theta\right)f_{mn} + \ldots \no
A_m &=& a_m + \left(\theta\gamma_m w\right) - \frac{1}{8}\left(\theta\gamma_{mnp}\theta\right)f^{np} + \ldots \no
W^{\alpha} &=& w^{\alpha} + \frac{1}{4} \left(\theta\gamma^{mn}\right)^{\alpha}f_{mn} + \ldots \no
F_{mn} &=& f_{mn} - \left(\theta\gamma_{m}\left(\partial_{n}w +[a_n,w]\right)\right) + \left(\theta\gamma_{n}\left(\partial_{m}w +[a_m,w]\right)\right) + \ldots \, ,
\eqne
where we have used  the gauge freedom in the above expansion to set the $\theta$-independent piece of $A_{\alpha}$ to be zero. The field $f_{mn}$ is defined by $f_{mn} = \partial_m a_n -\partial_m a_n + [a_m,a_n]$.

%%%%%%%%%%%%%%%%%%%%%%%%%%%%%%%%%%%%%%%%%%%%%%%%%%

\end{document}